\documentclass[a4paper,10pt,prd,aps,preprintnumbers,nofootinbib,floatfix,superscriptaddress]{revtex4}

\usepackage{amssymb}
\usepackage{amsmath}
\usepackage{graphicx}
\usepackage{hyperref}
\usepackage{multirow}
\usepackage{orcidlink}
\usepackage{microtype}
\usepackage{slashed}
\usepackage{bbm}

\newcommand{\orcidauthorBENNETT}{0000-0002-1678-6701}
\newcommand{\orcidauthorDENGLER}{0000-0002-2305-8868}
\newcommand{\orcidauthorHONG}{0000-0002-3923-4184}
\newcommand{\orcidauthorHSIAO}{0000-0002-8522-5190}
\newcommand{\orcidauthorLEE}{0000-0002-4616-2422}
\newcommand{\orcidauthorLIN}{0000-0003-3743-0840}
\newcommand{\orcidauthorLUCINI}{0000-0001-8974-8266}
\newcommand{\orcidauthorMAAS}{0000-0002-4621-2151}
\newcommand{\orcidauthorPIAI}{0000-0002-2251-0111} 
\newcommand{\orcidauthorVADACCHINO}{0000-0002-5783-5602}
\newcommand{\orcidauthorZIERLER}{0000-0002-8670-4054}

\makeatletter
\def\@collaboration@present#1#2#3{%
 \par
 \begingroup
  \frontmatter@collaboration@above
  \@author@present{\ignorespaces#2\unskip}{#3}%
  \par
 \endgroup
 \set@listcomma@list#1%
}%
\makeatother

\begin{document}
\preprint{CTPU-PTC-26-13}
\preprint{TUM-EFT 218/26}

\title{Resonant scattering in two-flavored \texorpdfstring{$Sp(4)$}{Sp(4)} lattice gauge theories}

\author{Ed Bennett \orcidlink{\orcidauthorBENNETT}}
\email{e.j.bennett@swansea.ac.uk}
\affiliation{Centre for Quantum Fields and Gravity, Faculty  of Science and Engineering, Swansea University, Singleton Park, SA2 8PP, Swansea, United Kingdom}
\affiliation{Swansea Academy of Advanced Computing, Swansea University (Bay Campus), Fabian Way, Swansea SA1 8EN, United Kingdom}

\author{Yannick Dengler \orcidlink{\orcidauthorDENGLER}}
\email{yannick.dengler@uni-graz.at}
\affiliation{University of Graz, Universitätsplatz 5, 8010 Graz, Austria}

\author{Deog~Ki Hong \orcidlink{\orcidauthorHONG}}
\email{dkhong@pusan.ac.kr}
\affiliation{Department of Physics, Pusan National University, Busan 46241, Korea}
\affiliation{Extreme Physics Institute, Pusan National University, Busan 46241, Korea}

\author{Ho Hsiao \orcidlink{\orcidauthorHSIAO}}
\email{hohsiao@ccs.tsukuba.ac.jp}
\affiliation{Center for Computational Sciences, University of Tsukuba, Tsukuba, Ibaraki 305-8577, Japan}

\author{Jong-Wan Lee \orcidlink{\orcidauthorLEE}}
\email{j.w.lee@ibs.re.kr}
\affiliation{ Particle Theory  and Cosmology Group, Center for Theoretical Physics of the Universe, Institute for Basic Science (IBS), Daejeon, 34126, Korea }

\author{C.-J. David Lin \orcidlink{\orcidauthorLIN}}
\email{dlin@nycu.edu.tw}
\affiliation{Institute of Physics, National Yang Ming Chiao Tung University, 1001 Ta-Hsueh Road, Hsinchu 30010, Taiwan}
\affiliation{Centre for High Energy Physics, Chung-Yuan Christian University, Chung-Li 32023, Taiwan}
\affiliation{Physics Division, National Centre for Theoretical Sciences, Taipei 106319, Taiwan}

\author{Biagio Lucini \orcidlink{\orcidauthorLUCINI}}
\email{b.lucini@swansea.ac.uk}
\affiliation{Swansea Academy of Advanced Computing, Swansea University (Bay Campus), Fabian Way, Swansea SA1 8EN, United Kingdom}
\affiliation{Department of Mathematics, Faculty of Science and Engineering, Swansea University (Bay Campus), Fabian Way, SA1 8EN Swansea, United Kingdom}
\affiliation{School of Mathematical Sciences, Queen Mary University of London, Mile End Road, London, E1 4NS, UK}

\author{Axel Maas\,\orcidlink{\orcidauthorMAAS}}
\email{axel.maas@uni-graz.at}
\affiliation{University of Graz, Universitätsplatz 5, 8010 Graz, Austria}

\author{Maurizio Piai\,\orcidlink{\orcidauthorPIAI}}
\email{m.piai@swansea.ac.uk}
\affiliation{Centre for Quantum Fields and Gravity, Faculty  of Science and Engineering, Swansea University, Singleton Park, SA2 8PP, Swansea, United Kingdom}
\affiliation{Department of Physics, Faculty  of Science and Engineering, Swansea University, Singleton Park, SA2 8PP, Swansea, United Kingdom}

\author{Davide Vadacchino \orcidlink{\orcidauthorVADACCHINO}}
\email{davide.vadacchino@plymouth.ac.uk}
\affiliation{Centre for Mathematical Sciences, University of Plymouth, Plymouth, PL4 8AA, United Kingdom}

\author{Fabian Zierler \orcidlink{\orcidauthorZIERLER}}
\email{fabian.zierler@tum.de}
\affiliation{Technical University of Munich, TUM School of Natural Sciences, Physics Department, James-Franck-Straße 1, 85748 Garching, Germany}
\affiliation{Centre for Quantum Fields and Gravity, Faculty  of Science and Engineering, Swansea University, Singleton Park, SA2 8PP, Swansea, United Kingdom}
\affiliation{Department of Physics, Faculty  of Science and Engineering, Swansea University, Singleton Park, SA2 8PP, Swansea, United Kingdom}

\collaboration{(on behalf of the TELOS collaboration) \vspace{4pt}\\ 
\href{https://telos-collaboration.github.io}{ \includegraphics[height=1cm]{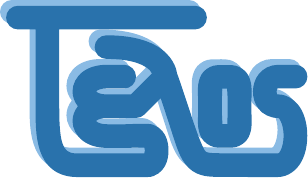} } }
\noaffiliation

\begin{abstract}

We apply L\"uscher's method to the vector channel of the scattering amplitude of Pseudo-Nambu-Goldstone-Bosons (PNGBs), in the  $Sp(4)$ lattice gauge theory coupled to $N_f=2$ flavors of Wilson-Dirac fundamental fermions. 
We generalize existing algorithms and numerical implementations of the method, to adapt them to this  prominent candidate for the completion of proposed extensions of the Standard Model (SM). We present the first ab initio measurements of key properties of the vector resonances in the theory, including the coupling to the PNGBs,  that are relevant to direct and indirect new physics searches, both for composite Higgs models (CHMs), as well as for strongly interacting massive particle (SIMP) realizations of dark matter.
We also present a global update of the spectroscopy of the mesons in the theory, improving both the statistics and analysis systematics in respect to previous lattice measurements reported in the literature.

\end{abstract}
\maketitle
\newpage
\tableofcontents

\section{Introduction}
\label{sec:introduction}

In a large class of strongly coupled field theories, which includes Quantum Chromodynamics (QCD), the lightest stable particles are Pseudo-Nambu-Goldstone-Bosons (PNGBs), arising from the spontaneous breaking of approximate continuous global symmetries. Other heavier composite states (resonances) decay into final states containing multiple PNGBs, through processes dictated by the conservation laws of the theory.  Modern lattice field theory technologies provide unprecedented opportunities to study these unstable bound states. The L\"uscher method, originally proposed in Ref.~\cite{Luscher:1986pf,Luscher:1990ux,Luscher:1991cf}, and further developed in Refs.~\cite{Rummukainen:1995vs,Kim:2005gf,Christ:2005gi,Bernard:2010fp,Luu:2011ep,Hansen:2012tf} and~\cite{Lellouch:2000pv, Lin:2001ek, Lin:2002nq, Leskovec:2012gb,Briceno:2012yi,Gockeler:2012yj,Briceno:2014oea,Briceno:2017tce,Romero-Lopez:2018zyy}, allows measuring the properties of such resonances as a function of the fundamental parameters of the theory, and of the kinematical regions of interest. Operationally, one first estimates the energy shift of multi-particle states in finite lattice volumes. Following, for example, Refs.~\cite{Guo:2012hv,Alexandrou:2017mpi}, one can do so by adopting a basis that includes states made of multiple PNGBs, and performing a variational analysis of the combined data sets.  One then uses this information to reconstruct the momentum dependence of PNGB scattering amplitudes in the infinite volume. Next,  from the PNGB scattering amplitude one extracts information about the resonances that appear as intermediate states, such as mass, width, and momentum dependence of their contribution to the scattering amplitude.  By modeling the latter (e.g., as a Breit-Wigner function), one can finally measure such quantities  as the coefficients of the low-energy effective field theory (EFT) description of the PNGBs, or, alternatively,  their effective coupling to the resonances.

Much attention has been devoted to applying this method to the study of p-wave, $(\pi\pi)$-scattering amplitudes in lattice QCD, targeting the $\rho$ meson, the lightest resonance carrying isospin. This isospin triplet has quantum numbers $I^G(J^{PC})=1^+(1^{--})$, mass $M_{\rho}\sim {\cal O}(775\,\,{\rm MeV})$, width $\Gamma_{\rho}\sim {\cal O}(150\,\,{\rm MeV})$, and it predominantly decays to final states with two PNGBs~\cite{ParticleDataGroup:2024cfk}. Lattice numerical results compare favorably with the experimental measurements---see for instance the reviews in Refs.~\cite{Briceno:2017max,Morningstar:2017spu,Lee:2021kfn,Mai:2022eur} and references therein, including for example Refs.~\cite{Feng:2010es,CS:2011vqf,Dudek:2012xn,Bali:2015gji,Bulava:2016mks,Fu:2016itp}.  Lattice systematic uncertainties, connected with taking the continuum and infinite volume limits while using realistic values of the quark masses, can be treated with established techniques.  Additional systematics affecting the numerical implementation of  L\"uscher's method can also be ameliorated. For the resolution of power-law and exponential  finite-volume effects, see for instance Refs.~\cite{Bedaque:2006yi,Albaladejo:2012jr,Romero-Lopez:2018rcb}; for discussions of subtleties in the choice of operator basis, see, e.g., Refs~\cite{Briceno:2017max, Wilson:2015dqa, Padmanath:2017oya}; and for the optimization of the range of applicability of the generalized Eigenvalue Problem (GEVP) analysis, see also Refs.~\cite{Feng:2010es}.

Motivated by its successes in lattice QCD,  the first results of the implementation of L\"uscher's method in other gauge theories, relevant to extensions of the Standard Model (SM) of particle physics, have appeared.  For example, lattice studies of PNGB scattering have been performed in the $SU(2)$ theory coupled to $N_f=2$ Dirac fermions transforming in the fundamental representation of the gauge group, focusing on both scalar and vector resonances~\cite{Drach:2020wux,Drach:2021uhl}, in the $N_f=8$, $SU(3)$ theory, focusing on scalar resonances~\cite{LatticeStrongDynamicsLSD:2021gmp},  as well as in the $N_f=2$, $Sp(4)$ theory~\cite{Dengler:2024maq} and other interesting theories~\cite{Arthur:2014zda, Baeza-Ballesteros:2022azb, DeGrand:2024lvp, Baeza-Ballesteros:2025iee,Gerhold:2011mx,Jenny:2022atm}, for non-resonant channels. In this paper, we present the first lattice measurements entering the analysis of the p-wave scattering amplitude in the aforementioned $Sp(4)$ theory. In the rest of this introduction, we motivate our choice of theory and observables, contextualizing it as part of the ongoing lattice explorations of strongly coupled candidates for new physics.
 
Despite the success of the Standard Model as a predictive theory of elementary interactions, on both theoretical and phenomenological grounds it is expected to be incomplete. 
Among the former arguments, the fact that its renormalization group flow does not approach a 
known fixed point at short distances (see, e.g., Refs.~\cite{Aizenman:2019yuo,Luscher:1987ay,Luscher:1987ek, Luscher:1988uq,Bulava:2012rb, Molgaard:2014mqa, Chu:2018ldw}) suggests the existence of a cutoff scale, $\Lambda$, above which the Standard Model is completed by a more fundamental theory. On phenomenological grounds, this new theory should fill current gaps in accounting for observational evidence. For example, it should explain the hierarchy of scales that appear in the SM electroweak, Higgs and fermion sectors, provide a candidate for dark matter (see, e.g., the review in Ref.~\cite{Cirelli:2024ssz}, and its extensive bibliography), and yield a phase transition in the early universe strong enough to meet the out-of-equilibrium condition required for baryogenesis to account for the observed matter-antimatter asymmetry in the universe~\cite{Sakharov:1967dj}---a property not realized by the Standard Model~\cite{Kajantie:1996mn, Laine:2012jy}.

The paradigm of compositeness is a promising way to achieve these goals, as it provides a short-distance completion leading to new phenomena, while naturally preserving the SM long-distance physics and its successes. It requires postulating the existence of a new fundamental theory, dynamically explaining the formation, at scale $\Lambda$ and below, of composite particles and effective interactions that one  identifies with the SM one---see the reviews in Refs.~\cite{Panico:2015jxa,Witzel:2019jbe,Cacciapaglia:2020kgq}, and references therein.

The $Sp(2N)$ gauge theories (with $N>1$), coupled to an admixture of fermion matter fields transforming as the fundamental or two-index antisymmetric representations of the gauge group, play a central role in the context of  composite models of new physics---see the review in Ref.~\cite{Bennett:2023wjw}, and references therein. They provide  a rich and diverse class of calculable theories, acting as realistic case studies for many of the aforementioned scenarios. They can complete composite Higgs Models (CHMs)~\cite{Kaplan:1983fs,Georgi:1984af,Dugan:1984hq} (such as those in Refs.~\cite{Barnard:2013zea,Ferretti:2013kya,Ferretti:2016upr,Cacciapaglia:2019bqz} and references therein) while also implementing flavor physics within the top partial compositeness (TPC) paradigm~\cite{Kaplan:1991dc} (see also Refs.~\cite{Grossman:1999ra,Gherghetta:2000qt,Chacko:2012sy}). They provide dark matter candidates in the form of strongly interacting massive particles (SIMPs), as discussed for example in Refs.~\cite{Hochberg:2014dra,Hochberg:2014kqa,Hochberg:2015vrg,Bernal:2017mqb,Berlin:2018tvf,
Bernal:2019uqr,Tsai:2020vpi,Kondo:2022lgg,Chu:2024rrv,Chu:2025hga}---see also Refs.~\cite{Arthur:2016ozw,Maas:2021gbf,Zierler:2021cfa,Kulkarni:2022bvh,Bennett:2023rsl,Bennett:2024wda,Pomper:2024otb,Appelquist:2024koa,Kolesova:2025ghl}. 
They can give rise, in the early universe, to new phase transitions, and an associated,  detectable stochastic background of relic gravitational waves (GWs)~\cite{Witten:1984rs,Kamionkowski:1993fg, Allen:1996vm,Schwaller:2015tja, Croon:2018erz, Christensen:2018iqi, Caprini:2019egz,Maggiore:2019uih}---see Refs.~\cite{Huang:2020crf, Halverson:2020xpg, Kang:2021epo, Reichert:2021cvs, Reichert:2022naa, Pasechnik:2023hwv}, as well as Refs.~\cite{Holland:2003kg,Bruno:2024dha} for studies of the Yang-Mills $Sp(2N)$ theories at finite temperature.

To take advantage of these possibilities, it is necessary to gain quantitative control over the underlying strong-coupling  dynamics. To this purpose, the TELOS collaboration has been pursuing an extensive programme of studies of the $Sp(2N)$ gauge theories on the lattice~\cite{Bennett:2017kga,Lee:2018ztv, Bennett:2019jzz, Bennett:2019cxd, Bennett:2020hqd, Bennett:2020qtj, Bennett:2022yfa, Bennett:2022gdz, Bennett:2022ftz, Bennett:2023wjw, Bennett:2023gbe, Bennett:2023mhh, Bennett:2023qwx, Bennett:2024cqv, Bennett:2024wda, Bennett:2024tex,  TELOS:2025ash, Bennett:2025whm}---see also Refs.~\cite{Hong:2017suj,Kulkarni:2022bvh,Bennett:2023rsl, Dengler:2024maq, Bennett:2024bhy}. These studies provide non-perturbative information on a plethora of  observables relevant for model-building and phenomenological purposes, that include measurements of masses, decay constants, and matrix elements of bound states, as well as topological observables, and spectral densities.\footnote{Holography (gauge-gravity dualities) can also be used as a complementary tool to compute the spectrum of bound states of gauge theories coupled to matter fields that realize global symmetry patterns relevant to the CHM and SIMP paradigm---see for example Refs.~\cite{Erdmenger:2020lvq,Erdmenger:2020flu,Elander:2020nyd,Elander:2021bmt, Erdmenger:2023hkl,Elander:2023aow, Elander:2024lir,Erdmenger:2024dxf,Alfano:2024aek,Alfano:2025dch,Alfano:2025non,Alfano:2025ahq} and references therein---or to characterize finite-temperature first-order phase transitions and bubble dynamics in strongly coupled field theories---see, e.g., Refs.~\cite{Bigazzi:2020phm,Ares:2020lbt,Bea:2021zsu,
Bigazzi:2021ucw,Henriksson:2021zei,Ares:2021ntv,Ares:2021nap,Morgante:2022zvc,Bea:2021zol,
 Bea:2022mfb,Bea:2024xgv,Bea:2024bxu,Bea:2024bls}.
} The regime of fermion masses explored so far is such that the ground state particles are stable, as their decays to PNGBs are kinematically forbidden. Lowering the fermion mass and gathering  information about  the width and coupling of the bound states, from the relevant PNGB scattering amplitudes, is the natural next step  of this programme. The vector resonances are the lightest bound states, besides the PNGBs. They can be treated in EFT terms using  the language of Hidden Local Symmetry (HLS)~\cite{Bando:1984ej,Casalbuoni:1985kq,Bando:1987br,Casalbuoni:1988xm,Harada:2003jx}  (see also Refs~\cite{Georgi:1989xy,Appelquist:1999dq,Piai:2004yb,Piai:2010ma}, and~\cite{BuarqueFranzosi:2016ooy} in the CHM context). The couplings appearing in the HLS-EFT Lagrangian density have to be determined non-perturbatively within the underlying theory. In the CHM context, these couplings control the contribution of vector resonances  to electroweak precision observables, but also to the (total and partial) production and decay rates, relevant in collider physics.

The $Sp(4)$ theory coupled to $N_f=2$ Dirac fermions transforming in the fundamental representation provides also a suitable candidate for realizing the SIMP paradigm~\cite{Kulkarni:2022bvh}. The effective interactions among its bound states are important in our interpretation of the dark matter distribution in galaxies~\cite{Kaplinghat:2015aga}, and the existence and properties of its resonances determine our understanding of early-universe evolution~\cite{Bernreuther:2023kcg,Chu:2025hga,Chu:2024rrv}. The presence of resonances in non-scalar channels, especially the spin-1 states~\cite{Bernreuther:2023kcg}, in the range of parameter space in proximity of the kinematical  threshold for decay, could provide the necessary non-linear behaviour in the non-relativistic regime that is the subject of current observational astronomy~\cite{Kaplinghat:2015aga}. As we shall see later in the paper, our results show that in this theory we can tune the parameters so as to move a resonance across the threshold, depending on the dark matter mass, in a region of parameter space that is favored experimentally~\cite{Hochberg:2014kqa,Tsai:2020vpi,Kulkarni:2022bvh,Dengler:2024maq}. This mechanism could be decisive for making models of this class phenomenologically viable.

As we shall see, the $Sp(4)$, $N_f=2$  theory resembles QCD in essential qualitative respects, which makes it possible to use much of the existing lattice technology. Yet, the meson spectrum is classified in terms of a different, enhanced non-Abelian global symmetry coset, $SU(4)/Sp(4)$, and hence the basis of multi-particle states used in the scattering analysis is different. Furthermore, the range of parameters of the theory taken in isolation (before coupling to the other SM fields) that is of phenomenological relevance is complementary to the QCD one, as in both the CHM and SIMP contexts one expects the vector meson mass to be either below or in proximity of the decay threshold, as explicit symmetry breaking effects (the fermion mass) are not strongly suppressed.\footnote{The hierarchy between the scale of the resonances and the electroweak scale arises through vacuum misalignment \cite{Peskin:1980gc,Preskill:1980mz}, and is triggered by the interplay of the SM couplings with the underlying composite dynamics.} The study we perform therefore requires the development of lattice algorithms and their numerical implementation for a different theory, with distinct states and couplings, as well as their optimization for use in different kinematical and dynamical regimes. 

The paper is organized as follows. In Sect.~\ref{sec:glob_sym_and_scat_chan}, we introduce the continuum field theory of interest, describe its symmetry properties,  and define the observables of interest. We expand this last aspect in Sect.~\ref{sec:spectroscopy}, by specifying the observables and our analysis procedure for the correlation functions and L\"uscher's method in its lattice field theory implementation.
Section~\ref{Sec:spectra} provides an additional important set of results of the paper, in the form of  updated results of the spectroscopy of flavored mesons obtained with the  implementation of this theory on the lattice by means of (unimproved) Wilson fermions, with fermion masses large enough that the mesons of interest are all stable. The measurements we provide here, and the continuum extrapolation we perform, represent a major improvement in both statistics and analysis  strategy, in respect to the existing literature~\cite{Bennett:2019jzz}. We envision the adoption of alternative lattice formulations to be necessary in order to significantly improve upon these results in future studies. In Sec.~\ref{sec:results}, we present our main results, obtained from the analysis of the phase shifts leading to the reconstruction of the PNGB scattering amplitudes, for a restricted set of new ensembles in the low mass regimes. Some preliminary results are available in Ref.~\cite{TELOS:2026mim}. We devote the short Sect.~\ref{Sec:SIMP} to a discussion of the potential implications of our results for SIMP dark matter models. We conclude with a summary and outlook section, in Sect.~\ref{Sec:outlook}. We relegate to the Appendix some technical details, as well as tables and plots of intermediate numerical results that are used in the analysis presented in the main body of this work.

\section{Continuum action, global symmetries and scattering channels}
\label{sec:glob_sym_and_scat_chan}

We study the $Sp(4)$ gauge theory coupled to $N_f=2$ Dirac fermions, $Q^I$, transforming in the fundamental representation of the gauge group, with $I=1,\,2$. The Lagrangian density in Minkowski space-time in the continuum is given by the following expression
\begin{align}
    \mathcal L = -\frac{1}{2} {\rm Tr} \left[ G_{\mu\nu} \frac{}{}G^{\mu\nu} \right] + \sum_{I=1}^{N_f} \overline Q^I  \left( i \gamma^{\mu}D_{\mu} - m^{f} \right) Q^I\,, 
\end{align}
where  $m^f$ is a (degenerate) mass for the fermions, $\gamma^{\mu}$ are the gamma matrices,  and $D_{\mu}$ is the covariant derivative acting on $Q$. We omit color and spinor indices, that are  summed over. 

Besides the Lorentz and gauge symmetries, this Lagrangian density has global symmetries acting on the flavor indices of the fermions. Because of the pseudo-real nature of the fundamental representation, this symmetry is enhanced to $U(1)_A\times SU(2N_f=4)$. The chiral anomaly,  together with the mass term, breaks the $U(1)_A$ explicitly. The mass also  breaks $SU(4)$ to its $Sp(4)$ subgroup. These properties are common to all 
$Sp(2N)$ gauge theories with fundamental matter field content. For lattice studies of $Sp(2)\sim SU(2)$ with $N_f=2$, see for example Refs.~\cite{Lewis:2011zb,Detmold:2014kba,Hietanen:2013fya,Hietanen:2014xca,Arthur:2016dir,Arthur:2016ozw,Pica:2016zst,Lee:2017uvl,Drach:2017btk,Drach:2020wux,Drach:2021uhl,Bowes:2023ihh}.

The continuous global symmetries of the Lagrangian density can be made explicit by rewriting it in terms of two-component spinors, $q^i$, with $i=1,\,\cdots,\,4$~\cite{Kogut:2000ek,Lewis:2011zb,vonSmekal:2012vx,Francis:2018xjd,Bennett:2019cxd,Kulkarni:2022bvh}.
To this purpose, it is useful to write explicitly the two 4-component spinors in the form
\begin{align}
    Q^{I\,a} = \left(\begin{array}{c}
    q^{I\,a}\cr
    -\Omega^{ab}\left(\tilde{C}q^{I+2\,\ast}\right)_b
    \end{array}
    \right)\,,
\end{align}
where we now make explicit reference to the color indices, $a,\,b=1,\,\cdots,\,4$, and wrote the charge conjugation matrix in terms of the second Pauli matrix, as $\tilde{C}\equiv-i\tau^2$, introducing the symplectic matrix, $\Omega$, that can be written in terms of $2 \times 2$ matrix blocks as follows:
\begin{align}
    \Omega &\equiv  
    \begin{pmatrix}
        \mathbbm{O}_{2\times 2}  & \mathbbm{1}_{2\times 2} \\
        -\mathbbm{1}_{2\times 2} & \mathbbm{O}_{2\times 2}
    \end{pmatrix}.
\end{align}

After dropping a harmless total derivative, that does not appear in the equations of motion, and suppressing spinor and color indices that are summed over, the Lagrangian density can then be rewritten as
\begin{align}\label{eq:lagrangian_nambu_gorkov}
    \mathcal L = -\frac{1}{2} {\rm Tr} \left[ G_{\mu\nu} \frac{}{} G^{\mu\nu} \right] + 
    \frac{1}{2}\sum_{j=1}^{4}\left[i q^{j\,\dagger} \bar{\sigma}^{\mu} D_{\mu} q^j
    \,-\, i \left(D_{\mu}q^{j}\right)^{\dagger} \bar{\sigma}^{\mu} q^j\right]
    -\frac{1}{2}m_{f}\sum_{j,k=1}^{4}\Omega_{jk}\left[q^{j\,T}\Omega \tilde{C} q^k\,-\, q^{j\,\dagger} \Omega \tilde{C} q^{k\,\ast}\right]\,.
\end{align}
In this expression we see the appearance of a second, $4\times 4$, symplectic matrix, $\Omega_{jk}$, which  happens to be identical to $\Omega$, but acts on the global (flavor) symmetry space, rather than the gauge (color) one. The reader should try to keep them separated, despite the accidental fact that the exact global and gauge symmetry coincide, $Sp(2N)=Sp(4)=Sp(2N_f)$.
The $2\times 2$ matrices in spinor space, $\bar \sigma_\mu = \left(\mathbbm{1}_2, \tau^i \right)$, are generalized Pauli matrices, with $\tau^i$ being the three  Pauli matrices. The Lagrangian density written in the form in Eq.~\eqref{eq:lagrangian_nambu_gorkov} displays explicitly the property  that the kinetic part is invariant under the global $U(1)_A\times SU(4)$ global symmetry, and that the mass term explicitly breaks this symmetry down to the global $Sp(4)$ subgroup. In the absence of a mass term, one expects the dynamics to  spontaneously break the anomaly free $SU(4)$  global symmetry to $Sp(4)$, and the states of the theory to fill $Sp(4)$ irreducible representations.

\begin{table}
\caption{Labels and quantum numbers of the 
(flavored) meson operators, $\mathcal{O}_M$, and the associated lightest particles with spin $0$ and $1$, in the $Sp(4)$ gauge theory coupled to $N_f=2$ Dirac fermions transforming in the fundamental representation---for details, see also Refs.~\cite{Bennett:2019jzz,Bennett:2019cxd,TELOS:2025ash}. 
The fermions are degenerate in mass,
so that the non-anomalous, approximate, global $SU(4)$ symmetry breaks to its $Sp(4)$ subgroup---a more general classification of mesons can be found in Ref.~\cite{Bennett:2023rsl} and references therein.
Besides their schematic  structure (color and spinor indices are summed over and omitted), for each operator we report the space-time quantum number, $J^P$, and the dimension of the representation under the global  $Sp(4)$ symmetry (see also Ref.~\cite{Lewis:2011zb}).
We restrict attention to operators with (Dirac) flavor indices $I\neq J$, 
to isolate the non-trivial $Sp(4)$ representations and 
exclude the possibility of contamination in the correlation functions coming from flavor singlets.
To provide guidance to the reader,  for each operator  we name the corresponding particle
in the QCD classification of mesons, although the global symmetry pattern, and consequently the dimension of the multiplets, are different.
Two of the operators source the same particles (the $\rho$ meson).
}
\label{tab:spec_operators}
\begin{center}
\begin{tabular}{|c|c|c|c|c|}
\hline\hline
Label ($M$) & Interpolating operator ($\mathcal{O}_M$) & Meson & $J^{P}$
& $Sp(4)$ \cr
\hline
PS & $\overline{Q^I}\gamma_5 Q^J$ & $\pi$ & $0^{-}$ & $5$ \cr
S & $\overline{Q^I} Q^J$ & $a_0$ & $0^{+}$ & $5$ \cr
V & $\overline{Q^I}\gamma_\mu Q^J$ & $\rho$ & $1^{-}$ & $10$ \cr
T & $\overline{Q^I}\gamma_0\gamma_\mu Q^J$ & $\rho$ & $1^{-}$ & $10$  \cr
AV & $\overline{Q^I}\gamma_5\gamma_\mu Q^J$ & $a_1$ & $1^{+}$ & $5$ \cr
AT & $\overline{Q^I}\gamma_5\gamma_0\gamma_\mu Q^J$ & $b_1$ & $1^{+}$ & $10$ \cr
\hline\hline
\end{tabular}
\end{center}
\end{table}

\subsection{Mesons and scattering channels}

We focus our attention on flavored mesons. The global $SU(4)$ symmetry breaks to $Sp(4)$ both explicitly and spontaneously, due to the formation of fermion condensates. Hence, the spectrum governing long distance physics should contain five PNGBs. We summarize the meson operators sourcing flavored spin-0 and spin-1 states in Table~\ref{tab:spec_operators}, distinguishing them with labels, $\rm M = PS,\, S,\, V,\, T,\, AV,\, AT$, that refer to their gamma-matrix structure.  The flavor multiplets of mesons can be inferred from the fact that the fermions transform as the $\textbf{4}$ of the global $Sp(4)$ symmetry group. The decomposition of the tensor product of irreducible representations of the fermions is~\cite{Drach:2017btk,Bennett:2017kga,Feger:2019tvk} 
\begin{equation}
    \textbf{4}\otimes\textbf{4} = \textbf{1}\oplus\textbf{5}\oplus\textbf{10}\,.
\end{equation}
The PNGBs are sourced by the operator  ${\cal O}_{\rm PS}$, transforming as a $\textbf{5}$. The next-to-lightest mesons are expected to be the spin-1 states transforming as $\textbf{10}$ of $Sp(4)$,   sourced by  meson operators ${\cal O}_{\rm V}$ and ${\cal O}_{\rm T}$~\cite{Barnard:2013zea,Ferretti:2013kya,Drach:2017btk,Bennett:2017kga,Kulkarni:2022bvh}.\footnote{Both a scalar and pseudoscalar meson exist as singlets of $Sp(4)$. The spectrum of the pseudoscalar singlet  has been studied in Refs.~\cite{Bennett:2023rsl}, which, for the scalar singlet, reported systematic lattice artifacts to be sizable. By contrast, there are no vector meson singlets built of  just two fermions in this theory~\cite{Bennett:2023rsl}. For examples, the equivalent of the $\omega$ meson in QCD is part of the $\textbf{10}$ of $Sp(4)$. These relations can be understood in terms of the enhancement of the global symmetry  $U(1)_A\times U(1)_B  \times SU(2)_L\times SU(2)_R \subset  U(1)_A \times SU(4) $. } For completeness, we tabulate also the flavored meson operators, ${\cal O}_{\rm S}$, ${\cal O}_{\rm AV}$, and ${\cal O}_{\rm AT}$, albeit they play a limited role in this investigation---see also the detailed studies in Refs.~\cite{Bennett:2019jzz,Kulkarni:2022bvh}.

The  $g_{\rm VPP}$ coupling between one vector meson (${\rm V}\sim \textbf{10}$) and two PNGBs (${\rm PS}\sim\textbf{5}$)~\cite{Janowski:2019svg,Drach:2020wux} provides an important contribution to the scattering of two  PNGBs, as can be seen from the decomposition into  $Sp(4)$ irreducible representations of the product:
\begin{align}\label{eq:two-pion_multiplets}
    \textbf{5}\otimes\textbf{5} = \textbf{1}\oplus\textbf{10}\oplus\textbf{14}\,.
\end{align}
All resulting channels have been analyzed on the lattice for the closely related $SU(2)$ gauge theory with $N_f=2$~\cite{Arthur:2014zda,Drach:2017btk, Feger:2019tvk, Drach:2020wux, Drach:2021uhl}. Earlier work on the $Sp(4)$, $N_f=2$ theory focused on the $\textbf{14}$ channel (corresponding to the isospin-2 channel in QCD)~\cite{Dengler:2024maq, Dengler:2025ulb, Dengler:2023szi}. The scattering amplitudes and scattering lengths have been calculated within chiral perturbation theory in Ref.~\cite{Bijnens:2011fm}. The flavor-singlet channel, $\textbf{1}$, is affected by technical challenges and high computational costs, and we leave it to future studies. We focus this work  on the \textbf{10}, and the vector resonances appearing in this channel.

This theory provides a prominent microscopic realization of the SIMP paradigm for dark matter models, that rely on number-lowering processes in the dark sector~\cite{Hochberg:2014dra, Hochberg:2014kqa}. Given the pseudo-real nature of the fundamental representation, and the consequent enhancement of the global symmetries with $N_f=2$ flavors, the Wess-Zumino-Witten term in chiral perturbation theory provides a natural origin for  $3\to 2$, number-lowering,  interactions~\cite{Hochberg:2014dra,Hochberg:2014kqa,Kulkarni:2022bvh}.

By examining the $Sp(4)$ multiplets entering the product of  three PNGBs, which are~\cite{Feger:2019tvk}
\begin{align}\label{eq:three-pion_multiplets}
\begin{split}
   \textbf{5}\otimes\textbf{5} \otimes\textbf{5} &= (\textbf{1} \oplus\textbf{10} \oplus\textbf{14} )\otimes\textbf{5}  \\
   &= \textbf{5} \oplus (\textbf{5}\oplus\textbf{10}\oplus\textbf{35} )\oplus (\textbf{5}\oplus\textbf{30}\oplus\textbf{35})    \\
   &=( 3\times \textbf{5}) \oplus \textbf{10} \oplus \textbf{30} \oplus (2 \times \textbf{35})\,,
\end{split}
\end{align}
one sees that the only final state that can be connected to the product $\textbf{5}\otimes\textbf{5}$, to realize the $3\to 2$ process between PNGBs, involves the $\textbf{10}$, the channel that contains the vector meson, ${\rm V}$. The lattice investigation into the vector meson channel of the PNGB scattering amplitude  is hence necessary to set the stage for future studies of the $3\to 2$ multi-particle scattering process. We notice that the analysis of the $3 \to 2$ process would require the use of the three particle quantization condition~\cite{Hansen:2014eka,Hansen:2019nir,Briceno:2017max,Dawid:2025doq}, but this is beyond the aims of this work.

We use L\"uscher's formalism to relate the energy levels of a multi-particle system in a finite box to the scattering phase shift in an infinite volume \cite{Luscher:1985dn,Luscher:1986pf,Luscher:1990ux}. In particular, we use the extension to p-wave scattering that is required for the vector meson channel \cite{Rummukainen:1995vs}---see also Ref.~\cite{Prelovsek:2011nk} for a brief introduction to the scattering formalism on lattice.
The key input to this formalism is the measurement of the energy levels obtained with a lattice having finite volume.  We perform these calculations by a variational analysis of multiple interpolating operators in the same channel, to include the scattering states. Given a basis consisting of $n$ different states,  we can extract up to  $n$ eigenvalues of the energy in the given channel. We can supplement this measurement with information obtained with external momenta, to access additional energy levels that can  be used in the scattering analysis. We detail our determination of the energy levels in Sect.~\ref{sec:spectroscopy}
Once we have determined all accessible energy levels for a given ensemble, we can reconstruct the scattering phase shift. By fitting the discrete results for the phase shift to known analytic forms, we can then ascertain whether resonances are present. 
We provide details and numerical results in Sect.~\ref{sec:results}.

\section{Spectroscopy and scattering on the lattice}
\label{sec:spectroscopy}

\begin{table}[t]
    \centering
    \caption{Ensembles used throughout this work. The volume in four space-time dimensions is denoted as $\tilde{V}=(aN_t)\times(aN_s)^3$, the inverse gauge coupling as $\beta$, and the bare fermion mass of the Wilson fermions, expressed in lattice units,  as $am_0$.  In the body of the paper, we refer to the three sets of ensembles sharing the same lattice parameters, but for the different spatial volume, as heavy, medium, and light, respectively. While we reuse the ensembles with $\beta=6.9$ generated for previous works~\cite{Bennett:2019jzz, Kulkarni:2022bvh,Bennett:2023rsl,Dengler:2024maq}, we add to them the ensembles with $\beta=7.05$, obtained by adopting the Hasenbusch mass-splitting technique implemented in the HiRep code~\cite{Bussone:2018mzi}. 
    For each ensemble, we explicitly report the number of configurations, $N_{\rm conf}$. We further report the number of trajectories skipped, $N_{\rm skip}$, between adjacent configurations,  as well as the average plaquette, $\langle P \rangle$, and topological charge, $Q$. We also report the respective residual integrated autocorrelation times, expressed in units of the configurations retained in the ensembles for the physical analysis. The autocorrelation time is defined in Appendix \ref{app:ensembles}, and we notice that it is small but non-vanishing, ${\cal O}(1\div3)$. We remove potential residual autocorrelations from Lüscher's analysis by binning the correlation matrix with a bin size of two before we extract the energy levels.}
    \begin{tabular}{|l|c|c|c|c|c|c|c|c|c|c|}
    \hline \hline
    label & $\beta$ & $-am_0$ & $N_t$ & $N_s$ & $N_{\rm config.}$ & $N_{\rm skip}$ & $\langle P \rangle$ & $Q$ & $\tau^{\langle P \rangle}$ & $\tau^{Q}$ \\
	\hline\hline
	light & 6.9 & 0.85 & 32 & 16 & 100 & 24 & 0.546753(52) & 1.06(96) & 2.0(1.0) & 1.48(36) \\
	light & 6.9 & 0.87 & 32 & 16 & 100 & 24 & 0.550525(61) & 0.98(85) & 1.86(87) & 1.8(1.2) \\
	light & 6.9 & 0.89 & 32 & 16 & 100 & 24 & 0.554785(58) & -3.12(86) & 2.6(1.4) & 1.45(36) \\
	light & 6.9 & 0.9 & 32 & 16 & 75 & 32 & 0.556961(74) & -0.38(81) & 2.2(1.4) & 1.57(57) \\
	light & 6.9 & 0.91 & 32 & 16 & 435 & 30 & 0.559353(27) & -0.20(39) & 3.1(1.3) & 2.2(1.3) \\
	\hline\hline
	heavy & 6.9 & 0.92 & 32 & 16 & 288 & 20 & 0.562073(34) & 0.43(38) & 2.4(1.1) & 1.93(92) \\
	heavy & 6.9 & 0.92 & 32 & 20 & 360 & 40 & 0.562042(22) & -0.50(47) & 2.5(1.2) & 1.50(19) \\
	heavy & 6.9 & 0.92 & 32 & 24 & 588 & 28 & 0.562072(12) & -0.80(50) & 2.5(1.1) & 1.58(51) \\
	\hline\hline
	light & 6.9 & 0.924 & 32 & 24 & 782 & 12 & 0.563217(11) & 2.47(43) & 7.5(4.0) & 2.18(43) \\
	light & 7.05 & 0.835 & 36 & 20 & 100 & 20 & 0.575269(29) & -0.11(76) & 2.0(1.2) & 2.1(1.1) \\
	light & 7.05 & 0.85 & 36 & 24 & 100 & 24 & 0.577371(24) & -0.10(87) & 2.00(94) & 2.3(1.5) \\
	light & 7.05 & 0.857 & 36 & 32 & 210 & 20 & 0.5783099(99) & -2.34(81) & 2.1(1.0) & 1.66(36) \\
	\hline\hline
	medium & 7.05 & 0.863 & 36 & 16 & 386 & 28 & 0.579321(21) & 0.32(19) & 1.89(49) & 1.8(1.1) \\
	medium & 7.05 & 0.863 & 36 & 20 & 372 & 20 & 0.579317(15) & -0.19(26) & 2.8(1.5) & 2.0(1.2) \\
	medium & 7.05 & 0.863 & 36 & 24 & 344 & 20 & 0.579245(12) & -0.75(34) & 2.06(61) & 1.8(1.1) \\
	medium & 7.05 & 0.863 & 36 & 36 & 298 & 20 & 0.5792137(74) & -0.21(74) & 3.1(2.0) & 2.2(1.2) \\
	\hline\hline
	light & 7.05 & 0.867 & 36 & 16 & 365 & 20 & 0.580074(22) & -0.12(16) & 3.2(1.6) & 1.79(44) \\
	light & 7.05 & 0.867 & 36 & 24 & 251 & 28 & 0.579940(14) & 0.53(36) & 2.17(80) & 1.7(1.0) \\
	light & 7.05 & 0.867 & 36 & 36 & 361 & 24 & 0.5798251(60) & -0.79(60) & 2.8(1.8) & 1.8(1.1) \\
	light & 7.2 & 0.76 & 36 & 16 & 200 & 12 & 0.587666(25) & 0.08(32) & 1.81(70) & 3.6(1.6) \\
	light & 7.2 & 0.77 & 36 & 24 & 200 & 12 & 0.588460(12) & 1.20(45) & 2.2(1.4) & 3.5(2.1) \\
	light & 7.2 & 0.78 & 36 & 24 & 508 & 12 & 0.5892779(85) & 1.54(32) & 1.79(65) & 3.6(2.1) \\
	light & 7.2 & 0.79 & 36 & 24 & 500 & 12 & 0.5901269(86) & 0.30(29) & 3.2(2.0) & 3.8(1.4) \\
	light & 7.2 & 0.794 & 36 & 28 & 504 & 12 & 0.5904516(67) & 0.30(32) & 2.6(1.2) & 4.0(2.4) \\
	light & 7.2 & 0.799 & 40 & 32 & 451 & 12 & 0.5908623(53) & -0.43(39) & 2.15(72) & 3.4(2.2) \\
	light & 7.2 & 0.803 & 42 & 36 & 266 & 12 & 0.5912395(63) & 1.01(59) & 3.2(2.0) & 4.3(2.0) \\
	light & 7.2 & 0.808 & 48 & 24 & 400 & 12 & 0.5917041(80) & 0.53(23) & 1.97(83) & 3.4(1.4) \\
	light & 7.2 & 0.808 & 48 & 36 & 300 & 16 & 0.5916970(56) & 2.45(45) & 3.9(2.3) & 2.6(1.2) \\
	light & 7.2 & 0.808 & 48 & 42 & 334 & 12 & 0.5916657(38) & -1.55(60) & 3.6(2.3) & 3.4(1.9) \\
	light & 7.2 & 0.813 & 48 & 24 & 325 & 16 & 0.5921669(92) & -0.26(23) & 3.4(1.8) & 2.33(55) \\
	light & 7.2 & 0.813 & 48 & 36 & 500 & 12 & 0.5921347(38) & 0.16(34) & 2.7(1.7) & 3.3(1.6) \\
	light & 7.4 & 0.74 & 48 & 16 & 660 & 8 & 0.606112(11) & -0.335(65) & 2.49(94) & 16.1(8.7) \\
	light & 7.4 & 0.74 & 48 & 24 & 574 & 8 & 0.6060586(58) & -0.54(16) & 1.91(67) & 12.0(5.6) \\
	light & 7.4 & 0.74 & 48 & 36 & 342 & 8 & 0.6060474(44) & 1.24(38) & 1.54(42) & 17(11) \\
	light & 7.4 & 0.75 & 48 & 24 & 501 & 8 & 0.6066056(66) & 0.41(11) & 2.1(1.3) & 13.4(8.5) \\
	light & 7.4 & 0.75 & 48 & 32 & 800 & 8 & 0.6066494(34) & -0.45(16) & 2.2(1.3) & 17(11) \\
	light & 7.4 & 0.755 & 48 & 24 & 850 & 8 & 0.6069551(55) & -0.168(64) & 3.4(2.1) & 11.4(7.2) \\
	light & 7.4 & 0.755 & 48 & 42 & 217 & 8 & 0.6069055(43) & -1.94(27) & 2.1(1.0) & 6.1(3.9) \\
	light & 7.5 & 0.71 & 48 & 36 & 250 & 8 & 0.6128190(49) & 1.13(33) & 1.7(1.1) & 15.2(9.7) \\
	light & 7.5 & 0.72 & 48 & 36 & 399 & 8 & 0.6132577(36) & -0.46(20) & 1.74(73) & 12.4(6.5) \\
    \hline \hline
\end{tabular}

    \label{tab:ensembles}
\end{table}

In this section, we describe our strategy for the spectroscopy measurements, including our implementation of L\"uscher's formalism. We start by describing the lattice field theory ensembles we use in the study. We define the basis of operators used in the finite-volume analysis, and then provide technical information about the variational analysis we use to extract correlation functions and energy levels. We then describe the partial wave decomposition of the scattering amplitudes, and how we use its results to extract information about the phase shifts.
    
\subsection{The lattice theory}
\label{sec:lattice}

We discretize the (Wick rotated) Euclidean action by introducing a hypercubic lattice having $N_s$ sites along the three spatial directions, with periodic boundary conditions for all fields, and $N_t$ sites in the time direction, with periodic boundary conditions for the bosons and anti-periodic for the fermions. The lattice spacing is denoted by $a$, and we define the spatial and temporal lattice extent as $L \equiv a N_s$ and $T \equiv a N_t$, respectively.
We adopt the standard Wilson gauge action for the gauge fields and the Wilson fermion action for the fermion fields \cite{Wilson:1974sk}.

The gauge configurations have been generated using the HiRep code~\cite{DelDebbio:2008zf,HiRepSUN} extended to treat symplectic gauge groups~\cite{HiRepSpN} on machines with CPU architecture. Some of the ensembles analyzed have been used in previous investigations~\cite{Bennett:2019jzz, Kulkarni:2022bvh,Bennett:2023rsl,Dengler:2024maq}. We have supplemented them with new ensembles generated using the Hybrid Monte Carlo (HMC) algorithm~\cite{Duane:1987de} with Hasenbusch acceleration~\cite{Hasenbusch:2001ne,Bussone:2018mzi}. For the measurements, we employ $Z_2\times Z_2$ stochastic sources~\cite{Boyle:2008rh} with spin-dilution and use the inverted operator as a source for the next inversion---commonly known as sequential sources~\cite{Foley:2005ac,Alexandrou:2017mpi}---for the analysis of ensembles used in our implementation of  L\"uscher's analysis.

We show the ensembles studied by means of L\"uscher's method in Tab.~\ref{tab:ensembles}. They are grouped in three sets, within each of which the parameters in the lattice action ($\beta$ and $a m_0$) are the same, but we repeat our measurements for ensembles with different space-time extents and number of configurations. 
In all three cases, we first perform a measurement of the energy levels of the vector state transforming as a $\textbf{10}$,  by considering only two-point functions that use as source and sink the single-meson operators
denoted as ${\rm V}$ and ${\rm T}$ in Tab.~\ref{tab:spec_operators}, and performing a fit of the correlation functions at large Euclidean time. In the first set, dubbed heavy (because of the heavy choice of bare mass $am_0=-0.92$, with inverse coupling $\beta=6.9$), the analysis of the single meson operators~\cite{Bennett:2019jzz} shows that the ${\rm V}$-meson ground state is clearly below the threshold for decay to two PNGBs (${\rm PS}$ mesons). In this case, the result of the single-meson analysis should provide a good estimate of the mass of the resulting, stable bound state, and we will verify this by the more general analysis that includes also two-PNGB operators.

We also analyze ensembles with $am_0=-0.863$ and $\beta=7.05$ (called medium), for which the single-meson analysis identifies a  ${\rm V}$-meson state with mass close to the kinematical  threshold for decay to two PNGBs. Finally, we consider ensembles with  the same inverse coupling $\beta=7.05$   but $am_0=-0.867$ (light), in which case the ${\rm V}$-meson state identified by the single-meson analysis is heavier than the two-PNGB threshold, and hence the particle would be unstable. In these two, lighter cases, the preliminary single-meson analysis is not sufficient, in general, to establish the nature of the spectrum and its states, but must be complemented by a detailed analysis of the energy levels that includes also two-PNGB states, as we shall show.

When measuring individual energy levels we estimate the uncertainty of the ensemble average using a standard jackknife analysis. We resample the result for the energy levels by a Gaussian distribution, when comparison of different ensembles is considered---for example when different lattice volumes are used in a single fit. Prior to doing so, we verify that the distributions coming from the different configurations follow Gaussian distributions.
In performing L\"uscher's analysis we take into account the potential residual autocorrelations visible in Tab.~\ref{tab:ensembles} by binning the correlation matrix with a bin size of two configurations, before we extract the energy levels.

\subsection{Interpolating operators}

In the analysis of the vector channel of the scattering between two PNGBs, we adopt a basis consisting of two types of operators. Besides the flavored fermion bilinears representing mesons, with fermion indices contracted to identify the relevant spin-1, $\textbf{10}$ channel (${\rm V}$), we also implement operators combining two individual PNGB operators ($2{\rm PS}$), following the approach used in QCD~\cite{Alexandrou:2017mpi}. We measure the spectrum of energy levels in lattices with different extents of the finite volume and in different reference frames, by employing operators with finite (non-zero) momentum. As a consequence, the symmetry is reduced from the octahedral group, $O_h$, characterizing the hypercubic lattice in the center-of-mass frame,  to its appropriate little group (LG)~\cite{Gockeler:2012yj,Prelovsek:2016iyo,Leskovec:2012gb,Rummukainen:1995vs,Boyle:2024grr}. The spectrum then depends on which irreducible representation ($R$) of the resulting lattice symmetry group we analyze. In the analysis of the numerical data, we project on the irreducible representations the results of using the basis of operators sourcing ${\rm V}$ and $2{\rm PS}$ states~\cite{Gockeler:2012yj,Prelovsek:2016iyo,Alexandrou:2017mpi,Boyle:2024grr}. Our choices of momenta, little groups and irreducible representations are tabulated in Tab.~\ref{tab:little_groups}. We focus on operator choices that have non-trivial overlap with the spin-1 channel transforming as a $\textbf{10}$ of the unbroken global $Sp(4)$, and  neglect contributions from higher partial waves, as was done for QCD in Ref.~\cite{Alexandrou:2017mpi,Wang:2025hew}. 

\begin{table}[t]
\caption{External momenta, $\vec P=\frac{2\pi}{L} \vec{d}$, corresponding little groups, LG, and irreducible representations, $R$, that are used in L\"uscher's analysis. We follow the notation of Ref.~\cite{Wang:2025hew}. For the ${\cal O}_{\rm V}$ operators, we also show the projections used in Eq.~(\ref{eq:projection_formula})~\cite{Alexandrou:2017mpi,Gockeler:2012yj}. We include in the analysis the  ${\cal O}_{2{\rm PS}}$ operators only in the $A_1$ irreducible representation. We restrict attention to the case in which the total net momentum is carried by one of the two PNGBs in ${\cal O}_{2{\rm PS}}$.}
\begin{tabular}{|c|c|c|c|c|}
\hline
 $\vec{d}$ & LG & $R$ & ${\cal O}^R_{\rm V}$ & ${\cal O}^R_{2{\rm PS}}$ \\ \hline \hline
 (0,0,0) & $O_h$ & $T_1$ & ${\cal O}_{{\rm V}_x}+{\cal O}_{{\rm V}_y}+{\cal O}_{{\rm V}_z}$  & - \\ \hline
 \multirow{2}{*}{(0,0,1)} & \multirow{2}{*}{$C_{4v}$} & $A_1$ & ${\cal O}_{{\rm V}_z}$ & ${\cal O}_{\rm PS}(\vec{d}\,){\cal O}_{\rm PS}(0)$ \\
                          &  & $E$  & ${\cal O}_{{\rm V}_x}+{\cal O}_{{\rm V}_y}$ & - \\ \hline
 \multirow{2}{*}{(1,1,0)} & \multirow{2}{*}{$C_{2v}$} &$A_1$ & ${\cal O}_{{\rm V}_x}+{\cal O}_{{\rm V}_y}$ & ${\cal O}_{\rm PS}(\vec{d}\,){\cal O}_{\rm PS}(0)$ \\
                          & & $B_1$ & ${\cal O}_{{\rm V}_z}$ & - \\ \hline
 \multirow{2}{*}{(1,1,1)} & \multirow{2}{*}{$C_{3v}$} & $A_1$ & ${\cal O}_{{\rm V}_x}+{\cal O}_{{\rm V}_y}+{\cal O}_{{\rm V}_z}$ & ${\cal O}_{\rm PS}(\vec{d}\,){\cal O}_{\rm PS}(0)$ \\
                          & & $E$ & $2{\cal O}_{{\rm V}_x}-{\cal O}_{{\rm V}_y}-{\cal O}_{{\rm V}_z}$ & - \\ \hline
\end{tabular}
\label{tab:little_groups}
\end{table}

Following Ref.~\cite{Alexandrou:2017mpi}, we write the local  ${\rm V}$-meson operators, ${\cal O}_{\rm V}$, as follows
\begin{align}
    \label{eq:rho_A1_operator}
    \mathcal{O}_{\rm V}(t,\vec{P}) &= \sum_{\vec{x}} \overline{Q^I}(t,\vec{x})\left( \vec \gamma \cdot \hat P \right)  Q^J(t,\vec{x}) e^{i \vec{x} \cdot \vec P}\,,
\end{align}
with $I\neq J$ the flavor indices, and where a finite three-momentum, 
\begin{align}
    \vec P\equiv \frac{2\pi}{L} \vec{d}\,,
\end{align}
(with $L\equiv N_s a$) is added via a Fourier transform, with $\hat P \equiv \vec P/\left| P \right|$. The same states are also sourced by the 
operators
\begin{align}
    \label{eq:tensor_A1_operator}
    \mathcal{O}_{\rm T}(t,\vec{P}) =& \sum_{\vec{x}} \overline{Q^I}(t,\vec{x})\gamma_0 \left( \vec \gamma \cdot \hat P \right) Q^J(t,\vec{x}) e^{i \vec{x} \cdot \vec P}.
\end{align}
As described in Sect.~\ref{sec:glob_sym_and_scat_chan}, the scattering channel of interest is the totally antisymmetric combination of two PNGB operators, ${\cal O}_{2{\rm PS}}$. We construct it from the single PNGB operators 
\begin{align}\label{eq:pip_pim_operator}
   {\cal O}_{\rm PS}^{21}(t,\vec{p}) \equiv \sum_{\vec{x}} \overline{Q^2}(t,\vec{x})\gamma_5 Q^1(t,\vec{x}) \text{e}^{i \vec{x}\cdot\vec{p}}\,, \\
   {\cal O}_{\rm PS}^{12}(t,\vec{p}) \equiv \sum_{\vec{x}} \overline{Q^1}(t,\vec{x})\gamma_5 Q^2(t,\vec{x}) \text{e}^{i \vec{x}\cdot\vec{p}}\,,
\end{align}
so that the $\textbf{10}$, with spin-1, negative parity, and overlap with  $A_1$ is given by~\cite{Drach:2020wux}
\begin{align}\label{eq:pipi_operator}
  \mathcal{O}_{2{\rm PS}}^{A_1}(t,\vec{P}) \equiv \frac{1}{\sqrt{2}}\left[ {\cal O}_{\rm PS}^{21}(t,\vec{p_1})    {\cal O}_{\rm PS}^{12}(t,\vec{p_2}) - {\cal O}_{\rm PS}^{12}(t,\vec{p_1})  {\cal O}_{\rm PS}^{21}(t,\vec{p_2})\right]\,,
\end{align}
where $\vec{P}=\vec{p_1}+\vec{p_2}$. In our numerical analysis, we will only consider the special choices $\vec p_2=0$ and $\vec p_1 = \vec P$, to reduce numerical costs, as it requires fewer inversion of the lattice Dirac operator. 

The projection of operators into the desired irreducible representations, $R$, can be represented as follows~\cite{Gockeler:2012yj,Alexandrou:2017mpi,Boyle:2024grr}:
\begin{align}\label{eq:projection_formula}
    \mathcal{O}_{\rm CM}^{R}(\vec{p},t)=\frac{\dim(R)}{N_{\rm LG}}\sum_i \chi_R(\hat{S}_i) \,\hat{S}_i\, \mathcal{O}_{\rm CM}(\vec{p},t)
\end{align}
where $\hat{S}_i$ are the elements of the little group, LG, of order $N_{\rm LG}$, and $\chi_R(\hat{S}_i)$ are the characters of the irreducible representation, $R$, which has dimension $\dim(R)$~\cite{Gockeler:2012yj,Dresselhaus}. The subscript CM indicates that the operator is defined in the center-of-mass frame, where the sum of the momenta is equal to zero. On the lattice, we employ the operator in a different frame, which is obtained by adding the total momentum $\vec{P}$ to the momenta in the CM frame. 

We consider three choices of  non-vanishing external momenta, with $\vec d = (0,0,1)$, $(1,1,0)$, and $(1,1,1)$. The corresponding operators and the associated little groups are given in Tab.~\ref{tab:little_groups}.
The implementation of two-PNGB operators with higher momenta that probe different irreducible representations is numerically more costly and outside the scope of this exploratory study. In light of the lattice sizes available, higher momenta would also likely suffer from stronger discretization artifacts.

\subsection{Variational analysis and Wick contractions}
\label{Sec:GEVP}

We extract the energy levels in the spin-1, $\textbf{10}$ channel (${\rm V}$), projected in the $A_1$ irreducible representation listed in Table~\ref{tab:little_groups}, by first constructing a Hermitian correlation matrix with entries 
\begin{align}\label{eq:correlation_matrix_elements}
    \mathcal{C}_{ij}(t-t') = \langle \mathcal{O}_i(t)\mathcal{O}_j^\dagger(t') \rangle\,,
\end{align}
where $\mathcal O_i$ stands for the operators considered in the channel. We perform a variational analysis, and extract the eigenvalues, $\lambda_i(t-t')$, of the correlation matrix, $\mathcal{C}(t-t')$. At large Euclidean times, the $i$-th eigenvalue is dominated by the energy of the $i$-th eigenstate, as long as the operators have sufficient overlap~\cite{Blossier:2009kd}. Subleading exponential terms, originating from contamination from higher energy states, are suppressed by the energy difference, $\Delta E> 0$, as
\begin{align} \label{eq:large-t-correlator}
    \lambda^{(i)}(t) \propto e^{-E_i t} \left( 1 + \mathcal{O}\left(e^{-\Delta E t}\right) \right).
\end{align}
We build a $3\times3$ cross correlation matrix using the operators given in Eqs.~\eqref{eq:rho_A1_operator}, \eqref{eq:tensor_A1_operator}, and \eqref{eq:pipi_operator}. For the other irreducible representations, we only have access to the single vector  operators, ${\cal O}_{\rm V}$, defined in Table~\ref{tab:little_groups}.

\begin{figure}
    \includegraphics[scale=0.85]{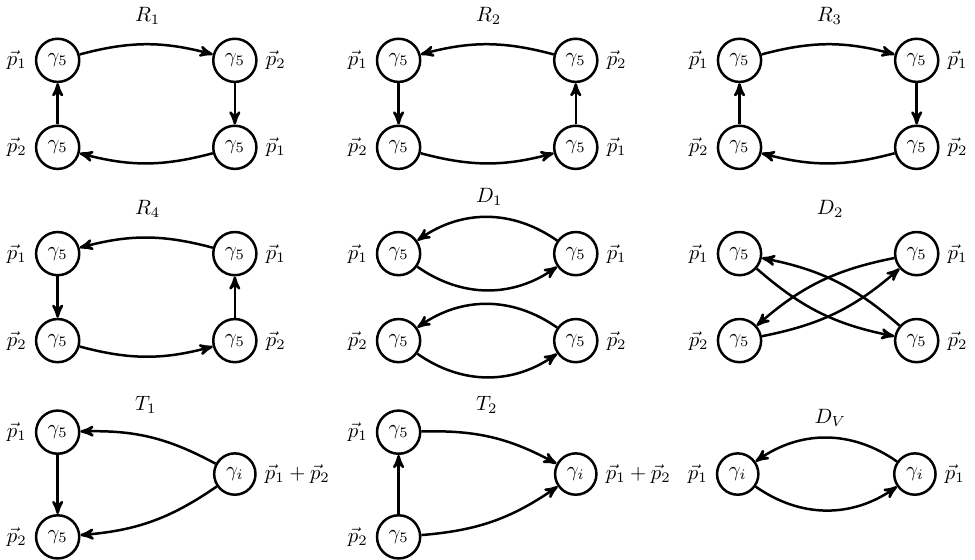}
    \caption{Wick diagrams entering the correlation matrix elements in Eq.~\eqref{eq:correlation_matrix_elements}. We label the diagrams as $R_1$, $R_2$, $R_3$, $R_4$, $D_1, D_2, T_1, T_2$, and $D_V$ (left to right, top to bottom). Additional diagrams of the form of $T_1$, $T_2$, and $D_V$, but involving the operator $\mathcal{O}_{T}$ replacing $\mathcal{O}_{V}$, are obtained by substituting $\gamma_i \to \gamma_0 \gamma_i$ as in Eq.~\eqref{eq:tensor_A1_operator}. The arrows indicate the direction of the Wilson lines, with respect to the spatial indices.
    \label{fig:wick_contractions}}
\end{figure}

In order to evaluate the elements of the correlation matrix  in Eq.~\eqref{eq:correlation_matrix_elements}, we perform the Wick contractions and obtain the diagrams to be evaluated on the lattice. We find one diagram for $C_{{\rm V}\to {\rm V}}$, two diagrams for $C_{{\rm V} \to 2{\rm PS}} = - C_{2{\rm PS} \to {\rm V}}$ and six diagrams for $C_{2{\rm PS}  \to 2{\rm PS} }$~\cite{Alexandrou:2017mpi, Drach:2020wux, Fu:2016itp}.  We show the diagrams in Fig.~\ref{fig:wick_contractions} and label them so that
\begin{align}
        C_{{\rm V}\to{\rm V}} &= D_{{\rm V}}\,, \\
        C_{2{\rm PS}\to{\rm V}} &= -C_{{\rm V}\to 2{\rm PS}} = T_1 - T_2\,, \\
        C_{2{\rm PS}\to 2{\rm PS}} &= D_1-D_2 +R_1+R_2-R_3-R_4\,.
\end{align}
We restrict our analysis to the case $\vec{p}_2=0$, as anticipated. Diagrams $T_1$ and $T_2$ are purely imaginary and therefore only differ by a sign, as the correlation matrix is Hermitian. The mixing diagrams, $D_1$ and $D_2$, can be constructed by multiplying two individual pion correlation functions. The diagram $D_2$ probes the correlation of creating a pion with momentum and annihilating it without momentum. The momentum transfer is allowed due to gluon exchange, but numerically the diagram is negligible. For the other diagrams, we observe that $R_1 \approx R_2 \gg R_3 \approx R_4$ within statistical errors. We give the explicit expression for the diagrams depicted in Fig.~\ref{fig:wick_contractions} in 
Appendix~\ref{app:wick_contractions}.

We test two approaches to perform the variational analysis: the standard eigenvalue problem (EVP) and the generalized eigenvalue problem (GEVP). The former consists of a direct diagonalization of the correlation matrix in Eq.~\eqref{eq:correlation_matrix_elements}. The latter requires defining generalized eigenvalues $\lambda(t,t_0)$, with eigenstates, $v(t,t_0)$, obeying the relation
\begin{align} \label{eq:gevp}
    \mathcal{C}(t)v(t,t_0) = \lambda(t,t_0) \mathcal{C}(t_0) v(t,t_0)\,,
\end{align}
and  the introduction of a reference time, $t_0$, in the GEVP. This introduces another systematic due to the choice of the value of $t_0$. It can be shown that as long as both $t_0$ and $t$ are large, with $t > t_0$,  these systematic effects are exponentially suppressed~\cite{Blossier:2009kd}.

\subsection{Correlation functions}\label{ssec:fitting}

We denote the lattice data to be fitted as a time sequences, $C(t)$, in our fitting strategy for both correlation functions and eigenvalues. In the case of $R=A_1$, in order to remove around-the-world effects from the two-PNGB correlation functions which give rise to a constant contribution~\cite{Umeda:2007hy, Prelovsek:2008rf}, we additionally apply a numerical derivative using a central difference scheme
\begin{align} \label{eq:correlator_derivative}
    \tilde C(t) &\equiv \frac{C(t+1)-C(t-1)}{2}\,.
\end{align}

We extract the energy states in every channel by fitting the eigenvalues (or the correlation function) at large Euclidean times to its expected decay, for which we must choose suitable fit functions and reliable fit ranges. 
As a first step, we examine the effective mass, defined as 
\begin{align}
\label{Eq:meff}
    m_{\rm eff} = \log \left( \frac{C(t)}{C(t+1)} \right),
\end{align}
which is expected to exhibit a plateau equal to the energy level, if the correlation function is dominated by a single exponential, as in Eq.~\eqref{eq:large-t-correlator}. When performing a variational analysis the correlation function, $C(t)$ in Eq.~\eqref{Eq:meff} is replaced by the corresponding eigenvalue, $\lambda(t)$,  obtained from the correlation matrix, $\tilde C(t)$.

We use the effective mass, $m_{\rm eff}$,   only to identify the presence of said plateau, not to extract the energy levels themselves. Instead, we use the fact that a generic diagonal correlation matrix element (and the eigenvalues arising from  the variational analysis) can be written as a positive sum of exponentially decaying functions. Taking the lattice periodicity into account, we expect
\begin{align}\label{eq:correlator_form}
     C(t) = \langle O(t^\prime)\bar{O}(t)\rangle = \sum_k \vert \langle 0| O|k\rangle \vert^2 \left( e^{-t E_k} \pm e^{-(T-t) E_k} \right) = \sum_k A_k \left( e^{-t E_k} \pm e^{-(T-t) E_k} \right),
\end{align}
where the sign depends on the symmetry of the correlation function (or eigenvalue) to be fitted. We can exploit the structure of Eq.~(\ref{eq:correlator_form}) in the fitting procedure by following the approach outlined in Refs.~\cite{Lepage:2001ym, Hornbostel:2011hu, Bouchard:2014ypa} and implemented in the \textsc{corrfitter} software package \cite{Lepage:zenod_corrfitter}. This uses a Bayesian approach to include an arbitrary number of exponentials in the fit function and further exploits the positivity of the coefficients and the positivity of energy difference between states. The fit is stabilized by an additional prior term added to the $\chi^2$ function to be minimized:
\begin{align}
    \chi^2 = \chi^2_{\rm prior} + \sum_{t,t'} \left( C(t) - C_{\rm fit}(t) \right) {\rm Cov}^{-1}(t,t') \left( C(t') - C_{\rm fit}(t') \right)\,.
\end{align}
The covariance matrix, ${\rm Cov}(t,t')$, measured on the data, takes correlations between time slices into account in the definition of $\chi^2$. For the details of the prior, we refer to Refs.~\cite{Lepage:2001ym} and \cite{Lepage:zenod_corrfitter}. The use of more than one exponential also sidesteps the issue of fixing a lower end of the fit interval, since most time slices can be included in the fit.   

\subsection{Scattering analysis}\label{sec:scattering}

This subsection describes the core of the analysis, starting from the partial wave decomposition of the relevant cross-sections, to the application of L\"uscher's formalism and the extraction of the continuum observables.

\subsubsection{Partial-wave decomposition}

We write the total cross-section and the differential cross-section for the scattering of two particles as
\begin{align} \label{eq:total_cross_section_omega}
    \sigma_{\text{tot}} = \int \frac{d\sigma}{d\Omega}d \Omega\,,
\end{align}
where $d\Omega=\sin \theta\, d\theta\, d\phi$ is the solid angle element. The differential cross-section is related to the scattering matrix, $\mathcal{M}$:
\begin{align} \label{eq:diff_cross_section}
    \frac{d\sigma}{d\Omega} = \frac{1}{64\pi^2s}|\mathcal{M}|^2\,.
\end{align}
The matrix  $\mathcal{M}$ can be written in terms of the partial-wave scattering amplitudes, $f_{\ell}$, as
\begin{align} \label{eq:matrix_element}
    \mathcal{M} &= 32 \pi \sum_{\ell=0}^\infty (2\ell+1) f_\ell(s) P_\ell(\cos\theta)\,, 
\end{align}
where $P_\ell$ denotes the Legendre polynomials, with $s$  the center-of-mass energy squared, 
\begin{align}
    s \equiv 4(m^2 + p^2)\,,
\end{align}
computed for  two scattering particles with equal mass, $m$, and back-to-back scattering momentum, $\vec p$, of magnitude $p$. 

The partial-wave $S$-matrix elements, $S_\ell$,  transition matrix elements, $t_\ell$, and phase shifts, $\delta_\ell$, are defined in terms of the partial-wave scattering amplitudes, $f_{\ell}$, by the following relations:
\begin{align}
        S_\ell &= e^{2i\,\delta_\ell} = 1+2i\,t_\ell \label{eq:s_matrix}\,, \\
        t_\ell &= e^{i\delta_\ell}\sin{\delta_\ell} = \left(\cot(\delta_\ell) -i\right)^{-1}\,,\label{eq:transition_matrix} \\
        f_\ell &= \frac    {\sqrt{s}}{2p}t_\ell\,. \label{eq:partial_wave_amplitude}
\end{align}
Because of  the orthogonality of the Legendre polynomials, different partial waves do not mix, and we can rewrite the total cross-section, $\sigma_{\rm tot}$, in terms of the phase shifts, $\delta_\ell$, or the partial-wave cross-sections, $\sigma_\ell$, as
\begin{align}
    \sigma_{\rm tot} = \frac{4}{s} \int |(2\ell+1)f_\ell(s)P_\ell(\cos \theta)|^2 d\Omega = \frac{4\pi}{p^2} \sum_{\ell=0}^{\infty} (2\ell+1)\sin^2(\delta_\ell) \equiv \sum_{\ell=0}^{\infty}\sigma_\ell\,.
\end{align}

\subsubsection{Scattering Parameterizations}

The nature of scattering states and resonances appearing in the scattering of two PNGBs can be parameterized, or modelled, in several ways, that can serve different theoretical and phenomenological purposes, and are most effective in different kinematical regions. For example, the effective range expansion (ERE)~\cite{Blatt:1949zz,Bethe:1949yr} describes the behaviour of the phase shift close to the elastic threshold as an expansion in powers of the scattering momentum squared, $p^{2}$. One then defines the scattering length, $a_\ell$, and the effective range, $r_\ell$, for the partial wave $\ell$, as
\begin{align}
    p^{2\ell+1} \cot(\delta_\ell) &= -\frac{1}{a_\ell^{2\ell+1}} + \frac{p^{2}}{2r_\ell^{2\ell-1}} + \mathcal{O}(p^{4})\,,\label{eq:ERE} 
\end{align}
and writes the partial-wave cross-section as
\begin{align}  
    \sigma_{\ell}^{\rm ERE} &= \frac{4 \pi  (2 \ell+1) p^{4 \ell}}{\left(\frac{1}{2} p^{2} r_\ell^{1-2 \ell} - a_\ell^{-2 \ell-1}\right)^2+p^{4 \ell+2}}\label{eq:partial_wave_cross_section_ERE}\,.
\end{align}
In the limit $p\to 0$, and with a finite scattering length, the cross-section in any partial wave with $\ell\geq 1$ vanishes. 

\begin{table}
    \centering
    \caption{The $\ell=1$ phase shift, $\delta_1$, in different irreducible representations, $R$,  of the little groups (LGs) associated with the total lattice momentum, $\vec{P}=\frac{2\pi}{L}\vec{d}$, expressed in terms of the functions $w_{lm}$ defined in  Eq.~\eqref{eq:def_wlm}, obtained from the quantization condition, by taking the symmetries of the lattice and LGs into account~\cite{Gockeler:2012yj, Alexandrou:2017mpi, Rummukainen:1995vs}.}
    \label{tab:phase_shift_formulas}
    \begin{tabular}{|c|c|c|c|}
    \hline 
    LG & $\vec{d}$ & $R$ & $\cot (\delta_1)$ \\ \hline \hline
    \multirow{2}{*}{$C_{4v}$} & \multirow{2}{*}{(0,0,1)} & $A_1$ & $w_{00}+2w_{20}$\\
    & & $E$ & $w_{00}-w_{20}$ \\ \hline
    \multirow{2}{*}{$C_{2v}$} & \multirow{2}{*}{(1,1,0)} & $A_1$ & $w_{00}-w_{20}-i\sqrt{6}w_{22}$ \\
    & & $B_1$ & $w_{00}+2w_{20}$ \\ \hline
    \multirow{2}{*}{$C_{3v}$} & \multirow{2}{*}{(1,1,1)} & $A_1$ & $w_{00}-w_{20}+i\sqrt{6}w_{22}$ \\
    & & $E$ & $w_{00}+i\sqrt{6}w_{22}$ \\ \hline
    \end{tabular}
\end{table}

As a complementary approach, the partial-wave phase shift, $\delta_\ell$, in the vicinity of a (narrow) resonance, ${\cal R}$, can be described by a {Breit-Wigner} function,
 parameterized by the {resonance position}, $m_{\cal R}$, and the {resonance width}, $\Gamma(s)$, as
\begin{equation}
    \cot(\delta_\ell) = \frac{m_{\cal R}^2-s}{\sqrt{s}\,\Gamma(s)}\,. \label{eq:phase_shift_breit_wigner}
\end{equation}
One can further relate $\Gamma(s)$  to the coupling, $g_{{\cal R}{\rm PP}}$, that controls the decay of the resonance ${\cal R}$ to two PNGBs (${\rm PS}$):
\begin{equation}
    \Gamma(s) = \frac{g_{{\cal R}{\rm PP}}^2}{6\pi}\frac{{p}^3}{s}\,.\label{eq:resonance_width_breit_wigner}
\end{equation}

\subsubsection{L\"uscher's formalism}

We relate the energy levels obtained from spectroscopic calculations performed on finite volume lattices to infinite volume scattering phase shifts by applying  L\"uscher's formalism, for which we follow closely Ref.~\cite{Alexandrou:2017mpi}. In the absence of interactions, the energy level of two-PNGB eigenstates in a cubic lattice of size $L^3$, expressed in the lattice frame, is
\begin{align} \label{eq:non_int}
    E_\text{non-int} &= \sqrt{m_{\rm PS}^2+|\vec{p}_1|^2} + \sqrt{m_{\rm PS}^2+|\vec{p}_2|^2}\,.
\end{align}
In the presence of an interaction, the energy levels are shifted, in respect to Eq.~\eqref{eq:non_int}, and we
call $E$ the resulting interacting energy level, measured on the lattice. The center-of-mass energy squared is then 
\begin{align}
    E_{\rm CM}^2 = E^2 - \vec{P}^2 \equiv 4 \left( m_{\rm PS}^2 + {p^{\ast}}^2 \right),
\end{align}
in which we implicitly define the scattering momentum, $p^\ast$, in the second step, as the magnitude of the back-to-back momentum in the center-of-mass frame. Note that $p^\ast$ takes continuous values, not related to multiples of $2\pi/L$. 

\begin{table}
\caption{%
\label{tab_spec:fit_results-0}
Low energy coefficients entering the analysis of the continuum extrapolations, inspired from NLO Wilson chiral perturbation theory, performed by fitting Eqs.~(\ref{eq:fit_m}) and~(\ref{eq:fit_f}) to measurements of masses (top table) and decay constants (bottom), extracted from two-point functions for single-meson operators as source and sink. We present also the value of reduced chi-square, $\chi^2/{\rm N_{\rm d.o.f}}$, associated with the individual fits for each observable.}
\begin{center}
\begin{tabular}{|c|c|c|c|c|c|}
\hline\hline
M & $\hat{m}_{M,\,\chi}^2$ & $L_{\rm M}^m$ & $W_M^m$ & $R_M^m$ & $\chi^2/{\rm N_{d.o.f}}$ \\
\hline\hline
$\rm V$ & $0.315(12)$ & $2.85(14)$ & $-0.161(24)$ & $0.017(13)$ & $1.47$ \\
$\rm T$ & $0.326(16)$ & $2.75(17)$ & $-0.173(32)$ & $0.020(17)$ & $1.68$ \\
$\rm S$ & $0.672(63)$ & $2.59(45)$ & $-0.08(16)$ & $-0.03(14)$ & $1.35$ \\
$\rm AV$ & $0.758(55)$ & $2.15(29)$ & $-0.01(15)$ & $-0.03(11)$ & $0.86$ \\
$\rm AT$ & $0.793(63)$ & $2.31(34)$ & $-0.07(17)$ & $-0.02(13)$ & $1.09$ \\
\hline\hline
\end{tabular}

\end{center}
\begin{center}
\begin{tabular}{|c|c|c|c|c|c|}
\hline\hline
M & $\hat{f}_{M,\,\chi}^2$ & $L_{\rm M}^f$ & $W_M^f$ & $R_M^f$ & $\chi^2/{\rm N_{d.o.f}}$ \\
\hline\hline
$\rm PS$ & $0.00533(35)$ & $3.66(26)$ & $-0.00235(90)$ & $0.00107(55)$ & $0.80$ \\
$\rm V$ & $0.0196(14)$ & $1.25(16)$ & $0.0039(34)$ & $-0.0015(20)$ & $1.00$ \\
$\rm AV$ & $0.0397(76)$ & $0.52(16)$ & $-0.058(24)$ & $0.034(17)$ & $1.03$ \\
\hline\hline
\end{tabular}

\end{center}
\end{table}

The L\"uscher quantization condition can be rewritten in terms of the phase shifts, $\delta_{\ell}$, and quantities that can be determined from the energy levels. From here on, we neglect partial waves with $\ell>1$, which have been shown to be negligible in QCD~\cite{Dudek:2012xn,Estabrooks:1975cy}, and focus entirely on $\ell=1$. The cotangent of the phase shift, $\cot(\delta_{1})$, is in general a linear combination of the $w_{lm}$ functions~\cite{Rummukainen:1995vs,Alexandrou:2017mpi,Gockeler:2012yj}, with $l$ and $m$ the angular momentum quantum numbers (or, equivalently, the labels of the irreducible representation and row in $SO(3)$, respectively), that are defined as follows: 
\begin{align}\label{eq:def_wlm}
    w_{lm} \equiv \frac{\mathcal{Z}_{lm}(1;q^2)}{\pi^{3/2}\sqrt{2l+1}\gamma q^{l+1}}\,,
    \quad   {\rm where} \quad 
    q \equiv \frac{p^\ast L}{2\pi}\,,
\end{align}
and where
$\gamma=E/\sqrt{s}$ is the Lorentz factor, while 
\begin{align}
    \mathcal{Z}_{lm}(s;q^2) &\equiv \sum_{\vec{r}\in P_d}\frac{\mathcal{Y}_{lm}(\vec{r})}{\left(\vec{r}^2-q^2\right)^s}\,, \\
    P_d&\equiv\left\{\vec{r} | \vec{r} = \gamma^{-1}\vec{n} \right\},\,\,\vec{n}\in \mathbb{Z}^3\,.\label{eq:def_zeta_function}
\end{align}
In these expressions, $\mathcal{Z}_{lm}(s;q^2)$ is the {L\"uscher Zeta function}, while $\mathcal{Y}_{lm}(\vec{r}) \equiv x^lY_{lm}(\theta,\phi)$ are defined in terms of the spherical harmonics associated to $l$ and $m$. The L\"uscher Zeta function contains a sum over the space, $P_d$, of all allowed momenta in the center-of-mass frame~\cite{Leskovec:2012gb}. Examples of numerical implementations can be found in Refs.~\cite{Jenny:2022atm, Leskovec:2012gb,Leskovec:2017uqb}. 
We report the linear combinations of $w_{lm}$ that give the $\ell=1$ partial-wave phase shift, $\cot(\delta_1)$, in Tab.~\ref{tab:phase_shift_formulas}.

\begin{figure*}[t]
\begin{center}
\includegraphics[width=0.8\linewidth]{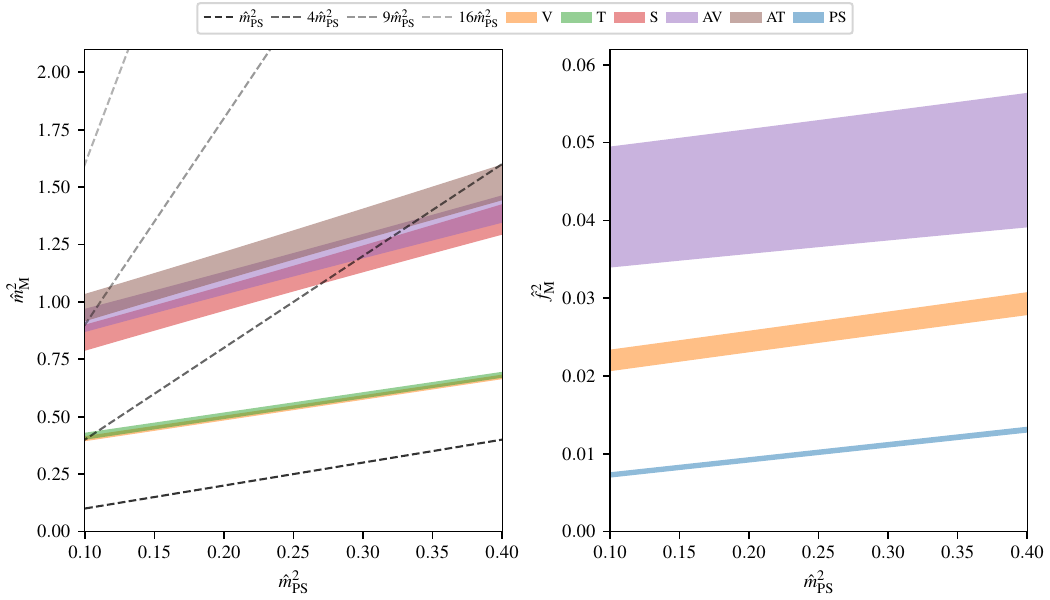}
\caption{%
\label{fig:mass_con_summary-0}%
Continuum extrapolations, performed with the simplified analysis (inspired by NLO-W$\chi$PT) defined in Eqs.~(\ref{eq:fit_m}) and~(\ref{eq:fit_f}), of spectroscopy measurements involving two-point functions of single meson operators. The square of the meson masses, $\hat{m}_{\rm M}^2$ (left panel), and the decay constants,  $\hat{f}_{\rm M}^2$ (right panel), are expressed in units of the gradient flow scale, $w_0$,
and plotted as a function of the mass squared of the PNGBs, $\hat{m}_{\rm PS}^2$. The acronyms in the legend, M$=$V, T, AV, AT, S and PS, refer to the single-meson operator basis described in Tab.~\ref{tab:spec_operators}. The dashed lines correspond to the square of $1$, $2$, $3$, and $4$ times the mass of the PNGBs, and are shown to illustrate the relevant decay thresholds of the heavy mesons. The extrapolation of the masses of V and T states agree, as expected, since the two operators source the same set of states. The extrapolation of S, AV, and AT masses also overlap, within the fitting errors, yet these are three distinct states.}
\end{center}
\end{figure*}

\begin{figure*}[t]
\begin{center}
\includegraphics[width=0.8\linewidth]{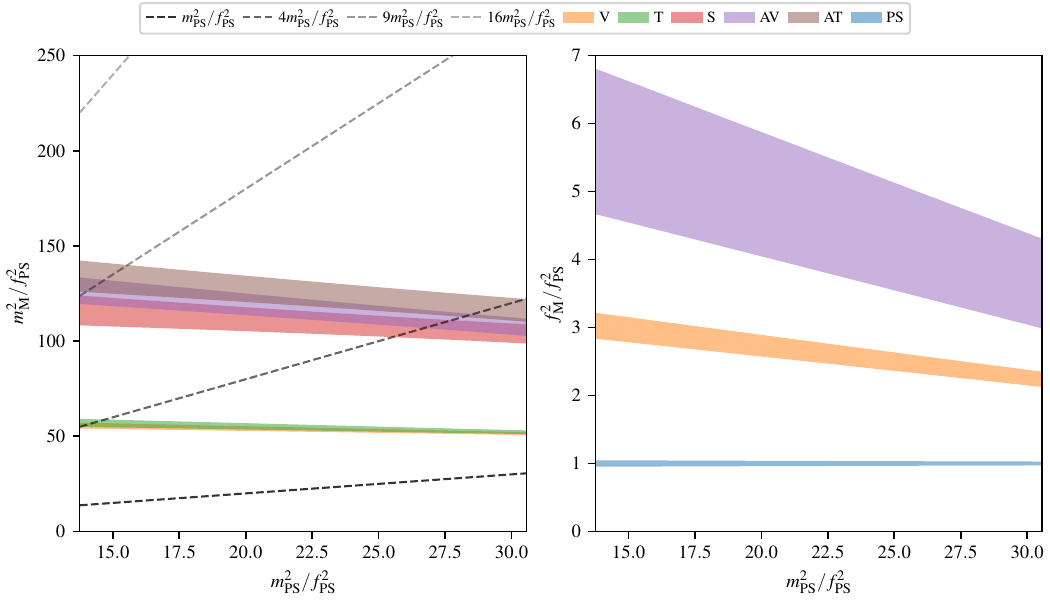}
\caption{%
\label{fig:mass_con_summary-1}%
Continuum extrapolations, performed with the simplified analysis (inspired by NLO-W$\chi$PT) defined in Eqs.~(\ref{eq:fit_m}) and~(\ref{eq:fit_f}), of spectroscopy measurements involving two-point functions of single meson operators. The square of the meson masses, ${m}_{\rm M}^2$ (left panel), and the decay constants,  ${f}_{\rm M}^2$ (right panel), are expressed in units of the decay constant of the PNGBs, $f_{\rm PS}$, and plotted as a function of the mass squared of the PNGBs, ${m}_{\rm PS}^2/{f}_{\rm PS}^2$. The acronyms in the legend, M$=$V, T, AV, AT, S and PS, refer to the single-meson operator basis described in Tab.~\ref{tab:spec_operators}. The dashed lines correspond to the square of $1$, $2$, $3$, and $4$ times the mass of the PNGBs, and are shown to illustrate the relevant decay thresholds of the heavy mesons. The extrapolation of the masses of V and T states agree, as expected, since the two operators source the same set of states. The extrapolation of S, AV, and AT masses also overlap, within the fitting errors, yet these are three distinct states.
}
\end{center}
\end{figure*}

\section{Meson spectroscopy, single meson analysis}
\label{Sec:spectra}

The first step of our analysis, before we implement  L\"uscher's method, consists of a broad exploration of the 
space of lattice coupling, $\beta$, and Wilson-Dirac bare mass, $a m_0$, for which purpose we generated the ensembles listed in Tab.~\ref{tab:spec_ens}. The algorithms adopted, and the properties of the ensembles, such as the gradient flow scale, $w_0/a$, the topological charge, $Q$, and multiple estimates of the autocorrelation, are  described in Appendix~\ref{app:ensembles}. This part of the study provides an update in respect to the earlier Ref.~\cite{Bennett:2019jzz}, by including ten new ensembles in the region of smaller PNGB mass, but also by increasing the statistics of existing ensembles.

We perform a study of two-point correlation functions defined by using as sink and source the (flavored) single meson operators listed in Tab.~\ref{tab:spec_operators}. Beside the aforementioned improvements in statistics and range of lattice parameters explored provided by the updated ensembles, we improved also the analysis, by implementing both APE~\cite{APE:1987ehd,Falcioni:1984ei} and Wuppertal~\cite{Gusken:1989qx,Roberts:2012tp,Alexandrou:1990dq} smearing techniques in the measurements, and by extracting spectroscopy information with a generalized eigenvalue problem (GEVP)~\cite{Blossier:2009kd}, along the lines of the analysis presented in Refs.~\cite{Bennett:2023mhh,Bennett:2024wda,TELOS:2025ash}. Technical details are reported in Appendix~\ref{app:meson_masses}, which focuses on the PNGB and vector masses.  We also investigate the size of residual finite volume effects and conclude that the corresponding errors are well expected to be negligible in the spectroscopic measurements using the ensembles in Tab.~\ref{tab:spec_operators}.

We express all dimensional quantities in units of the gradient flow scale, $w_0$, measured with the reference scale  $W_0=0.28125$, following the conventions described in Ref.~\cite{Bennett:2022ftz}.\footnote{We note that this convention differs from our previous studies in Refs.~\cite{Bennett:2019jzz} and~\cite{Bennett:2019cxd}, for the dynamical and quenched calculations, respectively.} 
For the lattice spacing, we write $\hat{a}\equiv a/w_0 = 1/w_0^{\rm lat}$, and analogous expressions for masses and decay constants.  To perform the extrapolation towards the continuum limit, and remove discretization artifacts, we borrow ideas from Wilson chiral perturbation theory, at the next-to-leading order (NLO-W$\chi$PT)~\cite{Sheikholeslami:1985ij,Rupak:2002sm,Sharpe:1998xm,Symanzik:1983dc, Luscher:1996sc}, but we note that our measurements are not close to the limit of massless PNGBs.
We hence analyze the measurements of meson masses and decay constants using simplified fit functions, written as (polynomial) expansions in power of the PNGB mass squared, the lattice spacing, and their combination, without including logarithms. For the mass of  mesons, we write the fitting function as 
\begin{align}
\label{eq:fit_m}
\hat{m}^{2, {\rm NLO}}_M = \hat{m}_{M,\,\chi}^2(1+L_{M}^m \hat{m}_{\rm PS}^2)+W_{M}^{m} \hat{a}+R_{M}^{m} \hat{a}^2 \,,
\end{align}
where $\hat{m}_{M,\,\chi}$, $L^{m}_{M}$, $W^{m}_{M}$, and $R_{M}^{m}$ are the low-energy constants (LECs) determined by our fits.
Similarly, for the decay constants, our ansatz reads
\begin{align}
\hat{f}^{2, {\rm NLO}}_M = \hat{f}_{M,\,\chi}^2(1+L_{M}^f \hat{m}_{\rm PS}^2 )+W_{M}^f \hat{a}+R_{M}^f \hat{a}^2\,,
\label{eq:fit_f}
\end{align}
with  $\hat{f}_{M,\,\chi}$, $L^{f}_{M}$, $W^{f}_{M}$, and $R_{M}^{f}$ the corresponding LECs.
We notice here that the specific truncation presented in Eqs.~(\ref{eq:fit_m}) and~(\ref{eq:fit_f}) of what should be an infinite series in powers of $\hat{a}$ and $\hat{m}_{\rm PS}^2$ is motivated \emph{a posteriori}, by checking explicitly that the addition of subleading terms does not affect significantly (within error) the central values of the coefficients that have been retained. More details and discussion of the analysis is relegated to Appendix~\ref{app:continuum}, where we explicitly make a comparison to the fit results using the other ansatz containing an additional term of $\hat{a}\hat{m}_{\rm PS}^2$. Yet, we stress here that our extrapolations provide a valid approximation of the continuum theory only within the mass range of the available measurements, $\hat{m}_{\rm PS}^2 \leq 0.4$, while for larger masses we expect non-negligible effects to appear, due to terms ${\cal O}(\hat{a} \hat{m}_{\rm PS}^2)$ and higher, that we neglect.

As discussed in Appendix~\ref{app:ensembles} (see also Table~\ref{tab:spec_ens}), we perform spectral measurements for five different values of the lattice coupling, $\beta=6.9,\,7.05,\,7.2,\,7.4,\,7.5$. The fit results, corresponding to the low-energy constants in our simplified treatment of NLO-W$\chi$PT, are reported in Table~\ref{tab_spec:fit_results-0}. In Figs.~\ref{fig:mass_con_summary-0} and~\ref{fig:mass_con_summary-1}, we summarize our findings by displaying the continuum limit of all accessible spectral quantities, meson masses squared, and decay constants squared, as a function of the mass of the PNGBs. We show these physical quantities in units of the gradient flow scale $w_0$, that has been used in the extrapolation, but also in units of the decay constant of the PNGBs $f_{\rm PS}$, which may be of more direct use for phenomenological purposes. Furthermore, as a guidance for the potential opening of meson decay channels, we present the two-, three- and four-PNGBs thresholds as dashed lines. 

Aside from the PNGBs, the lightest states are the spin-1 states forming a $\textbf{10}$ of ${\rm Sp}(4)$, that are sourced by both the ${\cal O}_V$ and ${\cal O}_T$ operators in Tab.~\ref{tab:spec_operators}. As can be seen in the figures, over the portion of parameter space included in these extrapolations, these states are below the threshold for decay into two PNGBs, and hence they are stable. A few  measurements at the lowest end of the available range, that yield results close to or below threshold, have been excluded from the fits, and will be analyzed in a more rigorous way in the next section.
The other meson states are heavier, and affected by larger statistical and systematic uncertainties. The S, AV, and AT states are distinct, having different quantum numbers, yet given current uncertainties we cannot clearly distinguish their masses from one another, except that we note that their masses are generally below the thresholds for decay into three (and four) PNGBs.

\begin{figure}
    \centering
    \includegraphics[width=0.39\linewidth,page=1]{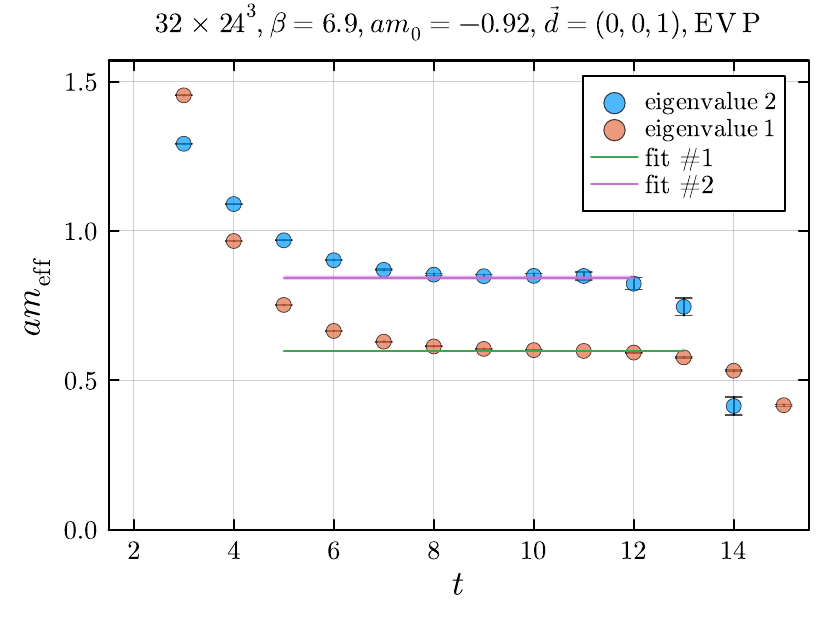}
    \includegraphics[width=0.39\linewidth,page=3]{assets/scattering/plots/effective_masses_comparison_evp_gevp.pdf}
    \includegraphics[width=0.39\linewidth,page=5]{assets/scattering/plots/effective_masses_comparison_evp_gevp.pdf}
    \includegraphics[width=0.39\linewidth,page=6]{assets/scattering/plots/effective_masses_comparison_evp_gevp.pdf}
    \includegraphics[width=0.39\linewidth,page=16]{assets/scattering/plots/effective_masses_comparison_evp_gevp.pdf}
    \includegraphics[width=0.39\linewidth,page=18]{assets/scattering/plots/effective_masses_comparison_evp_gevp.pdf}
    \includegraphics[width=0.39\linewidth,page=20]{assets/scattering/plots/effective_masses_comparison_evp_gevp.pdf}
    \includegraphics[width=0.39\linewidth,page=21]{assets/scattering/plots/effective_masses_comparison_evp_gevp.pdf}
    \caption{Comparison of the variational analysis based on a standard eigenvalue problem (EVP) and the generalized eigenvalue problem (GEVP), for representative choices of ensembles and, in the GEVP case, of the starting point, $t_0$. We show the effective masses of the two lowest lying states in the spin-1 channel, transforming as $\mathbf{10}$ of $Sp(4)$. The measurements have been obtained with the three-dimensional variational basis described in the main body of the paper. The first two rows show the effective mass in lattice units, $a m_{\rm eff}$, defined in Eq.~(\ref{Eq:meff}), for the heavy ensembles, and last two rows for the light ensembles, respectively.}
    \label{fig:gevp_vary_t0}
\end{figure}

For phenomenological purposes, the coupling between the vector states (analogous to the $\rho$ meson in QCD) and two PNGBs, $g_{\rm VPP}$, is of particular interest.
A naive way to estimate its size, subject to unknown and potentially large systematic uncertainties, relies on extrapolating beyond QCD the phenomenological Kawarabayashi-Suzuki-Riazuddin-Fayyazuddin (KSRF) relation of the second kind~\cite{Kawarabayashi:1966kd,Riazuddin:1966sw}, and furthermore applying it to the extrapolations to the continuum and massless limits.
With our numerical results, we find  $g_{\rm VPP}^{{\rm KSRF}}=\hat{m}_{\rm V}/\sqrt{2}\hat{f}_{\rm PS}=
5.43(20)
$, which updates our previous estimation made with ensembles limited to heavy fermions and coarse ensembles~\cite{Bennett:2019jzz}. 
The error we quote here is the fitting error, obtained from the continuum and massless extrapolations, by including only statistical uncertainties in the spectroscopy measurements. 
Not only are the aforementioned methodological systematics intrinsic to this estimate, but moreover, the single-meson operator analysis used in this section has no applicability in the low mass region, in which the vector becomes a resonance, and the $g_{\rm VPP}$ coupling enters its decay rate. We therefore warn the reader to take this result with some caution. We will return to this discussion after performing the more rigorous analysis exposed in the next section.

\section{Scattering Results}
\label{sec:results}

This section contains our main results for the scattering analysis. We start with a short subsection to discuss a numerical test of our implementation of the variational analysis in the lattice spectroscopy measurements. We then present our measurement of the energy levels in the available finite-volume ensembles, and show how we use them to reconstruct the phase shifts, and perform the extrapolations to recover the physics results. 

\subsection{Variational Analysis and Fitting Strategy}

The numerical quality of our measured correlation functions allows us to extract the two lightest eigenvalues of the energy in the vector meson channel.  We fit the eigenvalues following the approach outlined in Sec.~\ref{ssec:fitting}, in order to test the two alternative approaches to performing the variational analysis, called EVP and GEVP, that we introduced in Sect.~\ref{Sec:GEVP}.  The GEVP introduces a spurious dependence on the choice of initial time, $t_0$, that appears in Eq.~\eqref{eq:gevp},  the ensuing systematic effects associated with the choice of $t_0$ being  ${\cal O}(\exp{\left(-\Delta E t_0\right)})$~\cite{Blossier:2009kd}. Thus, when using the GEVP methodology, $t_0$ needs to be chosen to be sufficiently large.

We compare the results of EVP and GEVP analysis in Fig.~\ref{fig:gevp_vary_t0}, for representative choices of ensemble and initial time, $t_0$. We highlight that we solve the $3\times 3$ variational problem in all cases considered. Furthermore, we extract the energy levels by fitting the correlation functions, yet in the plots we display the effective mass, $a m_{\rm eff}$, as it provides a better visual illustration of the fit range.  In Fig.~\ref{fig:gevp_vary_t0}, we compare the effect of different choices of $t_0$ for the largest available volumes in the heavy and light ensembles.  We further display the fit range with colored bands, and hold it fixed for all choices of $t_0$, for illustration purposes. Due to our use of multi-exponential fits, the results are only marginally sensitive to our choice of the lower bound $t_{\min}$ of the fit range.  

\begin{figure}[t]
    \centering
    \includegraphics[scale=0.4]{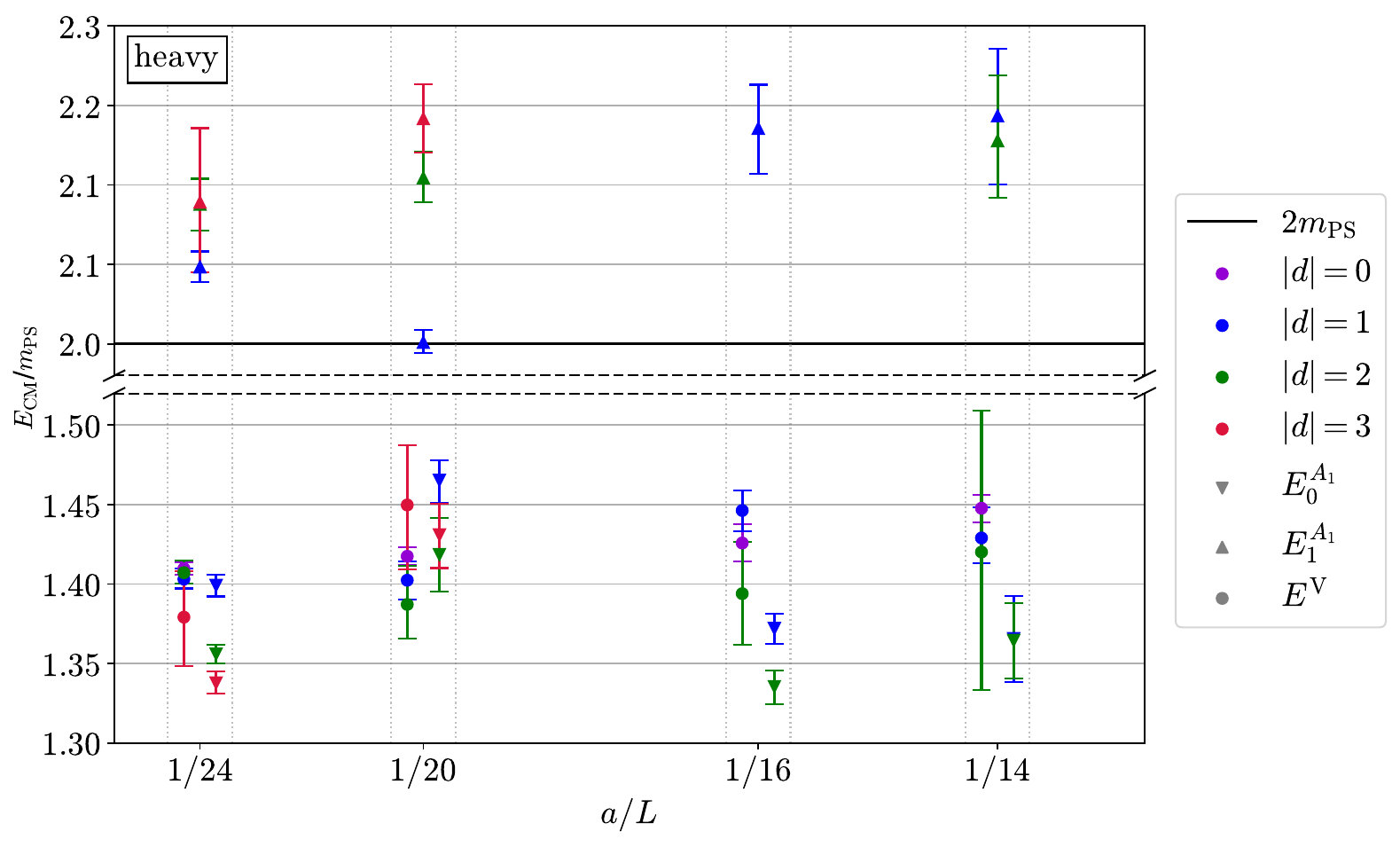}
    \caption{Energy levels, $E_{\rm CM}/m_{\rm PS}$, in the spin-1, $\textbf{10}$ channel, displayed against the inverse lattice extent, $a/L=1/N_s$, in the heavy ensembles ($\beta=6.9$ and $am_0=-0.92$). We split up the plot in two different energy ranges to improve visibility. For presentation reasons, we also introduce a horizontal off-set to distinguish energy levels found in different irreducible representations. The energy levels from the irreducible representations $T_1$, $E$, and $B_1$ are obtained by  single-meson operator analysis, and shown as circles, slightly shifted to the left. The energy levels from $A_1$  are obtained by solving a $3\times 3$ eigenvalue problem, and shown as upward (downward) facing triangles for the excited (ground) state level, with no (small, right) horizontal off-set. We distinguish by color the energy levels obtained  with operators with different total momentum. The two-PNGB threshold is shown as a solid horizontal line.}
    \label{fig:E_L_10_69_92}
\end{figure}

\begin{figure}[t]
    \centering
    \includegraphics[scale=0.4]{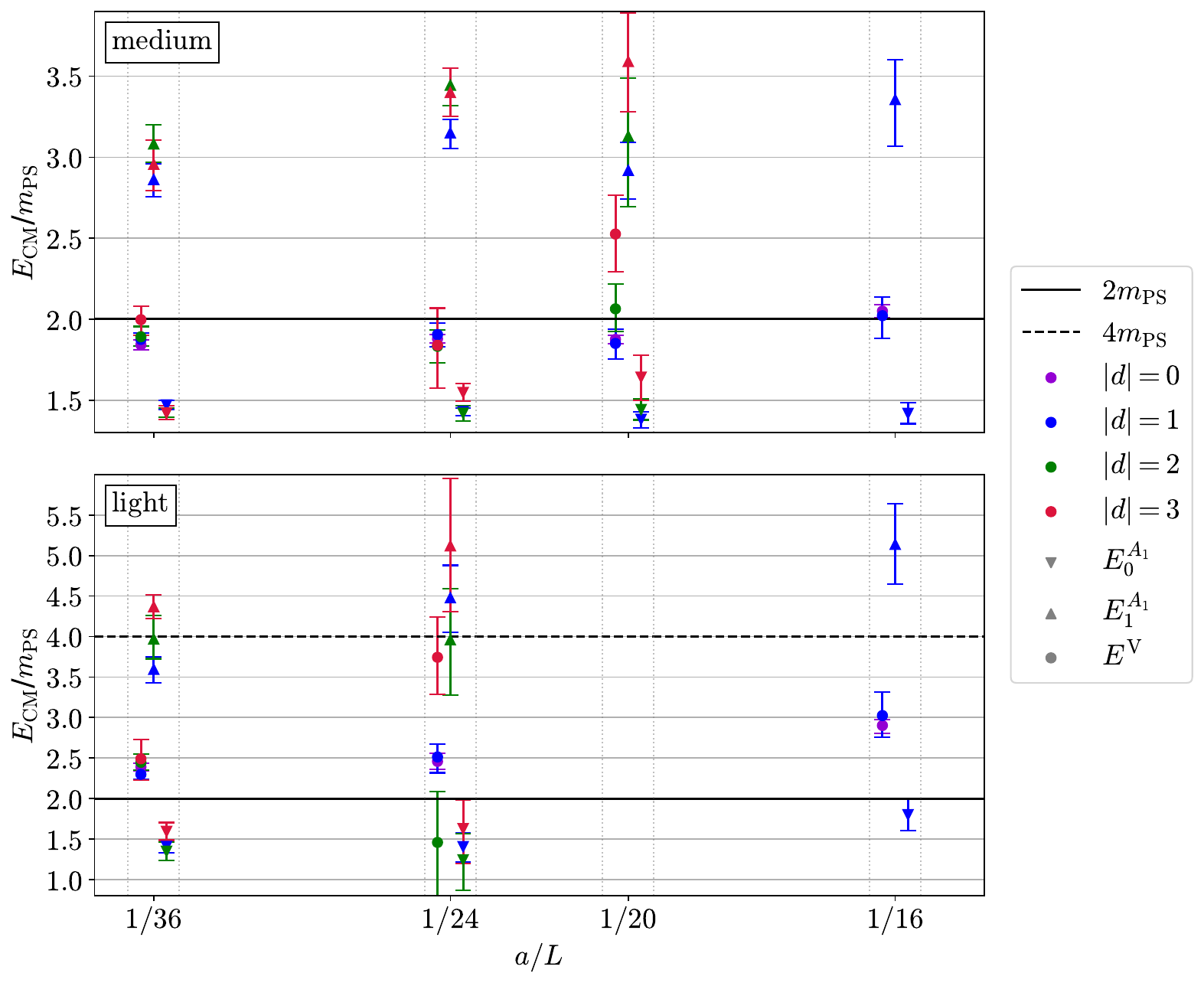}
    \caption{
   Energy levels, $E_{\rm CM}/m_{\rm PS}$, in the spin-1, $\textbf{10}$ channel, displayed against the inverse lattice extent, $a/L=1/N_s$, in the medium (top, $\beta=7.05$ and $am_0=-0.683$) and light (bottom, $\beta=7.05$ and $am_0=-0.867$) ensembles. We split up the plot in two different energy ranges to improve visibility. For presentation reasons, we also introduce a horizontal off-set to distinguish energy levels found in different irreducible representations. The energy levels from the irreducible representations $T_1$, $E$, and $B_1$ are obtained by  single-meson operator analysis, and shown as circles, slightly shifted to the left. The energy levels from $A_1$  are obtained by solving a $3\times 3$ eigenvalue problem, and shown as upward (downward) facing triangles for the excited (ground) state level, with no (small, right) horizontal off-set. We distinguish by color the energy levels obtained  with operators with different total momentum. The two-PNGB threshold is shown as a solid horizontal line, and the 4-PNGB one as a dashed horizontal line. }
    \label{fig:E_L_10_non_res_and_res}
\end{figure}

We generally observe a good quality of the numerical signal for the ground state, in all ensembles and for all choices of $t_0$. Closer inspection shows the presence of statistically significant discrepancies in the value of the energy levels obtained with the GEVP analysis, when varying the choice of $t_0$. We find that this effect depends also on which ensemble we consider.  For the smallest choice of $t_0$, the energy of the ground state for the light ensemble is substantially higher than those obtained with larger values of $t_0$. Conversely, the energy of the first excited state (second energy level) is lower than for larger $t_0$. We observe the convergence of the GEVP results towards the EVP ones, when increasing the value of $t_0$.  Unfortunately, this process is limited by the degradation of the signal available at larger Euclidean times, for larger values of $t_0$, which prevents us from exploring choices of $t_0 > 5$. We therefore use the energy levels obtained from the EVP method for the remainder of this analysis. 

\begin{table}[t]
    \caption{Energy levels and results of L\"uscher's analysis in the heavy ensembles. For each spatial extent, $N_s$,  and momentum, $|\vec{d}|$, we obtain one energy level in the irreducible representations denoted as $T_1$, $E$, and $B_1$ coming from a single ${\rm V}$-operator and two energy levels, labeled by $n$, in the $A_1$ irreducible representation, obtained from the variational analysis. We tabulate the energy in the lattice frame, $a E$, the center-of-mass energy, $a \sqrt{s}$, and the p-wave phase-shift, in those cases in which the center-of-mass energy falls within the kinematical window allowed by elastic scattering.  As indicated in the last column, only data points from the $A_1$ irreducible representation are used in the fit of the phase shift.}
    \begin{tabular}[t]{|c|c|c|c|c|c|c|c|}
    \hline 
    $N_L$ & $|\vec{d}|^2$ & $\Lambda$ & $n$ & $aE$ & $a\sqrt{s}$ & $\delta_1$ & Incl. \\ \hline \hline
    14 & 1 & A1 & 1 & 0.693(8) & 0.53(1) & - & no \\
    14 & 1 & A1 & 2 & 0.97(2) & 0.86(3) & 0(200) & yes \\
    14 & 1 & E & 1 & 0.712(5) & 0.552(7) & - & no \\
    14 & 2 & A1 & 1 & 0.825(6) & 0.527(9) & - & no \\
    14 & 2 & A1 & 2 & 1.06(2) & 0.85(2) & 17(3) & yes \\
    14 & 2 & B1 & 1 & 0.84(2) & 0.55(3) & - & no \\
    \hline
    16 & 0 & T1 & 1 & 0.551(4) & 0.551(5) & - & no \\
    16 & 1 & A1 & 1 & 0.660(3) & 0.530(4) & - & no \\
    16 & 1 & A1 & 2 & 0.94(2) & 0.85(2) & 200(200) & no \\
    16 & 1 & E & 1 & 0.683(4) & 0.559(5) & - & no \\
    16 & 2 & A1 & 1 & 0.758(3) & 0.516(4) & - & no \\
    16 & 2 & B1 & 1 & 0.774(9) & 0.54(1) & - & no \\
    \hline
\end{tabular}
\quad
\begin{tabular}[t]{|c|c|c|c|c|c|c|c|}
    \hline 
    $N_L$ & $|\vec{d}|^2$ & $\Lambda$ & $n$ & $aE$ & $a\sqrt{s}$ & $\delta_1$ & Incl. \\ \hline \hline
    20 & 0 & T1 & 1 & 0.548(2) & 0.548(2) & - & no \\
    20 & 1 & A1 & 1 & 0.648(5) & 0.566(6) & - & no \\
    20 & 1 & A1 & 2 & 0.835(4) & 0.774(4) & - & no \\
    20 & 1 & E & 1 & 0.626(4) & 0.542(5) & - & no \\
    20 & 2 & A1 & 1 & 0.706(7) & 0.548(9) & - & no \\
    20 & 2 & A1 & 2 & 0.945(8) & 0.834(9) & 18(4) & yes \\
    20 & 2 & B1 & 1 & 0.696(7) & 0.536(9) & - & no \\
    20 & 3 & A1 & 1 & 0.776(5) & 0.553(8) & - & no \\
    20 & 3 & A1 & 2 & 1.01(1) & 0.86(1) & 31(8) & yes \\
    20 & 3 & E & 1 & 0.78(1) & 0.56(2) & - & no \\
    \hline
    24 & 0 & T1 & 1 & 0.545(2) & 0.545(2) & - & no \\
    24 & 1 & A1 & 1 & 0.601(2) & 0.541(3) & - & no \\
    24 & 1 & A1 & 2 & 0.843(5) & 0.801(6) & 5(2) & yes \\
    24 & 1 & E & 1 & 0.602(2) & 0.542(3) & - & no \\
    24 & 2 & A1 & 1 & 0.642(2) & 0.524(2) & - & no \\
    24 & 2 & A1 & 2 & 0.903(9) & 0.824(10) & 16(7) & yes \\
    24 & 2 & B1 & 1 & 0.658(2) & 0.544(3) & - & no \\
    24 & 3 & A1 & 1 & 0.688(2) & 0.517(3) & - & no \\
    24 & 3 & A1 & 2 & 0.94(2) & 0.82(3) & 40(20) & yes \\
    24 & 3 & E & 1 & 0.700(9) & 0.53(1) & - & no \\
    \hline
\end{tabular}

    \label{tab:table_10_non_res}
\end{table}

\begin{table}[t]
    \centering
    \caption{Energy levels and results of L\"uscher's analysis in the medium ensembles. For each spatial extent, $N_s$,  and momentum, $|\vec{d}|$, we obtain one energy level in the irreducible representations denoted as $T_1$, $E$, and $B_1$ coming from a single ${\rm V}$-operator and two energy levels, labeled by $n$, in the $A_1$ irreducible representation, obtained from the variational analysis. We tabulate the energy in the lattice frame, $a E$, and the center-of-mass energy, $a \sqrt{s}$, as well as the p-wave phase-shift, in those cases in which the center-of-mass energy falls within the kinematical window allowed by elastic scattering. None of these measurements are included in the fit of the phase shifts.}
    \begin{tabular}[t]{|c|c|c|c|c|c|c|c|}
    \hline 
    $N_L$ & $|\vec{d}|^2$ & $\Lambda$ & $n$ & $aE$ & $a\sqrt{s}$ & $\delta_1$ & Incl. \\ \hline \hline
    16 & 1 & A1 & 1 & 0.489(8) & 0.29(1) & - & no \\
    16 & 1 & A1 & 2 & 0.79(5) & 0.69(6) & 140(20) & no \\
    16 & 1 & E & 1 & 0.57(2) & 0.42(3) & - & yes \\
    \hline
    20 & 0 & T1 & 1 & 0.386(5) & 0.386(5) & - & no \\
    20 & 1 & A1 & 1 & 0.424(7) & 0.28(1) & - & no \\
    20 & 1 & A1 & 2 & 0.68(3) & 0.60(4) & 140(20) & no \\
    20 & 1 & E & 1 & 0.49(1) & 0.38(2) & - & yes \\
    20 & 2 & A1 & 1 & 0.534(8) & 0.30(1) & - & no \\
    20 & 2 & A1 & 2 & 0.78(6) & 0.64(9) & 130(40) & no \\
    20 & 2 & B1 & 1 & 0.61(2) & 0.42(3) & - & yes \\
    20 & 3 & A1 & 1 & 0.64(1) & 0.34(3) & - & no \\
    20 & 3 & A1 & 2 & 0.92(5) & 0.74(6) & 130(20) & no \\
    20 & 3 & E & 1 & 0.75(3) & 0.52(5) & 90(40) & yes \\
    \hline
\end{tabular}
\quad
\begin{tabular}[t]{|c|c|c|c|c|c|c|c|}
    \hline 
    $N_L$ & $|\vec{d}|^2$ & $\Lambda$ & $n$ & $aE$ & $a\sqrt{s}$ & $\delta_1$ & Incl. \\ \hline \hline
    24 & 0 & T1 & 1 & 0.387(6) & 0.387(5) & - & no \\
    24 & 1 & A1 & 1 & 0.394(4) & 0.294(5) & - & no \\
    24 & 1 & A1 & 2 & 0.70(2) & 0.65(2) & 80(20) & no \\
    24 & 1 & E & 1 & 0.47(1) & 0.39(2) & - & no \\
    24 & 2 & A1 & 1 & 0.472(6) & 0.292(10) & - & no \\
    24 & 2 & A1 & 2 & 0.80(2) & 0.71(3) & 60(30) & no \\
    24 & 2 & B1 & 1 & 0.53(2) & 0.38(2) & - & no \\
    24 & 3 & A1 & 1 & 0.554(6) & 0.32(1) & - & no \\
    24 & 3 & A1 & 2 & 0.83(2) & 0.70(3) & 110(20) & no \\
    24 & 3 & E & 1 & 0.59(3) & 0.38(5) & - & yes \\
    \hline
    36 & 0 & T1 & 1 & 0.379(7) & 0.379(7) & - & no \\
    36 & 1 & A1 & 1 & 0.349(5) & 0.303(6) & - & no \\
    36 & 1 & A1 & 2 & 0.61(2) & 0.59(2) & 110(50) & no \\
    36 & 1 & E & 1 & 0.424(7) & 0.386(9) & - & no \\
    36 & 2 & A1 & 1 & 0.383(4) & 0.293(6) & - & no \\
    36 & 2 & A1 & 2 & 0.68(2) & 0.63(2) & 60(50) & no \\
    36 & 2 & B1 & 1 & 0.46(1) & 0.39(1) & - & no \\
    36 & 3 & A1 & 1 & 0.421(6) & 0.293(9) & - & no \\
    36 & 3 & A1 & 2 & 0.68(3) & 0.61(3) & 70(30) & no \\
    36 & 3 & E & 1 & 0.51(2) & 0.41(2) & - & no \\
    \hline
\end{tabular}

    \label{tab:table_10_close_res}
\end{table}

\begin{table}[t]
    \centering
    \caption{Energy levels and results of L\"uscher's analysis in the light ensembles.  For each spatial extent, $N_s$,  and momentum, $|\vec{d}|$, we obtain one energy level in the irreducible representations denoted as $T_1$, $E$, and $B_1$ coming from a single ${\rm V}$-operator and two energy levels, labeled by $n$, in the $A_1$ irreducible representation, obtained from the variational analysis. We tabulate the energy in the lattice frame, $a E$, and the center-of-mass energy, $a \sqrt{s}$, as well as the p-wave phase-shift, in those cases in which the center-of-mass energy falls within the kinematical window allowed by elastic scattering. As indicated in the last column, we only include measurements obtained with from single-meson operator,  in the irreducible representations labelled as $E$, $B_1$, and $T_1$, in the fit of the phase shift.}
    \begin{tabular}[t]{|c|c|c|c|c|c|c|c|}
    \hline 
    $N_L$ & $|\vec{d}|^2$ & $\Lambda$ & $n$ & $aE$ & $a\sqrt{s}$ & $\delta_1$ & Incl. \\ \hline \hline
    16 & 1 & A1 & 1 & 0.47(2) & 0.27(3) & - & no \\
    16 & 1 & A1 & 2 & 0.86(7) & 0.76(7) & - & no \\
    16 & 1 & E & 1 & 0.60(3) & 0.45(4) & 140(10) & yes \\
    \hline
    24 & 0 & T1 & 1 & 0.36(1) & 0.36(1) & 147(8) & yes \\
    24 & 1 & A1 & 1 & 0.33(2) & 0.21(3) & - & no \\
    24 & 1 & A1 & 2 & 0.71(6) & 0.66(6) & - & no \\
    24 & 1 & E & 1 & 0.46(2) & 0.37(3) & 140(10) & yes \\
    24 & 2 & A1 & 1 & 0.41(2) & 0.19(5) & - & no \\
    24 & 2 & A1 & 2 & 0.69(8) & 0.6(1) & - & no \\
    24 & 2 & B1 & 1 & 0.43(5) & - & - & no \\
    24 & 3 & A1 & 1 & 0.51(3) & 0.24(6) & - & no \\
    24 & 3 & A1 & 2 & 0.88(10) & 0.8(1) & - & no \\
    24 & 3 & E & 1 & 0.72(5) & 0.55(7) & - & no \\
    \hline
\end{tabular}
\quad
\begin{tabular}[t]{|c|c|c|c|c|c|c|c|}
    \hline 
    $N_L$ & $|\vec{d}|^2$ & $\Lambda$ & $n$ & $aE$ & $a\sqrt{s}$ & $\delta_1$ & Incl. \\ \hline \hline
    36 & 0 & T1 & 1 & 0.354(7) & 0.354(7) & 119(7) & yes \\
    36 & 1 & A1 & 1 & 0.272(9) & 0.21(1) & - & no \\
    36 & 1 & A1 & 2 & 0.56(2) & 0.53(2) & 80(40) & no \\
    36 & 1 & E & 1 & 0.382(8) & 0.340(9) & 137(9) & yes \\
    36 & 2 & A1 & 1 & 0.32(1) & 0.20(2) & - & no \\
    36 & 2 & A1 & 2 & 0.64(4) & 0.59(4) & - & no \\
    36 & 2 & B1 & 1 & 0.44(1) & 0.36(1) & 120(10) & yes \\
    36 & 3 & A1 & 1 & 0.384(10) & 0.24(2) & - & no \\
    36 & 3 & A1 & 2 & 0.71(2) & 0.65(2) & - & no \\
    36 & 3 & E & 1 & 0.48(3) & 0.37(4) & 60(60) & yes \\
    \hline
\end{tabular}

    \label{tab:table_10_res}
\end{table}

\subsection{Energy Levels}\label{ssec:energy_levels}

We display the finite-volume energy levels in the channel with spin-1, transforming as a $\textbf{10}$ multiplet of ${\rm Sp}(4)$, for the three sets of ensembles called heavy, medium, and light, in Figs.~\ref{fig:E_L_10_69_92} and~\ref{fig:E_L_10_non_res_and_res}, and tabulate them in Tabs.~\ref{tab:table_10_non_res}, \ref{tab:table_10_close_res}, and~\ref{tab:table_10_res}. We provide the energy in the lattice frame, $aE$, as well as the center-of-mass energy, $a\sqrt{s}$. If the latter is within the elastic scattering window, we determine and tabulate also the phase shift, $\delta_1$. We observe that we can only extract a phase shift for a subset of all channels, since a majority of the determined energy levels lie outside the elastic window. 

In the case of the heavy ensembles ($\beta=6.9$ and $am_0=-0.92$), Fig.~\ref{fig:E_L_10_69_92} shows a good agreement between the energy levels obtained with single-meson operators, labeled with $E^V$, and the ground state of the $A_1$ irreducible representation. These energy levels lie consistently below the kinematical threshold for decay to two PNGBs, as expected from the results of the mass estimates obtained by fitting 2-point functions involving single-meson operators. We can identify these states as stable vector-meson particles. The energy levels of the first excited state from the solutions of the EVP analysis lie just above the two-PNGB threshold, and we interpret them as scattering states. 

The top panel of Fig.~\ref{fig:E_L_10_non_res_and_res} shows the result for the channel with spin-1, transforming as a $\textbf{10}$, in the medium ensembles, for which the lattice action has $\beta=7.05$ and $am_0=-0.863$. While the energy levels obtained with single-meson operators are slightly below, or consistent with, the kinematical threshold for decay to two PNGBs, yet the ground state obtained from the analysis of the $A_1$ irreducible representation is significantly below the threshold. Especially at the largest available lattice, the errors are small, and this effect appears to be statistically significant. We interpret this result as first, preliminary evidence of the presence of a bound state, which went undetected in the single-meson analysis. 

A similar effect is visible also in the light ensembles ($\beta=7.05$ and $am_0=-0.867$), as shown in the bottom panel of Fig.~\ref{fig:E_L_10_non_res_and_res}, though it takes a different form. The ground state energies extracted from the $A_1$ irreducible representation are again significantly below threshold, albeit their measurements are affected by larger errors. In contrast to the medium ensembles, the energy levels obtained with single-meson operators are here above the kinematical threshold for decay to two PNGBs. We interpret these two states as a stable ground state and a resonant state, respectively. We shall test this conclusion on the grounds of  L\"uscher's analysis. 

\subsection{Phase Shifts}\label{ssec:phase_shift}

The phase shift, $\delta_1$, is obtained with the formulas given in Tab.~\ref{tab:phase_shift_formulas} for the different irreducible representations. For those ensembles in which the finite-volume energy is within the elastic scattering kinematical region, we report  $\delta_1$ in Tabs.~\ref{tab:table_10_non_res}, \ref{tab:table_10_close_res}, and \ref{tab:table_10_res}, for the heavy, medium, and light ensembles, respectively. We then use these measurements to parameterize the PNGB scattering cross-section, either in the ERE or Breit-Wigner form, as discussed in Sect.~\ref{sec:scattering}. We do not include all the measurements of the phase shift in this analysis, though, but only the subset indicated in  the last column of Tabs.~\ref{tab:table_10_non_res}-\ref{tab:table_10_res}. Before discussing our final results, we explain our measurement selection criteria.

\begin{figure}[b]
    \centering
    \includegraphics[scale=0.4]{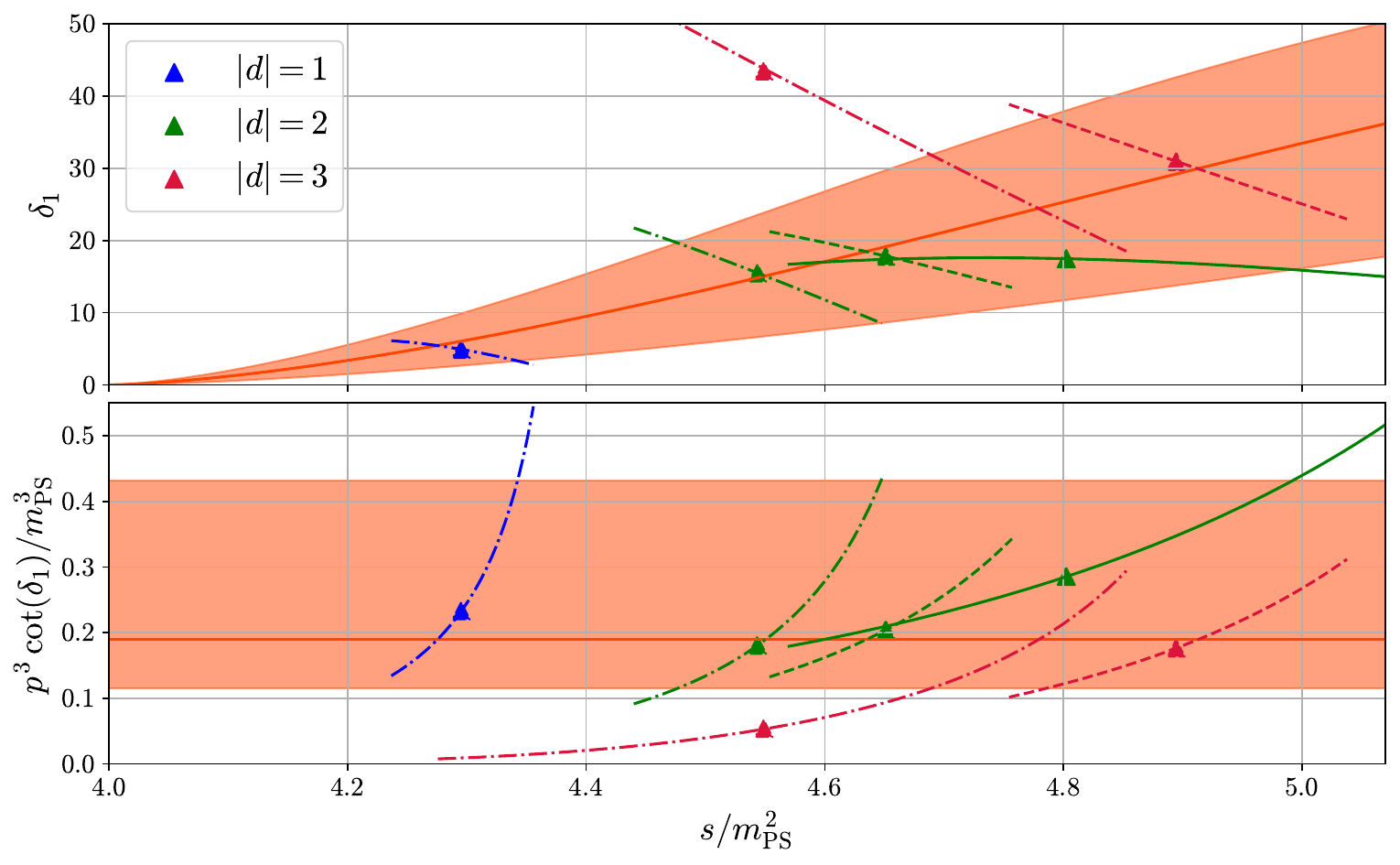}
    \caption{The results of our analysis, using L\"uscher's method, in the heavy ensembles. The p-wave phase shift, $\delta_1$, is shown in the top panel. The left-hand side of p-wave ERE from Eq.~(\ref{eq:ERE}) is shown in the bottom panel. In both panels, the horizontal axis is the center-of-mass energy squared, $s$, expressed in units of  the mass of the PNGBs, $m_{\rm PS}$. All data points are obtained as first excited states extracted from the EVP analysis of the operators in the $A_1$ irreducible representation. We indicate different lattice momenta by color. The orange line and band show the median and error estimate emerging from fit to a zeroth-order,  constant  ERE.}
    \label{fig:10_scat_res_non_res}
\end{figure}

\begin{figure}[t]
    \centering
    \includegraphics[scale=0.4]{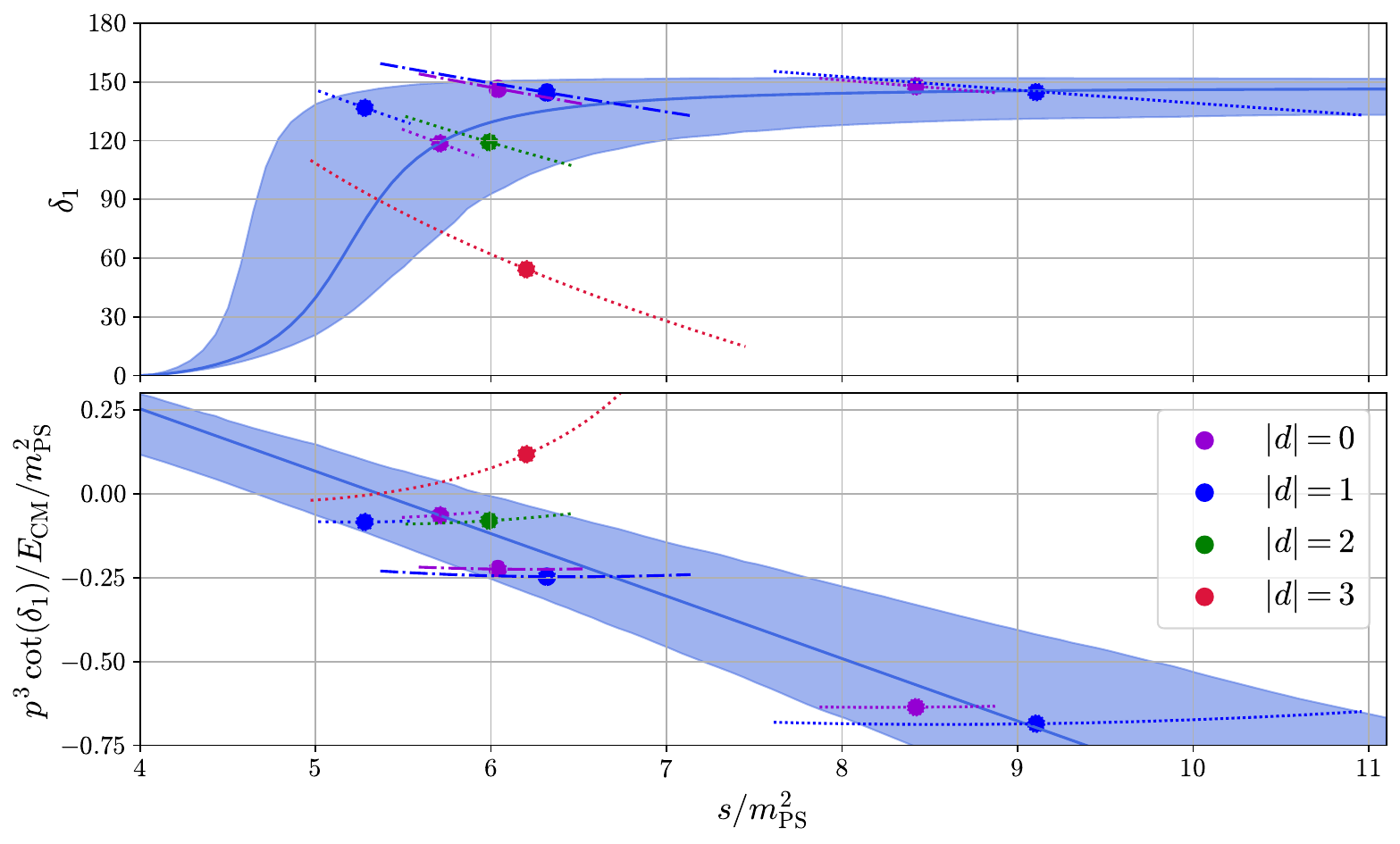}
    \caption{The results of our analysis, using L\"uscher's method, in the light  ensembles. The p-wave phase shift is shown in the top panel. The combination of quantities appearing  
  in the bottom panel is motivated by Eq.~(\ref{eq:breit_wigner_linear}), as a Breit-Wigner ansatz results in a straight line in this plot. In both panels, the horizontal axis is the center-of-mass energy squared, $s$, expressed in units of  the mass of the PNGBs, $m_{\rm PS}$. All data points are obtained from the analysis of single-meson operators, in different irreducible representations,  depending on the momentum employed. We indicate different momenta by color. The blue line and band show the median and error estimate of a fit to Breit-Wigner ansatz.}
    \label{fig:10_scat_res_res}
\end{figure}

Firstly, L\"uscher's analysis is highly sensitive to the presence of large errors in the measurement of the energy levels, hence we exclude measurements with large error bars from our final analysis.  Furthermore, as discussed in Sect.~\ref{sec:scattering}, the parameterizations we adopted can be trusted only in certain ranges of momenta. We further exclude measurements that lie outside those ranges; these include measurements with large energies, for ERE, or far away from the resonance position, in the Breit-Wigner case. The fits of the phase shift are then performed using a restricted set of measurements, and only these are shown in  the figures that illustrate this final part of the analysis, Figs.~\ref{fig:10_scat_res_non_res} and~\ref{fig:10_scat_res_res}. 

The upper panel of Fig.~\ref{fig:10_scat_res_non_res} shows our measurements of phase shift, $\delta_1$, obtained in the heavy ensembles. We do not see a hint of a resonance, and therefore we examine the data points using p-wave ERE, given by Eq.~(\ref{eq:ERE}). We show the resulting fit in the lower panel of the figure. We find agreement with a constant fit using zeroth-order ERE, in which the only parameter is the scattering length, $a_1$. The mean and error of the fit are shown as the orange line and band, respectively. We extract the scattering length to be $a_1m_{\rm PS}=-(1.76^{+0.11}_{-0.47})$, with  $\chi^2/N_{\rm d.o.f.} = 0.35$. The negative sign of the scattering length corresponds to an attractive interaction. 

The inclusion of higher-order terms in the fit of the same measurements yield values for the scattering length that are compatible with the zeroth-order fit. At the same time, the errors on the effective range, $r_1$, or on coefficients appearing at even higher orders are larger than their values, making such fit inconclusive. The finding that we cannot constrain well  higher-order terms agrees with the fact that the data do not show an explicit dependence on $s$. We can use these results to calculate the partial-wave cross-section via Eq.~(\ref{eq:partial_wave_cross_section_ERE}). We return to this task in Sec.~\ref{Sec:SIMP}, in order to compare with the cross-section reported elsewhere in the literature, in the repulsive, $\textbf{14}$ scattering channel~\cite{Dengler:2024maq}.

The upper panel of Fig.~\ref{fig:10_scat_res_res} shows our measurements of the phase shift, $\delta_1$, in the light ensembles. As we expect a resonance in this channel, we examine the data with a Breit-Wigner ansatz.  Because of the preliminary evidence we collected that  a lighter state is present, this resonance should be compared in QCD to the excited $\rho(1450)$ state, rather than the $\rho(770)$. For this reason,  we refer to it as ${\rm V}^{\prime}$. We rewrite Eq.~(\ref{eq:phase_shift_breit_wigner})  as 
\begin{align}
    \frac{p^{3}\cot \delta_1}{\sqrt{s}} = \frac{6\pi}{g_{\rm V^{\prime}PP}^2} (m_{\rm V^{\prime}}^2-s). \label{eq:breit_wigner_linear}
\end{align}
We use the left-hand side of this relation, normalized to the mass squared of the PNGBs, to parameterize our measurements, and display it  in the bottom panel of Fig.~\ref{fig:10_scat_res_res}, as in this form we expect the energy dependence to take a linear form, if the Breit-Wigner ansatz holds. 
The linear fit using Eq.~(\ref{eq:breit_wigner_linear}) yields $m_{\rm V^{\prime}}/m_{\rm PS}=2.31^{+0.18}_{-0.10}$ and $g_{\rm V^{\prime}PP}=10.3^{+1.6}_{-1.0}$, with an associated  $\chi^2/N_{\rm d.o.f.}=1.5$. 

Unfortunately, the experimental data for the $\rho(1450)$ in QCD is not good enough to reliably estimate $g_{\rho(1450)\pi\pi}$, given the large number of open channels \cite{ParticleDataGroup:2024cfk}, and hence a direct comparison is not possible at this stage.
An improved study that  includes a variational analysis in all irreducible representations, and uses an extended set of ensembles, including  lower values for the center-of-mass energy, would greatly improve the reach of this exploratory work, and allow a direct comparison with other theories. For example, studies of this type in the $SU(2)$~\cite{Drach:2020wux}  and $SU(3)$~\cite{Alexandrou:2017mpi} gauge theories coupled to two flavors of fundamental fermions yield values of $g_{\rm VPP}= 7.8\text{ and } 5.76$, respectively.

\section{Implications for a SIMP dark matter model}
\label{Sec:SIMP}

In this section, we focus our attention on the potential implications of our results for models of dark matter in the SIMP framework~\cite{Hochberg:2014dra, Hochberg:2014kqa,Kulkarni:2022bvh}. Arguments based on astrophysical observations and numerical simulations make self-interacting dark matter models potentially viable, provided they exhibit non-trivial physics at very low, non-relativistic energies~\cite{Eckert:2022qia, Sagunski:2020spe, Andrade:2020lqq, Kaplinghat:2015aga,Adhikari:2022sbh}. While this observation is still the subject of current investigations, if confirmed, such a behaviour would require non-trivial structures to emerge in close proximity to the elastic threshold. In particular, in channels other than the $s$-wave, a non-trivial contribution to the relevant cross-sections is only possible if a resonance appears close to threshold, to avoid angular momentum suppression. Thus, it is of special interest to investigate theories that admit the existence of such resonances, in view of their potential as dark matter models~\cite{Hochberg:2014dra, Hochberg:2014kqa,Kulkarni:2022bvh}.

The theory investigated in this study, the ${\rm Sp}(4)$ theory coupled to $N_f=2$ Dirac fermions transforming in the fundamental representation of the gauge group, is one prominent example of this possibility. The range of lattice parameter space we explored, in particular with the ensembles summarized in Table~\ref{tab:ensembles}, includes the most relevant one for the aforementioned scenario. Starting from the ensembles analyzed in Sect.~\ref{Sec:spectra}, we saw that by dialling the (Wilson) fermion mass, we can lower the mass of both the PNGBs and the spin-1 (${\rm V}$) mesons transforming as a  $\textbf{10}$ of Sp$(4)$ (that correspond to the $\rho$ mesons of QCD). At the same time, when doing so the ratio of their masses grows from $m_{\rm V}/m_{\rm PS}\sim 1.4\div 1.6$, up towards the threshold for decay of the ${\rm V}$ meson to two PNGBs, at which point the spectral analysis based on single-meson operators needs to be replaced. The analysis we conducted in Sects.~\ref{ssec:energy_levels} and~\ref{ssec:phase_shift}, which is based on L\"uscher's method, and considers the ensembles in Table~\ref{tab:ensembles}, while preliminary and subject to methodological caveats~\cite{Kulkarni:2022bvh}, confirms this trend. Furthermore, the results of section~\ref{ssec:phase_shift} strongly suggests that a vector resonance close to the elastic threshold is present  in our light ensembles,\footnote{With our current data we cannot distinguish unambiguously whether this resonance may have moved down in mass with decreasing fermion mass, or appears newly below a critical fermion mass. A larger number of data sets would be needed to test this behaviour explicitly. Nevertheless, the comparison with the  energy levels measured with single-meson operators suggests that the resonance appears once this level crosses the elastic threshold.} confirming that, by dialling the fermion mass, we may be able to tune  the mass of the vector resonance and bring it close to threshold. 

\begin{figure}[t]
    \centering
    \includegraphics[scale=0.4]{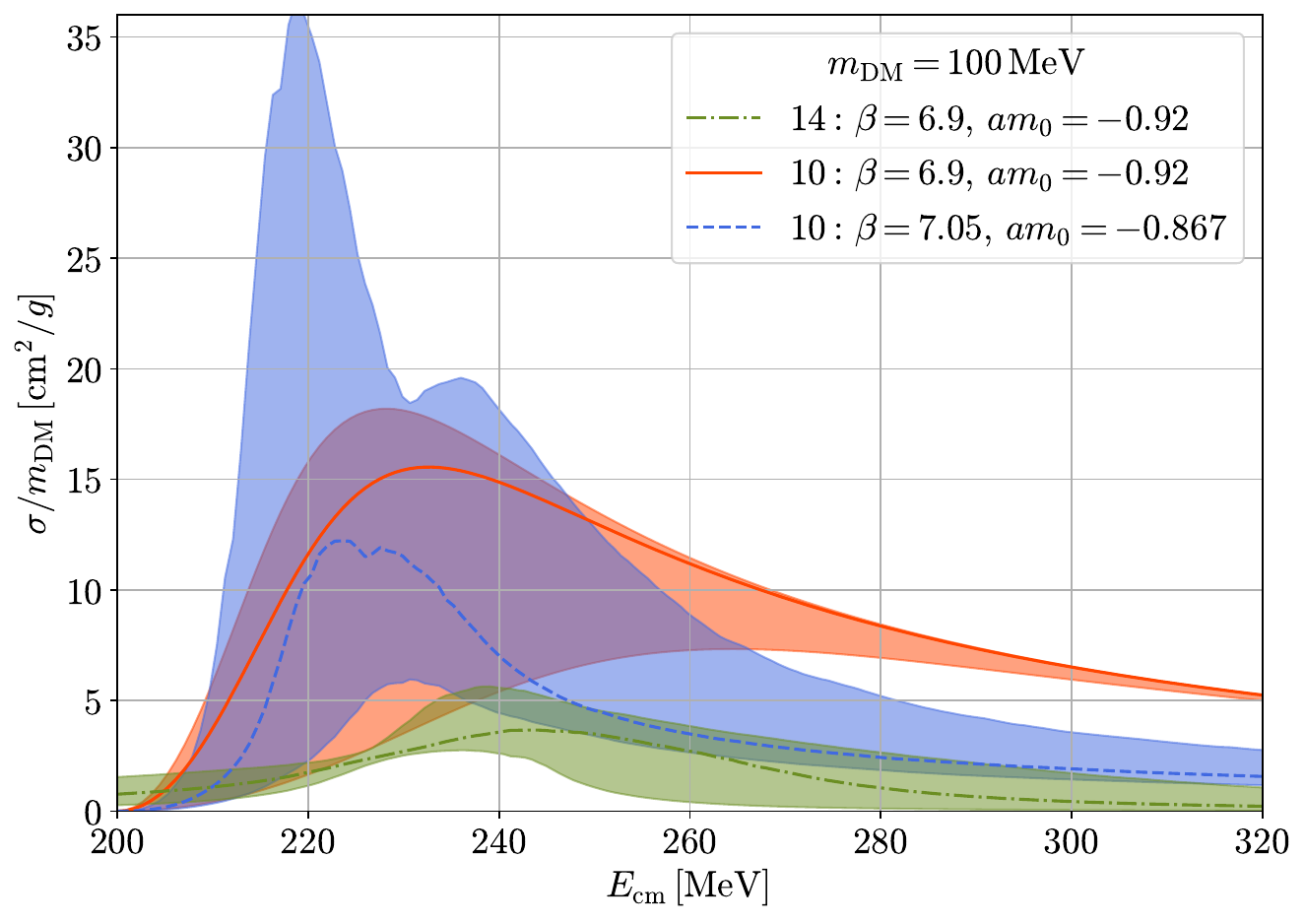}
    \caption{The cross-sections derived for the heavy ($\beta=6.9$, $a m_0=-0.92$) and light  ($\beta=7.05$, $a m_0=-0.867$) ensembles in Table~\ref{tab:ensembles}, labelled as $\textbf{10}$ of Sp$(4)$, as described in the main text. The curves labelled as $\textbf{14}$ are result for the maximal-isospin channel that is free of resonances analyzed in Ref.~\cite{Dengler:2024maq}. For the sole purpose of making this illustrative comparison, we set the units by assigning the mass of the PNGBs to be $100$ MeV, a typical scale consistent with astrophysical observations \cite{Kulkarni:2022bvh}. }
    \label{fig11:xs}
\end{figure}

To illustrate the potential phenomenological impact of the presence of resonances near threshold,  in Fig.~\ref{fig11:xs} we display  the cross-section, $\sigma_1$, in the spin-one channel, computed in terms of the phase shift, $\delta_1$, as in Sect.~\ref{sec:scattering}~\cite{Dengler:2024maq}:
\begin{equation}\label{eq:sigma(l)}
    \sigma_1(p) = 12\pi\frac{1}{p^{2}}\sin^2(\delta_1)\,.
\end{equation}
Inserting, for $\delta_1$, the results from Sect.~\ref{ssec:phase_shift}, and setting the PNGB mass as the dark matter candidate to $100$ MeV, which is a reasonable choice based on observational data~\cite{Dengler:2024maq,Kulkarni:2022bvh}, yields, for the heavy and light ensembles, the two main curves in Fig.~\ref{fig11:xs}. For comparison, in the figure we also display a representative example obtained by the analysis of the scalar, $\textbf{14}$ channel, taken from Ref.~\cite{Dengler:2024maq}. The two main trends discussed before are directly visible. At very low energies ($\lesssim 202$ MeV) the scalar channel dominates. Nevertheless, the resonance enhances the cross-section quite significantly, at energies a little higher than this. Within the existing ensembles, the location of the resonance in our lattice results is yet too far away from threshold to have a significant impact. Given that the relevant energies are around typical galactic escape velocities of $\sim 1000$ km/s$\approx 0.003c$, yielding $E_\text{CM}\lesssim 201$ MeV, the resonance would need to move within a few MeV to threshold to be effective in changing the total cross-section. Yet, as the figure shows, in this case the effect of the spin-1 channel would dominate the cross-section.

Our current exploratory results suggest that, at least in principle, and subject to dialling the mass of the resonance to tune its appearance near threshold, the cross-section could be enhanced to change the phenomenology significantly. In order to perform a lattice study in which the resulting tuning of the parameters would put this statement on firm grounds is quite challenging with current lattice methods, as it  would require the repeated tuning of the lattice parameters to ambitiously high precision, particularly in the process of extrapolating to the continuum limit. Yet,  further investigations on multiple ensembles in which the spin-1 meson transforming as a $\textbf{10}$ of $Sp(4)$ is just above and just below threshold might allow determining  the relevant parameters by interpolation. Using such a parameter set to constrain the typical low-energy effective theory in the spirit of \cite{Kulkarni:2022bvh} would yield a unique test of the viability of the present scenario for describing strongly-interacting dark matter.

\section{Summary and Outlook} 
\label{Sec:outlook}

We performed a detailed lattice investigation of the meson spectroscopy, and the first measurements of scattering amplitude and resonances in the spin-1 channel, in the ${\rm Sp}(4)$ gauge theory coupled to $N_f=2$ mass-degenerate flavors of Wilson-Dirac fermions transforming in the fundamental representation of the gauge group. The associated continuum theory is a leading candidate for extensions of the Standard Model in which composite particles arising from the underlying dynamics explain the physics of the Higgs boson and of dark matter. It provides a partial completion of the CHM paradigm, and a realization of the SIMP paradigm. Its main distinctive features are due to the pseudo-real nature of fundamental representation,  that leads to  the enhancement of the global symmetry to ${\rm SU}(4)$, broken explicitly and spontaneously to the ${\rm Sp}(4)$ maximal subgroup. Gauge invariant operators are built from even numbers of fermion fields, so that all bound states are bosons. The spectrum of this theory is expected to resemble that of the closely related two-flavor version of QCD,  in particular by giving rise to confinement and spontaneous symmetry breaking of the approximate global symmetries. 

We adopted the Wilson formulation of fermions without a clover term. We scanned a significant portion of the accessible lattice parameter space. We provided the most up-to-date results for the spectroscopy of flavored mesons in this theory, for choices of the (degenerate) fermion masses for which the decay to PNGBs is kinematically forbidden, using a basis of flavored, single-meson operators. This paper, in combination with the investigations of the unflavored, spin-zero channels reported in Ref.~\cite{Bennett:2023rsl}, and with the explicit flavor symmetry breaking effects studied in Ref.~\cite{Kulkarni:2022bvh}, provides a relatively complete picture of the low-energy spectroscopy  of the theory, in this kinematical region. 

We also conducted the first exploration of the low-mass regime, taking advantage of the Hasenbusch acceleration method to gain access to this complementary kinematical regime. We analyzed correlation functions with an extended basis, including multi-meson states. Besides extending the reach of our spectroscopy study,  we could access scattering processes and couplings between mesons. For the available ensembles with the lightest masses, we measured the scattering of two PNGBs, and the resonance structure, in the flavored spin-1, $\textbf{10}$,  channel, using L\"uscher's methodology. We found indications of a resonance emerging when the Wilson fermion mass is small enough.  In combination with the study of the scalar, $\textbf{14}$,  channel discussed in Ref.~\cite{Dengler:2024maq}, this study provides a proof of principle that precision measurements of spectroscopy, scattering amplitudes, and meson couplings can be carried out on the lattice also in the low-mass region of the lattice parameter space of this theory.

The potential phenomenological impact of the near-threshold and low-mass regimes of this theory, in the complementary CHM and SIMP contexts (and in the absence of direct access to experimental data) remains largely to be explored. To pursue the challenging programme of quantifying precisely the phenomenological properties of these theories, future studies would require to analyze an extended set of ensembles, and implement L\"uscher's analysis with a larger numbers of operators in the variational basis. In order to approach more effectively the physically interesting regions of parameter space in the continuum theory, it might be necessary to change the lattice action, by adopting a different type of fermion formulation, and improving the convergence to the continuum limit. This study sets the stage for such a future programme.

\acknowledgments

We thank Luka Leskovec for sharing his numerical implementation of the Zeta function and for the helpful discussions. We also thank Vincent Drach for helping us in utilizing the Hasenbusch acceleration method implemented in the HiRep code. 

EB is supported by the STFC Research Software Engineering Fellowship EP/V052489/1. 
EB, BL, MP and FZ are supported by the STFC Consolidated Grant No. ST/X000648/1.
BL is supported in part by the STFC Consolidator Grant No. ST/X00063X/1.
DKH is supported by Basic Science Research Program through the National Research Foundation of Korea (NRF) funded by the Ministry of Education (NRF-2017R1D1A1B06033701). 
J-WL is supported by IBS under the project code, IBS-R018-D1. 
HH and C-JDL acknowledge support from NSTC Taiwan, through grant number 112-2112-M-A49-021-MY3. 
C-JDL is also supported by the Taiwanese MoST grant 109-2112-M-009-006-MY3. 
C-JDL is supported by Grants No. 112-2639-M-002-006-ASP and No. 113-2119-M-007-013-.
DV is supported by STFC under Consolidated Grant No. ST/X000680/1.
YD is supported by the Austrian Science Fund research teams grant STRONG-DM (FG1).
FZ acknowledges support from the Advanced ERC grant ERC-2023-ADG-Project EFT-XYZ.

{\bf High performance computing}---This work used the DiRAC Data Intensive service (CSD3) at the University of Cambridge,  the DiRAC Data Intensive service (DIaL3) at the University of Leicester and the DiRAC Extreme Scaling service (Tursa) at the University of Edinburgh, managed respectively by the University of Cambridge University Information Services, the University of Leicester Research Computing Service and by EPCC on behalf of the STFC DiRAC HPC Facility (www.dirac.ac.uk). The DiRAC service at Cambridge, Leicester, and Edinburgh are funded by BEIS, UKRI and STFC capital funding and STFC operations grants. DiRAC is part of the UKRI Digital Research Infrastructure.

Numerical simulations have been performed on the Swansea SUNBIRD cluster (part of the Supercomputing Wales project) and on the NURION at KISTI. The Swansea SUNBIRD system is part funded by the European Regional Development Fund (ERDF) via Welsh Government. The NURION system is  supported by the National Supercomputing Center with supercomputing resources including technical support (KSC-2023-CRE-0549 and KSC-2024-CRE-0530).

Calculations were performed using supercomputer resources provided by the Austrian Scientific Computing center (ASC), in  particular using the Vienna Scientific Cluster (VSC4).

{\bf Research Data Access Statement}---The data generated and the analysis code for this manuscript can be downloaded from  Ref.~\cite{workflow_release,data_release}. 

{\bf Open Access Statement}---For the purpose of open access, the authors have applied a Creative Commons 
Attribution (CC BY) license to any Author Accepted Manuscript version arising.

\appendix

\section{Wick diagrams}
\label{app:wick_contractions}

In this Appendix, we provide explicitly the information encoded in the diagrams depicted in Fig.~\ref{fig:wick_contractions}.  These technical details are needed in order to reproduce our calculations of the energy levels, which we extracted from correlation functions built using a basis comprising  both single- and two-meson operators. The decomposition in irreducible representations of the little group is listed in Table~\ref{tab:little_groups}. 

The full inversion of the lattice Dirac operator, to obtain the fermion propagator, is prohibitively expensive. For the problem at hand, we further require all-to-all propagators for the two-PNGB operators. For our calculations, we use stochastic sources~\cite{Foley:2005ac} to sample the Dirac operator with a finite number, $N_{\rm src}$, of source vectors, $| \eta_i \rangle$, which satisfy the following condition for $N_{\rm src}\gg 1$,  
\begin{equation}
    \delta_{\alpha\beta}\delta^{ab}\delta(x,y)  = \lim_{N_{\text{src}} \to \infty} \frac{1}{N_{\text{src}}} \sum_{i=1}^{N_{\text{src}}} | \eta_i(x) \rangle^{a}_{\alpha} \langle \eta_i(y)|^{b}_{\beta} \,. \label{eq:stochastic_sources_delta}
\end{equation}
For the purposes of PNGB scattering, we follow Refs.~\cite{CP-PACS:2007wro, Drach:2020wux}, and employ stochastic noisy sources with spin dilution to provide a stochastic estimate of the fermion propagator.  A correlation function of interest is obtained as the average over multiple stochastic sources. We can sequentially invert the Dirac operator by taking the result of one inversion as the starting point  for the next inversion. 

A general box diagram, like those  appearing in Fig.~\ref{fig:wick_contractions}, is given by
\begin{equation}
    C_{\rm box}=\sum_{\substack{\vec{x}_1,\vec{x}_2 \\ \vec{y}_1,\vec{y}_2}} e^{i\vec{p_1}(\vec{x}_1-\vec{y}_1)}e^{i\vec{p_2}(\vec{x}_2-\vec{y}_2)} {\rm Tr}\left[ S(x_1|y_1) \Gamma S(y_1|y_2) \Gamma S(y_2|x_2) \Gamma S(x_2|x_1) \Gamma \right], \label{eq:wick_contract_four_quark}
\end{equation}
where we have omitted the color and Dirac indices (summed over) and $\Gamma$ could be an arbitrary gamma structure. For  this  Appendix, we write explicitly the results obtained with $\Gamma=\gamma_5$, which simplifies some of the algebra, by taking advantage of the $\gamma_5$-hermiticity property, $\gamma_5 S(x|y) \gamma_5 = S^\dagger(y|x)$, of the propagator.  The two spacetime coordinates $x_1$ and $x_2$ belong to the time slice of the sources where the state is generated, while $y_1$ and $y_2$ are located at the sink. The subscripts $1$ and $2$ label the two mesons and are used to identify which momentum they are connected to. In this study,  we only consider the case where the sink momenta are identical to the source momenta, but the formalism can be applied in more general cases. 

In order to evaluate Eq.~(\ref{eq:wick_contract_four_quark}), we insert a set of sources to rewrite the identity matrix as given in \eqref{eq:stochastic_sources_delta} and find the following expression, valid for a single stochastic source:
\begin{align}
C_{\rm box}(t)&=\sum_{\substack{\vec{x}_1,\vec{x}_2 \\ \vec{y}_1,\vec{y}_2,\vec{z}}} 
e^{i\vec{p_1}(\vec{y}_1-\vec{x}_1)}e^{i\vec{p_2}(\vec{y}_2-\vec{x}_2)} {\rm Tr}[S(x_1|y_1) \gamma_5 S(y_1|y_2)|\eta(y_2)\rangle \langle\eta(z)|  \gamma_5 S(z|x_2) \gamma_5 S(x_2|x_1) \gamma_5 ] \\
&=\sum_{\substack{\vec{x}_1,\vec{x}_2 \\ \vec{y}_1,\vec{y}_2,\vec{z}}} 
e^{i\vec{p_1}(\vec{y}_1-\vec{x}_1)}e^{i\vec{p_2}(\vec{y}_2-\vec{x}_2)}
{\rm Tr}[S(x_1|y_1) \gamma_5 S(y_1|y_2)|\eta(y_2)\rangle 
\langle\eta(z)| S^\dagger(z|x_2)\gamma_5 S^\dagger(x_2|x_1) ] \\
&=\sum_{\substack{\vec{x}_1,\vec{x}_2 \\ \vec{y}_1,\vec{z}}} 
e^{i\vec{p_1}(\vec{y}_1-\vec{x}_1)}e^{-i\vec{p_2}\vec{x}_2} 
{\rm Tr}[S(x_1|y_1) \gamma_5 Q_\eta(\vec{y}_1,t_s|\vec{q},t_s)
Q_\eta^\dagger(\vec{x_2},t_0|\vec{0},t_s) \gamma_5 S^\dagger(x_2|x_1) ] \\ 
&=\sum_{\vec{x}_1} e^{-i\vec{p_1}\vec{x}_1} {\rm Tr}[ W_\eta(\vec{x}_1,t|\vec{p_1},t_s|\vec{p_2},t_s)W_\eta^\dagger(\vec{x}_1,t|\vec{p_2},t_0|\vec{0},t_s)]\,,\label{eq:wick_contract_seq_sources}
\end{align}
where $t_0$ and $t_s$ are the source and sink time, respectively, while $t=|t_0-t_s|$. The correlation function is then estimated by averaging over all stochastic sources. In the final two equations, we have introduced the sequential sources, defined as follows~\cite{CS:2011vqf}:
\begin{align}
Q_\eta(\vec{y},t_2|\vec{q},t_3) &\equiv \sum_{\vec{z}} e^{i\vec{q}\vec{z}} S(\vec{y},t_2|\vec{z},t_3) |\eta(\vec{z},t_3)\rangle\,, \\
W_\eta(\vec{x},t_1|\vec{p},t_2|\vec{q},t_3) &\equiv \sum_{\vec{y}} e^{i\vec{y}\vec{p}} S(\vec{x},t_1|\vec{y},t_2)\gamma_5 Q_\eta(\vec{y},t_2|\vec{q},t_3)\,.\label{eq:def_seq_sources}
\end{align}

Our final expressions, for the diagrams involving two-PNGB operators, can be written as follows:
\begingroup
\allowdisplaybreaks
\begin{align}
    R_1&=\sum_{\vec{x}} e^{-i\vec{p_1}\vec{x}} {\rm Tr}\left[ W_\eta(\vec{x},t| \vec{p_1},t_s |\vec{0},t_s)  W_\eta^\dagger(\vec{x},t| \vec{p_2},t_0|-\vec{p_2},t_s)\right], \\
    R_2&=\sum_{\vec{x}} e^{-i\vec{p_1}\vec{x}} {\rm Tr}\left[ W_\eta(\vec{x},t|-\vec{p_2},t_0 |\vec{0},t_s)  W_\eta^\dagger(\vec{x},t|-\vec{p_2},t_s|-\vec{p_1},t_s)\right], \\
    R_3&=\sum_{\vec{x}} e^{-i\vec{p_1}\vec{x}} {\rm Tr}\left[ W_\eta(\vec{x},t| \vec{p_1},t_s |\vec{0},t_s)  W_\eta^\dagger(\vec{x},t| \vec{p_2},t_0|-\vec{p_2},t_s)\right], \\
    R_4&=\sum_{\vec{x}} e^{-i\vec{p_1}\vec{x}} {\rm Tr}\left[ W_\eta(\vec{x},t|-\vec{p_2},t_0 |\vec{0},t_s)  W_\eta^\dagger(\vec{x},t|-\vec{p_1},t_s|-\vec{p_2},t_s)\right], \\
    D_1&=\sum_{\vec{x},\vec{y}} e^{-i(\vec{p_1}\vec{x}+\vec{p_2}\vec{y})} {\rm Tr}\left[ Q_\eta^\dagger(\vec{x},t|\vec{0},t_s)Q_\eta(\vec{x},t|\vec{p_1},t_s)\right]{\rm Tr}\left[ Q_\xi^\dagger(\vec{y},t|\vec{0},t_s)Q_\xi(\vec{y},t|\vec{p_2},t_s)\right], \\
    D_2&=\sum_{\vec{x},\vec{y}} e^{-i(\vec{p_1}\vec{x}+\vec{p_1}\vec{y})} {\rm Tr}\left[ Q_\eta^\dagger(\vec{x},t|\vec{0},t_s)Q_\eta(\vec{x},t|\vec{p_2},t_s)\right]{\rm Tr}\left[ Q_\xi^\dagger(\vec{y},t|\vec{0},t_s)Q_\xi(\vec{y},t|\vec{p_2},t_s)\right], \\
    T_1&=\sum_{\vec{x}} e^{-i\vec{p_1}\vec{x}} {\rm Tr}\left[ W_\eta^\dagger(\vec{x},t|-\vec{p_1},t_s|\vec{0},t_s)(\gamma_5\gamma_i)Q_\eta(\vec{x},t|\vec{0},t_s)\right], \\
    T_2&=\sum_{\vec{x}} e^{-i\vec{p_1}\vec{x}} {\rm Tr}\left[ Q_\eta^\dagger(\vec{x},t|\vec{0},t_s)(\gamma_5\gamma_i)W_\eta(\vec{x},t|\vec{p_1},t_s|\vec{0},t_s)\right].
\end{align}
\endgroup

\section{Single meson analysis, technical details and intermediate numerical results}

For the purposes of this work, we extended the available data and improved the analysis in respect to previous studies of the spectroscopy of flavored mesons in the $N_f=2$, $Sp(4)$ theory~\cite{Bennett:2019jzz}, extracted from correlation functions involving only single-meson operators. We generated new ensembles with lighter dynamical fermions and finer lattices. Furthermore, we employed Wuppertal smearing~\cite{Gusken:1989qx,Alexandrou:1990dq,Roberts:2012tp} and APE smearing~\cite{APE:1987ehd,Falcioni:1984ei}, and extracted the single-meson energy levels with a GEVP variational method. The combination of these technological improvements reduces the effect of excited-state contamination in the measurement of masses and decay constants. The pertinent single-meson interpolating operators and their quantum numbers are listed in Tab.~\ref{tab:spec_operators}. We report in this Appendix technical details about the ensembles, and describe the process used for the measurement of masses and decay constants. We also discuss finite volume effects and present additional details about the continuum extrapolations of the lattice measurements, complementing our main results, displayed in the main body of the paper.

\subsection{Ensembles,  scale setting, and topology}
\label{app:ensembles}

\begin{table}
\caption{%
\label{tab:spec_ens}
List of ensembles generated for the single-meson spectroscopy study,  extending Table~1 in Ref.~\cite{Bennett:2019jzz}. Some of the new ensembles (marked by $\ast$),  have been generated by adopting Hasenbusch acceleration.  We tabulate for each the lattice bare coupling, $\beta$,  the fermion mass, $am_0$, the lattice temporal and spatial directions, $N_t$ and $N_s$, the number of configurations, $N_{\rm config.}$, the separation, in trajectories, between adjacent configurations, $\delta_{\rm traj.}$, and
the bin size in the bootstrap resampling method, $N_{\rm bin}$. The autocorrelation times,  $\tau_{\mathrm{int}}^{\langle P \rangle}$ and $\tau_{\mathrm{int}}^{w_0/a}$, are measured in units of number of trajectories. They are estimated using either the average plaquette, $\langle P \rangle$, or the gradient flow scale, $w_0/a$, respectively.
The mean and standard deviation, $Q_0$ and $\sigma_Q$, of the topological charge are computed with a Gaussian fit. 
We report the statistical uncertainties in parentheses.
}
\begin{center}
\begin{tabular}{|c|c|c|c|c|c|c|c|c|c|c|c|c|c|c|}
\hline\hline
Ensemble & $\beta$ & $m_0$ & $N_t$ & $N_s$ & $N_{\rm config.}$ & $\delta_{\rm traj.}$ & $N_{\rm bin}$& $\tau_{\mathrm{int}}^{\langle P \rangle}$ & $\tau_{\mathrm{int}}^{w_0/a}$ & $ \langle P \rangle$ & $w_0/a$ & $Q_0$ & $\sigma_Q$  \\
\hline\hline
DB1M1 & 6.9 & -0.85 & 32 & 16 & 100 & 24 & 2 & $25.6(6.0)$ & $18.2(5.4)$ & $0.546753(61)$ & $0.73285(64)$ & $1.4(1.5)$ & $8.9(1.3)$\\
DB1M2 & 6.9 & -0.87 & 32 & 16 & 100 & 24 & 2 & $20.9(5.6)$ & $38.2(6.9)$ & $0.550525(66)$ & $0.7800(10)$ & $0.7(1.2)$ & $7.5(1.0)$\\
DB1M3 & 6.9 & -0.89 & 32 & 16 & 100 & 24 & 4 & $38.6(7.0)$ & $32.9(6.6)$ & $0.554785(83)$ & $0.8445(17)$ & $-2.0(1.4)$ & $7.9(1.3)$\\
DB1M4 & 6.9 & -0.9 & 32 & 16 & 75 & 32 & 3 & $40.1(9.8)$ & $53(11)$ & $0.556961(90)$ & $0.8862(25)$ & $0.0(1.4)$ & $7.0(1.2)$\\
DB1M5 & 6.9 & -0.91 & 32 & 16 & 185 & 12 & 5 & $14.1(2.3)$ & $19.5(2.6)$ & $0.559520(43)$ & $0.9448(15)$ & $-0.41(62)$ & $6.49(52)$\\
DB1M6 & 6.9 & -0.92 & 32 & 24 & 498 & 28 & 6 & $33.6(3.3)$ & $59.5(4.1)$ & $0.562077(19)$ & $1.01683(81)$ & $-1.17(61)$ & $10.79(48)$\\
DB1M7 & 6.9 & -0.924 & 32 & 24 & 125 & 48 & 2 & $55(11)$ & $51(11)$ & $0.563183(30)$ & $1.0551(12)$ & $1.2(1.5)$ & $9.9(1.4)$\\
\hline
DB2M1 & 7.05 & -0.835 & 36 & 20 & 100 & 20 & 4 & $18.2(4.7)$ & $40.0(6.3)$ & $0.575269(34)$ & $1.1831(27)$ & $0.6(1.1)$ & $7.29(98)$\\
DB2M2 & 7.05 & -0.85 & 36 & 24 & 100 & 24 & 4 & $24.4(5.9)$ & $36.0(6.8)$ & $0.577371(31)$ & $1.2937(30)$ & $-0.4(1.4)$ & $7.8(1.2)$\\
DB2M3$^\ast$ & 7.05 & -0.857 & 36 & 32 & 175 & 24 & 8 & $22.5(4.4)$ & $64(41)$ & $0.578308(17)$ & $1.3576(13)$ & $-9(15)$ & $17(17)$\\
DB2M4$^\ast$ & 7.05 & -0.863 & 36 & 36 & 298 & 20 & 6 & $26.2(3.1)$ & $44.0(3.8)$ & $0.5792137(98)$ & $1.4341(12)$ & $-0.31(92)$ & $12.04(77)$\\
DB2M5$^\ast$ & 7.05 & -0.867 & 36 & 36 & 361 & 24 & 6 & $23.5(3.1)$ & $47.0(4.0)$ & $0.5798251(80)$ & $1.4995(14)$ & $-0.92(73)$ & $11.06(60)$\\
\hline
DB3M4 & 7.2 & -0.77 & 36 & 24 & 200 & 12 & 4 & $7.5(1.8)$ & $30.8(3.0)$ & $0.588460(13)$ & $1.4343(22)$ & $1.31(55)$ & $6.21(44)$\\
DB3M5 & 7.2 & -0.78 & 36 & 24 & 508 & 12 & 9 & $9.2(1.2)$ & $39.3(2.1)$ & $0.589278(11)$ & $1.4998(22)$ & $1.69(35)$ & $6.79(27)$\\
DB3M6 & 7.2 & -0.79 & 36 & 24 & 500 & 12 & 7 & $12.9(1.3)$ & $40.2(2.1)$ & $0.590127(10)$ & $1.5826(28)$ & $0.37(31)$ & $6.12(24)$\\
DB3M7 & 7.2 & -0.794 & 36 & 28 & 504 & 12 & 12 & $9.8(1.2)$ & $49.7(2.3)$ & $0.5904516(85)$ & $1.6177(29)$ & $0.11(36)$ & $6.88(28)$\\
DB3M8 & 7.2 & -0.799 & 40 & 32 & 451 & 12 & 10 & $11.7(1.4)$ & $56.6(2.6)$ & $0.5908623(65)$ & $1.6643(23)$ & $-0.33(44)$ & $7.93(36)$\\
DB3M9$^\ast$ & 7.2 & -0.803 & 42 & 36 & 133 & 24 & 4 & $16.6(4.5)$ & $49.7(6.7)$ & $0.5912427(93)$ & $1.7215(27)$ & $0.6(1.1)$ & $8.17(94)$\\
DB3M10$^\ast$ & 7.2 & -0.808 & 48 & 42 & 334 & 12 & 12 & $8.7(1.5)$ & $139.2(4.6)$ & $0.5916657(52)$ & $1.7896(31)$ & $0.04(69)$ & $9.52(59)$\\
DB3M11$^\ast$ & 7.2 & -0.813 & 48 & 36 & 500 & 12 & 10 & $9.0(1.2)$ & $41.5(2.1)$ & $0.5921347(50)$ & $1.8797(26)$ & $0.43(39)$ & $7.52(32)$\\
\hline
DB4M3$^\ast$ & 7.4 & -0.74 & 48 & 36 & 342 & 8 & 6 & $3.72(85)$ & $49.1(2.2)$ & $0.6060474(40)$ & $2.1787(32)$ & $1.06(44)$ & $6.74(38)$\\
DB4M4$^\ast$ & 7.4 & -0.75 & 48 & 32 & 780 & 8 & 10 & $7.22(68)$ & $141.5(2.4)$ & $0.6066482(42)$ & $2.3712(61)$ & $-0.50(17)$ & $4.50(13)$\\
DB4M5$^\ast$ & 7.4 & -0.755 & 48 & 42 & 215 & 8 & 10 & $4.4(1.1)$ & $77.8(3.5)$ & $0.6069061(52)$ & $2.4821(89)$ & $-1.83(32)$ & $4.01(26)$\\
\hline
DB5M3$^\ast$ & 7.5 & -0.71 & 48 & 36 & 250 & 8 & 16 & $4.6(1.0)$ & $128.0(4.1)$ & $0.6128190(56)$ & $2.397(11)$ & $1.29(43)$ & $5.32(33)$\\
DB5M4$^\ast$ & 7.5 & -0.72 & 48 & 36 & 399 & 8 & 10 & $4.91(85)$ & $43.2(1.9)$ & $0.6132577(36)$ & $2.5450(62)$ & $-0.40(22)$ & $4.02(18)$\\
\hline\hline
\end{tabular}

\end{center}
\end{table}

In order to improve the efficiency of our Monte-Carlo simulations, particularly in the region of small dynamical fermion mass, we supplement the ensembles generated with the HMC algorithm for the study in Ref.~\cite{Bennett:2019jzz}, with new ones that have been generated by exploiting the Hasenbusch mass-preconditioning method, in its existing  implementation in the HiRep code~\cite{Hasenbusch:2001ne,Bussone:2018mzi}. The ensembles used in this work are listed and characterized in Table~\ref{tab:spec_ens}.\footnote{
We keep the naming convention as in Ref.~\cite{Bennett:2019jzz}. We exclude some heavy ensembles that are not involved in the continuum extrapolations.} The asterisk beside an ensemble name indicates the application of the Hasenbusch method. 

We set the scale on the lattice by using the gradient flow of the Wilson gauge action method~\cite{Luscher:2010iy}. 
This process can be regarded as a diffusion equation (in Euclidean five-dimensional space) for a new gauge field, $B_{\mu}(t,x)$, at the fictitious flow time $t$, governed by the following equation: 
\begin{align}
\frac{\textrm{d}B_\mu (t,x)}{\textrm{d}t} = D_\nu G_{\nu\mu}(t,x),~\textrm{with}~B_\mu(0,x)=A_\mu(x)\,.
\label{Eq:gf}
\end{align}
Here, $D_\nu$ is the covariant derivative, written in terms of $B_\nu$, while 
$G_{\mu\nu}\equiv [D_\mu,D_\nu]$ is the field-strength tensor, and $A_\mu(x)$ is the four-dimensional non-Abelian gauge field, evaluated at the space-time coordinates $x$. 
After solving numerically the flow equation, we define the dimensionless observable, $\mathcal{W} (t)$, evaluated at a positive flow time, $t$,\footnote{In this subsection only, $t$ denotes the flow time, while everywhere else in the paper it stands for the lattice time coordinate.}  as
\begin{align}
  \label{eq:gradient_flow_definitions}
  \mathcal W(t) &\equiv \frac{\rm d }{\rm d \ln t}\left\{ t^2 \langle E(t) \rangle\right\}\,, 
\end{align}
where $\langle E(t)\rangle$ is the space and ensemble average of the quantity
\begin{align}
  \label{eq:gradient_flow_energy_density}
  E(t,x) &\equiv - \frac{1}{2} ~{\rm Tr} \, \left[G_{\mu\nu}(t,x) G_{\mu\nu}(t,x)\right]\,.
\end{align}

\begin{figure}[t]
    \centering
    \includegraphics[width=0.3\textwidth]{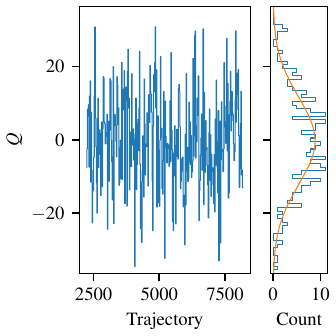} \hfill
    \includegraphics[width=0.3\textwidth]{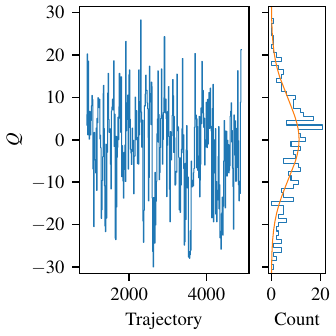} \hfill
    \includegraphics[width=0.3\textwidth]{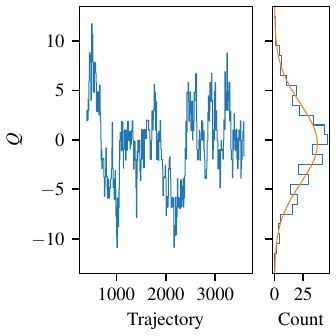}
    \caption{Left to right: representative examples of the history of the  topological charge, $Q$, and of its histogram, obtained with the HMC, for the ensembles DB2M4 ($\beta=7.05$, $am_0=-0.863$), DB3M10 ($\beta=7.2$, $am_0=-0.808$), and DB5M4 ($\beta=7.5$, $am_0=-0.72$). We include only measurements taken after discarding an initial simulation time interval, due to thermalization. The Gaussian fit of the resulting histograms are represented by solid orange lines.}
    \label{fig:topology}
\end{figure}

The gradient flow scale, $w_0$, is then set by imposing the condition
\begin{align}
    \mathcal{W}|_{t=w_0^2} \equiv \mathcal{W}_0\,,
\end{align}
for a conventionally chosen reference value of $\mathcal{W}_0$. In this work, we compute  $w_0$ for $\mathcal{W}_0 = 0.28125$.\footnote{
A different choice, $\mathcal{W}_0 = 0.35$ was used in Ref.~\cite{Bennett:2019jzz}; to understand the reasoning for changing to the convention used here, see the discussion in Ref.~\cite{Bennett:2022ftz}} In our lattice calculations, the field strength tensor is replaced by the four-plaquette clover operator. The lattice spacing, $a$, and all other dimensional quantities are expressed in units of the gradient flow scale. We write $\hat{a}\equiv a/w_0$, and similarly masses and decay constants are denoted as $\hat{m} \equiv m^{\rm lat} w_0^{\rm lat} = m w_0$ and $\hat{f} \equiv f^{\rm lat}w_0^{\rm lat}=f w_0$, respectively.

In order to monitor the extent of autocorrelation  resulting from our Markov chain updates, we measure the integrated autocorrelation time~\cite{Madras:1988ei,Wolff:2003sm,Luscher:2004pav}.
For a generic observable, $X$, the autocorrelation function, $\Gamma^{X}(\tau)$, is defined as
\begin{align}
\Gamma^X(\tau) = \sum_{i=1}^{N-\tau} \frac{\left( X_{i} - \bar X \right)\left(  X_{i+\tau} - \bar X  \right)}{N-\tau}\,,
\end{align}
where we denote by $X_i$ the measurements of the observable, $X$,  at Monte-Carlo time, $\tau=1,\,\dots,\,N$,  and by $\bar{X}$ their arithmetic mean. The integrated autocorrelation time,  $\tau_{\rm int}^X$, is then computed as follows:
\begin{align}
\tau^{X}_{\rm int} = \frac{1}{2} + \sum_{\tau=1}^{\tau_{\rm max}} \Gamma^{X}(\tau)\,.
\end{align}

During the generation of our ensembles, we monitor  the integrated autocorrelation time of the HMC by measuring the averaged plaquette value, $\langle P \rangle$, defined by the relation
\begin{align}
\langle P \rangle \equiv \frac{1}{12 N_t N_s^3}\sum_{x}\sum_{\mu>\nu} {\rm Re}\, {\rm Tr}\, \mathcal{P}_{\mu\nu}(x)\,,
\end{align}
where the local plaquette is $
\mathcal{P}_{\mu\nu}(x) \equiv U_\mu (x) U_\nu (x+\hat{\mu}) U^\dagger_\mu (x+\hat{\nu})U^\dagger_\nu(x)$. 
This quantity, $\tau^{\langle P \rangle}_{\rm int}$, dictates the separation between two subsequent Monte-Carlo trajectories, $\delta_{\rm traj.}$, that are retained as configurations in the spectral measurements, including the two-point functions and the gradient flow scale.  We then compute the integrated autocorrelation time of the gradient flow scale, $\tau^{w_0/a}_{\rm int}$, measured on the selected configurations. This second process determines the bin size, $N_{\rm bin}$, used in the bootstrap method for the error analysis, for which we generally require that  $\delta_{\rm traj.} \cdot N_{\rm bin} \gtrsim  2\tau^{w_0/a}_{\rm int}$. In a few cases of ensembles with a large estimate for $\tau^{w_0/a}_{\rm int}$,  in order to retain statistics, we determine the bin size by monitoring the error of $w_0/a$, while increasing $N_{\rm bin}$,  and select the value at which the error size reaches a plateau. We summarize our results for $\tau^{\langle P \rangle}_{\rm int}$, $\tau^{w_0/a}_{\rm int}$, $\delta_{\rm traj.}$ and $N_{\rm bin}$ together with lattice parameters in Tab.~\ref{tab:spec_ens}. We include only measurements taken after discarding an initial simulation time interval, due to thermalization of the HMC. 

Ergodicity requires sampling all topological sectors, hence we measure the topological charge, $Q$, over the HMC trajectories. A smoothing procedure is required to remove potentially large UV fluctuations with a magnitude larger than the desired signal. To this purpose, we apply the Wilson flow to the configurations, flowing to the flow time $t / a^2 = {N_t^2}/32$ (equivalent to $\sqrt{8t} = T/2$, where $T$ is the time extent of the lattice).
For each configuration, we compute $Q$ as 
\begin{align}
Q \equiv \frac{1}{32\pi^2} \sum_x \varepsilon^{\mu\nu\rho\sigma} \,\mathrm{Tr} \,\left\{ {\cal C}_{\mu\nu}(x) \frac{}{}{\cal C}_{\rho\sigma}(x)\right\}\,,
\end{align}
where ${\cal C}_{\mu\nu}(x)$ is the clover operator, defined using the same conventions as in Refs.~\cite{Sheikholeslami:1985ij,Hasenbusch:2002ai,Bennett:2022ftz}, computed with the flowed variables. We examine the distribution of $Q$, by fitting it to a Gaussian, with the functional form
\begin{align}
 n(Q) = A_n \exp \left( -\frac{(Q - Q_0)^2}{2\sigma_Q^2} \right)\,,
\label{eq:nQ}
\end{align}
where $Q_0$ and $\sigma_{Q}$ are the mean and the standard deviation of the Gaussian distribution, respectively.
We present the results of  the fit of these two parameters in Tab.~\ref{tab:spec_ens}, for all ensembles,  and show the HMC histories and distribution of representative ensembles in Fig.~\ref{fig:topology}. The histories and histograms in other ensembles display similar features.

\subsection{Meson masses and decay constants}
\label{app:meson_masses}

In order to measure the mass and decay constant of mesons, we define a general zero-momentum two-point correlation function at positive Euclidean time, $t$, as
\begin{align}
\label{eq:gen_corr}
C_{M}(t) &\equiv \sum_{\vec{x}} 
\langle 0 | \mathcal{O}^{\rm snk}_M(\vec{x},t) \mathcal{O}_M^{{\rm src},\dagger}(\vec{0},0) | 0\rangle \\
&= \sum_n \frac{1}{2E^M_n} \langle 0 \vert \mathcal{O}^{\rm snk}_{M} \vert \varphi^M_n \rangle \langle \varphi^M_n \vert   \mathcal{O}_M^{{\rm src},\dagger} \vert 0 \rangle \left[e^{-E^M_n t} + e^{-E^M_n (T-t)}\right] \,,
\end{align}
where ${O}_M^{{\rm src}}$ and $\mathcal{O}^{\rm snk}_M$ denote one of the single-meson, local interpolating operators listed in Tab.~\ref{tab:spec_operators}, sourcing meson $M$, inserted at the position of source and sink, respectively.
This correlation function can be decomposed in terms of a tower of states, $\varphi^M_n$, with different energy, $E^M_n$, of the $n$-th excited level of meson  $M$. For the ground state, we write $\varphi_0^M = M$. 
At large Euclidean time, $t$, correlation functions are dominated by the state with the lowest energy, $E_0$. At zero momentum,  the energy, $E_0^M=m_M$, provides an estimate of the mass, and the correlation functions behave as a hyperbolic cosine
\begin{align}
C_{M}(t)\xrightarrow{t\rightarrow \infty}\langle 0 |\mathcal{O}^{\rm snk}_M | M 
\rangle \langle 0 |\mathcal{O}^{\rm src}_M | M \rangle^* 
\frac{1}{2m_M} \left[e^{-m_M t}+e^{-m_M(T-t)}\right]\,.
\label{eq:corr_ground}
\end{align}

\begin{table}[t]
\caption{%
\label{tab_spec:extraction_1}
For each ensemble used in the single-meson correlation function analysis, we tabulate the parameters $\epsilon_{\rm W}$, $N_{\rm W}^{\rm diff}$, and $N_{\rm W}^{\rm max}$ indicating  the step size and the number of iterations associated with the Wuppertal smearing. The measurements of the PCAC mass, $am_{\mathrm{PCAC}}$, the pseudoscalar mass and decay constant, $am_{\mathrm{PS}}$ and $af_{\mathrm{PS}}$, in lattice units, are complemented by dimensionless products, $m_{\mathrm{PS}}L$ and $f_{\mathrm{PS}}L$, providing  guidance on assessing the physical size of available lattice volumes. }
\begin{center}
\begin{tabular}{|c|c|c|c|c|c|c|c|c|c|c|c|}
\hline\hline
Ensemble & $\beta$ & $m_0$ & $\varepsilon_{\rm W}$ & $N^{\mathrm{diff}}_{\rm W}$ & $N^{\mathrm{max}}_{\rm W}$ &  $am_{\mathrm{PCAC}}$ & $am_{\mathrm{PS}}$  & $af_{\mathrm{PS}} $ & $m_{\mathrm{PS}}L$ & $f_{\mathrm{PS}}L$\\
\hline\hline
DB1M1 & 6.9 & -0.85 & 0.1 & 30 & 90 & $0.14738(94)$ & $0.8375(13)$ & $0.1462(21)$ & $13.401(21)$ & $2.339(34)$ \\
DB1M2 & 6.9 & -0.87 & 0.1 & 40 & 80 & $0.11955(89)$ & $0.7353(26)$ & $0.1325(23)$ & $11.764(41)$ & $2.119(37)$ \\
DB1M3 & 6.9 & -0.89 & 0.1 & 60 & 120 & $0.08736(86)$ & $0.6274(26)$ & $0.1167(27)$ & $10.039(41)$ & $1.866(43)$ \\
DB1M4 & 6.9 & -0.9 & 0.12 & 50 & 100 & $0.07250(81)$ & $0.5582(29)$ & $0.1048(20)$ & $8.932(47)$ & $1.677(32)$ \\
DB1M5 & 6.9 & -0.91 & 0.12 & 60 & 120 & $0.05417(53)$ & $0.4843(16)$ & $0.0945(13)$ & $7.748(25)$ & $1.512(21)$ \\
DB1M6 & 6.9 & -0.92 & 0.12 & 70 & 140 & $0.03594(28)$ & $0.3857(11)$ & $0.08266(69)$ & $9.256(25)$ & $1.984(17)$ \\
DB1M7 & 6.9 & -0.924 & 0.16 & 80 & 160 & $0.02901(44)$ & $0.3414(15)$ & $0.0764(13)$ & $8.193(37)$ & $1.833(31)$ \\
\hline
DB2M1 & 7.05 & -0.835 & 0.1 & 40 & 80 & $0.05394(51)$ & $0.4397(16)$ & $0.0823(12)$ & $8.795(32)$ & $1.647(25)$ \\
DB2M2 & 7.05 & -0.85 & 0.18 & 50 & 100 & $0.03279(53)$ & $0.3285(24)$ & $0.0682(10)$ & $7.885(57)$ & $1.637(25)$ \\
DB2M3 & 7.05 & -0.857 & 0.18 & 60 & 120 & $0.02270(37)$ & $0.27526(89)$ & $0.06194(96)$ & $8.808(28)$ & $1.982(31)$ \\
DB2M4 & 7.05 & -0.863 & 0.16 & 90 & 180 & $0.01263(15)$ & $0.20483(97)$ & $0.05167(66)$ & $7.374(35)$ & $1.860(24)$ \\
DB2M5 & 7.05 & -0.867 & 0.18 & 90 & 180 & $0.00683(16)$ & $0.1509(11)$ & $0.04649(52)$ & $5.433(38)$ & $1.674(19)$ \\
\hline
DB3M4 & 7.2 & -0.77 & 0.12 & 60 & 120 & $0.05739(55)$ & $0.4224(12)$ & $0.07321(80)$ & $10.136(29)$ & $1.757(19)$ \\
DB3M5 & 7.2 & -0.78 & 0.14 & 60 & 120 & $0.04667(20)$ & $0.36890(83)$ & $0.06657(54)$ & $8.854(20)$ & $1.598(13)$ \\
DB3M6 & 7.2 & -0.79 & 0.16 & 60 & 120 & $0.03476(24)$ & $0.3113(11)$ & $0.05987(51)$ & $7.470(26)$ & $1.437(12)$ \\
DB3M7 & 7.2 & -0.794 & 0.18 & 70 & 140 & $0.02962(24)$ & $0.2875(11)$ & $0.05664(51)$ & $8.051(30)$ & $1.586(14)$ \\
DB3M8 & 7.2 & -0.799 & 0.18 & 80 & 160 & $0.02316(16)$ & $0.25348(83)$ & $0.05290(53)$ & $8.111(26)$ & $1.693(17)$ \\
DB3M9 & 7.2 & -0.803 & 0.18 & 90 & 180 & $0.01769(24)$ & $0.2206(17)$ & $0.04961(73)$ & $7.943(60)$ & $1.786(26)$ \\
DB3M10 & 7.2 & -0.808 & 0.2 & 90 & 180 & $0.01176(20)$ & $0.17591(65)$ & $0.04409(44)$ & $7.388(27)$ & $1.852(19)$ \\
DB3M11 & 7.2 & -0.813 & 0.24 & 90 & 180 & $0.00499(18)$ & $0.1161(12)$ & $0.03777(82)$ & $4.180(43)$ & $1.360(29)$ \\
\hline
DB4M3 & 7.4 & -0.74 & 0.18 & 80 & 160 & $0.02122(19)$ & $0.2154(11)$ & $0.04314(49)$ & $7.755(41)$ & $1.553(18)$ \\
DB4M4 & 7.4 & -0.75 & 0.2 & 80 & 160 & $0.00915(22)$ & $0.1404(11)$ & $0.03388(68)$ & $4.494(34)$ & $1.084(22)$ \\
DB4M5 & 7.4 & -0.755 & 0.24 & 90 & 180 & $0.00295(28)$ & $0.0849(31)$ & $0.0272(17)$ & $3.57(13)$ & $1.143(70)$ \\
\hline
DB5M3 & 7.5 & -0.71 & 0.2 & 90 & 180 & $0.02676(20)$ & $0.2253(10)$ & $0.04294(51)$ & $8.110(37)$ & $1.546(18)$ \\
DB5M4 & 7.5 & -0.72 & 0.24 & 90 & 180 & $0.01557(31)$ & $0.1678(19)$ & $0.03592(49)$ & $6.042(70)$ & $1.293(17)$ \\
\hline\hline
\end{tabular}

\end{center}
\end{table}

In order to improve the numerical quality of the signal, by reducing excited-state contamination in the ground state, and to access the first excited state, we apply a combination of smearing techniques and analyze the measurements with a variational method.  Our implementation of  Wuppertal smearing~\cite{Gusken:1989qx,Alexandrou:1990dq,Roberts:2012tp}  consists of  the following iterative diffusion process
\begin{align}
\label{eq:wuppertal}
q^{(n)}(x)\equiv \frac{1}{1+6\varepsilon_{\rm W} }\left[q^{(n-1)}(x)\frac{}{}+\frac{}{}\varepsilon_{\rm W} \sum_{\hat\mu} U_{\mu}(x) q^{(n-1)}(x+\hat{\mu})\right]\,,
\end{align}
where $\varepsilon_{\rm W}$ denotes the smearing step size, and the iterations are labelled by the integer $n=1,\cdots,N_{\rm W}$. In smearing the source, the starting point is a delta function localized at the position of the point-like source, so that  $q^{(0)}(x) \equiv \delta_{x,0}$ satisfies the Dirac equation, $\sum_{y, \beta,  b} D_{a \alpha, b \beta}(x,y) S^{b \beta}_{c \gamma} (y,0) = \delta_{x, 0} , \delta_{\alpha \gamma} ,\delta_{ac}$. In these expressions, $\alpha, \beta, \gamma$ denote spinor indices, while $a, b, c$ are color indices. The smearing of the sink acts iteratively by  applying Eq.~(\ref{eq:wuppertal}) to the fermion propagator, and by assigning $q^{(0)}(x) = S(x,0)$, at each position $x$.

We further implement  APE smearing~\cite{APE:1987ehd,Falcioni:1984ei} to smoothen the configurations, by smearing the gauge links in Eq.~(\ref{eq:wuppertal}) with a tunable parameter, $\alpha_{\textrm{APE}}$, entering the function
\begin{align}
\label{eq} U^{(m)}_{\mu}(x) \equiv \mathcal{P} \left( (1-\alpha_{\textrm{APE}}) U^{(m-1)}_{\mu}(x) + \frac{\alpha_{\textrm{APE}}}{6} S^{(m-1)}_{\mu}(x) \right)\,,
\end{align}
for $m = 1, \dots, N_{\rm APE}$. 
The initial condition, $U_{\mu}^{(0)} = U_{\mu}$, uses the gauge links, while the staple, $S_\mu$, is defined as $S_\mu^{(m)}(x) \equiv \sum_{\pm \nu \neq \mu} U_\nu^{(m)}(x)U^{(m)}_\mu(x+\hat{\nu})U^{(m),\dagger}_\nu(x+\hat{\mu})$.
The group projection, $\mathcal{P}$, ensures gauge invariance of closed loops after smearing. To incorporate the smearing in the two-point function, Eq.~(\ref{eq:gen_corr}) can be rewritten as
\begin{align}
C_{M}^{N_{\rm W}^{\rm src}, N_{\rm W}^{\rm snk} }(t) \equiv \sum_{\vec{x}} 
\langle 0 | \mathcal{O}^{N_{\rm W}^{\rm snk}}_M(\vec{x},t) \mathcal{O}_M^{{N_{\rm W}^{\rm src}},\dagger}(\vec{0},0) | 0\rangle \,,
\end{align}
which yields the correlation functions with $N_{\rm W}^{\rm src}$ iterations at the source and $N_{\rm W}^{\rm snk}$ iterations at the sink, all with the same smearing parameters of $\varepsilon_{\rm W}$, $\alpha_{\rm APE}$, and $N_{\rm APE}$. For each ensemble, we fix the smearing parameter, $\varepsilon_{\rm W}$, while the number of iterations in Wuppertal smearing for source, $N_{\rm W}^{\rm src}$, and sink, $N_{\rm W}^{\rm snk}$, are chosen from the set $\{ 0,~N^{\rm diff}_{\rm W},~2N^{\rm diff}_{\rm W},\cdots,~N^{\rm max}_{\rm W} \}$.
These tunable parameters, $\varepsilon_{\rm W}$, $N^{\rm diff}_{\rm W}$, and $N^{\rm max}_{\rm W}$ are reported in Tab.~\ref{tab_spec:extraction_1}.
The parameters controlling APE smearing are universal across all ensembles,  $(\alpha_{\rm APE}, N_{\rm APE}) = (0.4, 50)$.

\begin{table}[t]
\caption{%
\label{tab_spec:extraction_2}
For each ensemble used in the single-meson correlation function analysis, we tabulate the measurements of the vector and axial-vector masses, $am_{\mathrm{V}}$ and $am_{\mathrm{AV}}$, and their decay constants, $af_{\mathrm{V}}$ and $af_{\mathrm{AV}}$, in lattice units. The masses of the other mesons, $am_{\mathrm{T}}$, $am_{\mathrm{AT}}$, and $am_{\mathrm{S}}$, are also presented. The last column provides the energy eigenvalue of the first-excited state in the vector channel, $aE_1^{\rm V}$. The symbol “$\cdots$” indicates that no reliable signal could be extracted from the analysis.}
\begin{center}
\begin{tabular}{|c|c|c|c|c|c|c|c|c|c|c|}
\hline\hline
Ensemble & $\beta$ & $m_0$ & $am_{\mathrm{V}}$ & $af_{\mathrm{V}} $ & $am_{\mathrm{T}} $ & $am_{\mathrm{AV}}$  & $af_{\mathrm{AV}} $ & $am_{\mathrm{AT}}$  & $am_{\mathrm{S}}$ & $aE^{\mathrm{V}}_1$ \\
\hline\hline
DB1M1 & 6.9 & -0.85 & $0.9280(23)$ & $0.2365(40)$ & $0.9279(26)$ & $1.556(34)$ & $0.1687(32)$ & $1.55(10)$ & $1.531(43)$ & $\cdots$\\
DB1M2 & 6.9 & -0.87 & $0.8439(39)$ & $0.2237(44)$ & $0.8457(40)$ & $1.444(26)$ & $0.1717(67)$ & $1.402(23)$ & $1.466(40)$ & 1.54(10)\\
DB1M3 & 6.9 & -0.89 & $0.7412(64)$ & $0.2030(32)$ & $0.7458(66)$ & $1.286(32)$ & $0.1560(69)$ & $1.270(40)$ & $1.334(64)$ & 1.46(15)\\
DB1M4 & 6.9 & -0.9 & $0.6941(82)$ & $0.1908(37)$ & $0.6962(40)$ & $1.241(39)$ & $0.1539(75)$ & $1.233(31)$ & $1.189(36)$ & 1.376(95)\\
DB1M5 & 6.9 & -0.91 & $0.6283(55)$ & $0.1763(34)$ & $0.6282(26)$ & $1.096(19)$ & $0.1356(70)$ & $1.101(34)$ & $1.026(38)$ & 1.28(11)\\
DB1M6 & 6.9 & -0.92 & $0.5475(23)$ & $0.1597(12)$ & $0.5522(32)$ & $0.965(13)$ & $0.1300(67)$ & $0.962(21)$ & $0.869(46)$ & 1.117(29)\\
DB1M7 & 6.9 & -0.924 & $0.5107(44)$ & $0.1468(20)$ & $0.5143(50)$ & $0.931(70)$ & $0.1290(96)$ & $0.925(52)$ & $0.810(46)$ & 1.058(68)\\
\hline
DB2M1 & 7.05 & -0.835 & $0.5536(47)$ & $0.1412(17)$ & $0.5603(57)$ & $0.910(30)$ & $0.122(10)$ & $0.935(26)$ & $0.878(31)$ & 1.127(46)\\
DB2M2 & 7.05 & -0.85 & $0.4678(30)$ & $0.1242(14)$ & $0.4618(32)$ & $0.788(34)$ & $0.1150(96)$ & $0.791(20)$ & $0.767(23)$ & 0.838(47)\\
DB2M3 & 7.05 & -0.857 & $0.4188(28)$ & $0.1146(37)$ & $0.4192(56)$ & $0.709(12)$ & $0.1029(42)$ & $0.741(20)$ & $0.699(37)$ & 0.890(41)\\
DB2M4 & 7.05 & -0.863 & $0.3713(42)$ & $0.1015(24)$ & $0.3725(63)$ & $0.655(13)$ & $0.0959(21)$ & $0.665(11)$ & $0.643(21)$ & 0.830(38)\\
DB2M5 & 7.05 & -0.867 & $0.3438(41)$ & $0.0971(31)$ & $0.3421(40)$ & $0.570(23)$ & $0.0779(44)$ & $0.628(31)$ & $0.543(30)$ & 0.753(22)\\
\hline
DB3M4 & 7.2 & -0.77 & $0.5085(26)$ & $0.1204(12)$ & $0.5079(30)$ & $0.810(13)$ & $0.1042(22)$ & $0.817(15)$ & $0.759(16)$ & 0.915(44)\\
DB3M5 & 7.2 & -0.78 & $0.4648(15)$ & $0.1152(11)$ & $0.4659(19)$ & $0.762(16)$ & $0.1000(76)$ & $0.775(15)$ & $0.721(10)$ & 0.883(13)\\
DB3M6 & 7.2 & -0.79 & $0.4190(23)$ & $0.10449(96)$ & $0.4207(27)$ & $0.6744(99)$ & $0.0963(31)$ & $0.704(12)$ & $0.643(10)$ & 0.809(18)\\
DB3M7 & 7.2 & -0.794 & $0.3962(20)$ & $0.1014(11)$ & $0.4042(35)$ & $0.6412(78)$ & $0.0891(27)$ & $0.6584(83)$ & $0.611(12)$ & 0.797(21)\\
DB3M8 & 7.2 & -0.799 & $0.3764(17)$ & $0.0958(11)$ & $0.3780(22)$ & $0.6109(74)$ & $0.0833(41)$ & $0.6073(74)$ & $0.557(10)$ & 0.728(21)\\
DB3M9 & 7.2 & -0.803 & $0.3422(39)$ & $0.0896(21)$ & $0.3413(59)$ & $0.576(11)$ & $0.0787(25)$ & $0.571(20)$ & $0.530(24)$ & 0.722(28)\\
DB3M10 & 7.2 & -0.808 & $0.3181(47)$ & $0.0831(14)$ & $0.3189(19)$ & $0.520(13)$ & $0.0820(30)$ & $0.552(24)$ & $0.535(17)$ & 0.7036(98)\\
DB3M11 & 7.2 & -0.813 & $0.2796(41)$ & $0.0749(15)$ & $0.2815(57)$ & $0.435(19)$ & $0.0653(71)$ & $0.472(30)$ & $0.429(23)$ & 0.661(14)\\
\hline
DB4M3 & 7.4 & -0.74 & $0.3063(30)$ & $0.0741(13)$ & $0.3122(32)$ & $0.4913(91)$ & $0.0760(26)$ & $0.485(14)$ & $0.4715(65)$ & 0.652(18)\\
DB4M4 & 7.4 & -0.75 & $0.2487(43)$ & $0.0650(15)$ & $0.2452(78)$ & $0.4049(67)$ & $0.0645(26)$ & $0.4211(95)$ & $0.382(13)$ & 0.590(13)\\
DB4M5 & 7.4 & -0.755 & $0.2263(98)$ & $0.0546(36)$ & $0.2149(86)$ & $0.359(15)$ & $0.0577(96)$ & $0.383(24)$ & $0.385(74)$ & 0.627(16)\\
\hline
DB5M3 & 7.5 & -0.71 & $0.3009(23)$ & $0.0713(11)$ & $0.3041(22)$ & $0.4547(40)$ & $0.0677(58)$ & $0.4665(72)$ & $0.4409(53)$ & 0.553(15)\\
DB5M4 & 7.5 & -0.72 & $0.2622(40)$ & $0.0623(12)$ & $0.2588(59)$ & $0.4131(75)$ & $0.0644(29)$ & $0.424(11)$ & $0.410(23)$ & 0.556(14)\\
\hline\hline
\end{tabular}

\end{center}
\end{table}

\begin{figure}[t]
    \centering
    \includegraphics[width=0.4\textwidth]{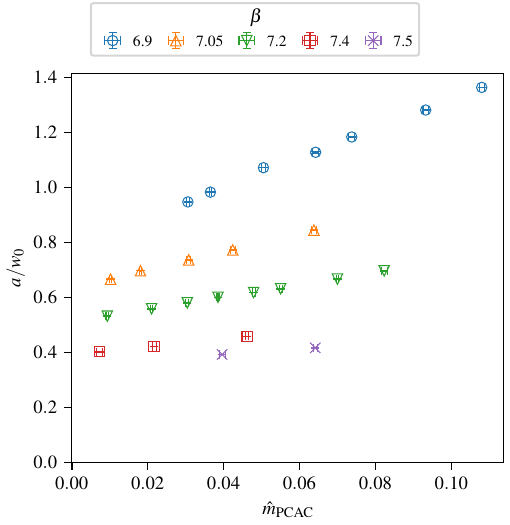}
    \includegraphics[width=0.4\textwidth]{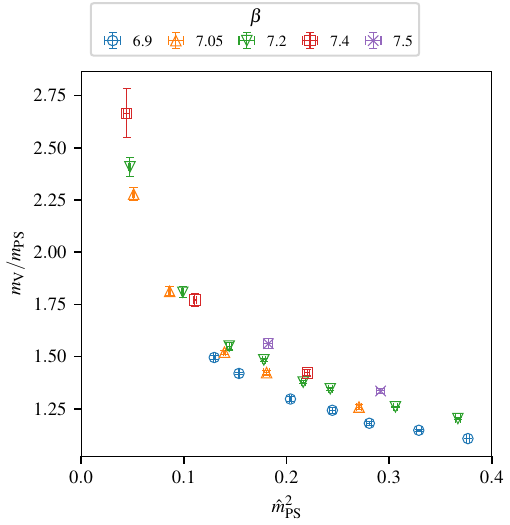}
    \caption{
    \label{fig_spec:latt_space}
    Left panel: inverse of the gradient flow scale, $a/w_0$, measured in all the ensembles considered in this work, as a function of the PCAC fermion mass, $\hat{m}_{\rm PCAC}\equiv m_{\rm PCAC}\, w_0$. The clover-leaf definition has been adopted in the energy density used for the gradient flow measurements, while the reference value $\mathcal{W}_0=0.28125$ has been used in the scale-setting procedure. Right panel: mass ratio between PS and V mesons as a function $\hat{m}_{\rm PS}^2$, in the same ensembles. }
\end{figure}

We measure the correlation functions in every single-meson channel by varying the choices  of $N_{\rm W}^{\rm src}$ and $N_{\rm W}^{\rm snk}$, and construct a correlation matrix, $\mathcal{C}_M(t)$, for each single meson operator, $M$.  We then define the GEVP, which can be written in terms of a square matrix with the size $(N^{\rm max}_{\rm W} /N^{\rm diff}_{\rm W} + 1)$, where $N^{\rm max}_{\rm W}$ is the maximum number of Wuppertal smearing steps, and $N^{\rm diff}_{\rm W} $ a choice of interval between smearing levels we retain. For each correlation matrix, $\mathcal{C}_M(t)$, the eigenvalues are computed by solving Eq.~(\ref{eq:gevp}), discussed in the body of the paper.
We extract the energy level of meson $M$ by fitting the corresponding eigenvalue with the functional form:
\begin{align}
\lambda_n^M(t) = A_n \left[e^{-E_n^M t} + e^{-E_n^M (T-t)}\right]\,.
\end{align}
The meson mass is obtained by fitting the lowest eigenvalue,  $m_M = E^M_0$.\footnote{As discussed in Sect.~\ref{sec:scattering}, the single -meson analysis described here has limitations.  For ensembles with light fermion mass, the scattering analysis is required to access the true ground state. We will comment in due time on our selection criteria in the choice of ensembles  for which the single-meson operator analysis is reliable, and hence the measurements can be included in subsequent analysis, for example in the extrapolation to the continuum limit.
See more details in Appendix~\ref{app:continuum}.}
We are also interested in the first excited state of the ${\rm V}$ meson, $E_1^{V}$, which is computed by fitting the next-to-lightest eigenvalue, $\lambda_1^M(t)$.

In the measurement  of the decay constants, we focus solely on PS, V, and AV mesons.
Their decay constants are defined by the following choice of parameterization of the matrix elements:
\begin{align}
\langle 0 |  \mathcal{O}_{\rm AV}| \text{PS} \rangle &= f_{\text{PS}} p_\mu, \\
\langle 0 | \mathcal{O}_{\rm V} | \text{V} \rangle &= f_{\text{V}} m_{\text{V}} \epsilon_\mu, \\
\langle 0 | \mathcal{O}_{\rm AV} | \text{AV} \rangle &= f_{\text{AV}} m_{\text{AV}} \epsilon_\mu,
\label{eq:decay_matrix_element}
\end{align}
where $p_\mu$ is the momentum of the meson, while $\epsilon_\mu$ denotes the polarization vector, which satisfies the transversality condition $\epsilon \cdot p = 0$. In the case of $f_{\rm PS}$, we introduce an additional two-point correlation function with different interpolating operators at the source and sink, 
\begin{align}
C_{\rm AV\text{-}PS}(t)&\equiv \sum_{\vec{x}} 
\langle 0 | [\overline{Q_1} \gamma_5 \gamma_\mu Q_2(\vec{x},t)]\,[\overline{Q_1} \gamma_5 Q_2(\vec{0},0)]^\dagger | 0\rangle, \\
&\xrightarrow{t\rightarrow\infty}
\frac{f_{\rm PS} \langle 0 | \mathcal{O}_{\rm PS} | {\rm PS} \rangle^*}{2}\left[
e^{-m_{\rm PS} t}-e^{-m_{\rm PS} (T-t)}\right].
\label{eq:axial_corr}
\end{align}
The matrix element $\langle 0 | \mathcal{O}_{\rm PS} | {\rm PS} \rangle^*$ can be obtained from $C_{\rm PS}(t)$ using Eq.~(\ref{eq:corr_ground}). 

In order to extract the matrix element defining the ${\rm PS}$ meson, we analyze two sets of correlation functions: one involving the smeared-source and smeared-sink operators, and one in which the smeared source is used with a point-like sink (with the same smearing level at the relevant source and sink). A simultaneous fit allows extracting the unsmeared matrix element. In our calculations, we consider $C_{\rm AV\text{-}PS}^{N_{\rm W}^{\rm max},0}$ and $C_{\rm PS}^{N_{\rm W}^{\rm max},N_{\rm W}^{\rm max}}$ for extracting the PS meson matrix element, while we measure $C_{M}^{N_{\rm W}^{\rm max},0}$ and $C_{M}^{N_{\rm W}^{\rm max},N_{\rm W}^{\rm max}}$  for $M={\rm V}$ and ${\rm AV}$.
We take into account multiplicative renormalization at the one-loop level in lattice perturbation theory for Wilson fermions, along with the tadpole improvement, as done in Ref.~\cite{Bennett:2019jzz} following the prescription of Ref.~\cite{Martinelli:1982mw} to
arrive at the renormalized decay constants.

We find it convenient to define the effective partially-conserved-axial-current (PCAC) mass, following Ref.~\cite{DelDebbio:2007wk}:
\begin{align}
\label{eq:eff_mpcac}
m^{\rm eff}_{\rm PCAC}(t) \equiv -\frac{m_{\rm PS}^{\rm eff}(t)}{\sinh m_{\rm PS}^{\rm eff}(t)}\left(
\frac{C_{\rm AV\text{-}PS}(t+1) - C_{\rm AV\text{-}PS}(t-1)}{4 C_{\rm PS}(t)}\right)\,,
\end{align}
where the effective PS meson mass, $m^{\rm eff}_{\rm PS}$, is given by
\begin{align}
m^{\rm eff}_{\rm PS}(t) = \cosh^{-1}\left({\frac{C_{\rm PS}(t+1)+C_{\rm PS}(t-1)}{2C_{\rm PS}(t)}}\right)\,.
\end{align}
For  this  calculation, we use the correlation functions involving point-like (not smeared) operators.  We perform a constant fit to the plateau appearing in $m_{\rm PCAC}^{\rm eff}(t)$ at large $t$, in order to extract the PCAC mass, $m_{\rm PCAC}$.

Extensive discussions  of our process in extracting  masses and matrix elements from two-point correlation functions can be found in Ref.~\cite{Bennett:2019jzz} and in Sect.~III of Ref.~\cite{TELOS:2025ash}. We summarize our results, expressed  in lattice units, in Tabs.~\ref{tab_spec:extraction_1} and~\ref{tab_spec:extraction_2}. To visualize the extent of the lattice parameter space explored in this work, we display  the PCAC mass, expressed  in units of the gradient flow, and the inverse of the gradient flow scale (the former a good proxy for the symmetry-breaking fermion mass, the latter for  the lattice spacing) in the left panel of Fig.~\ref{fig_spec:latt_space}, for each ensemble. The right panel of Fig.~\ref{fig_spec:latt_space} shows the mass ratio between the  V and PS mesons, as a function of $\hat{m}_{\rm PS}^2$, which provides an indication of  how close an ensemble is to the two-PNGB kinematic threshold, $\hat{m}_{\rm V}/\hat{m}_{\rm PS}=2$.

\begin{figure}
    \centering
    \includegraphics[width=0.3\textwidth]{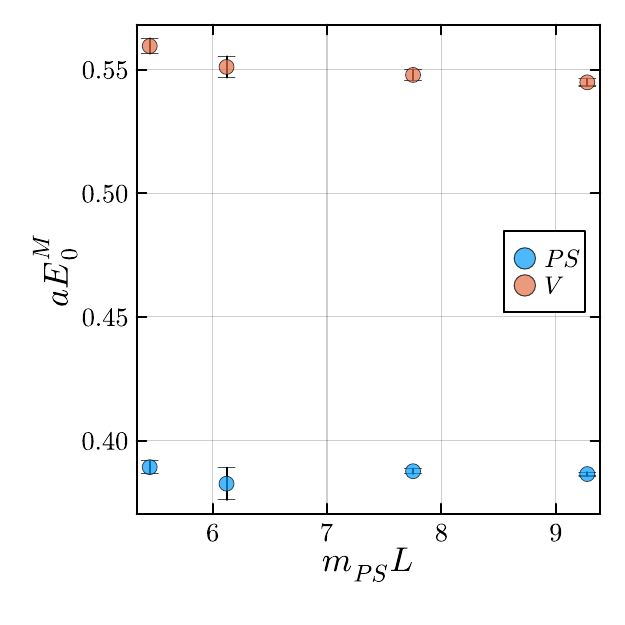}
    \includegraphics[width=0.3\textwidth]{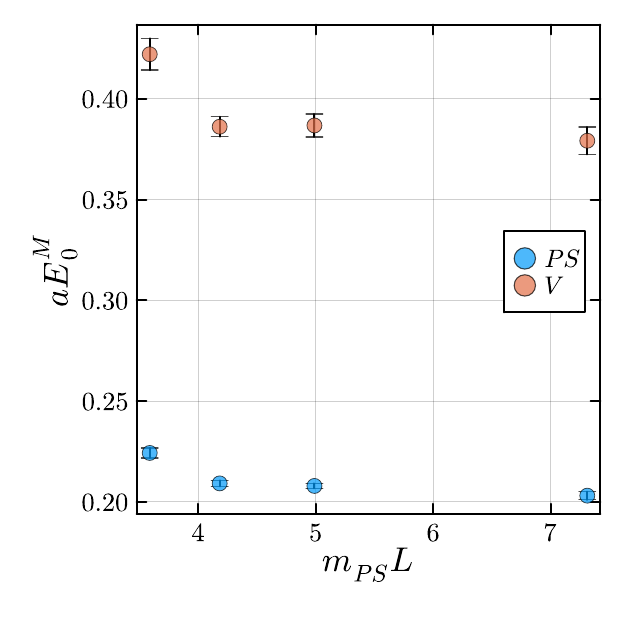}
    \includegraphics[width=0.3\textwidth]{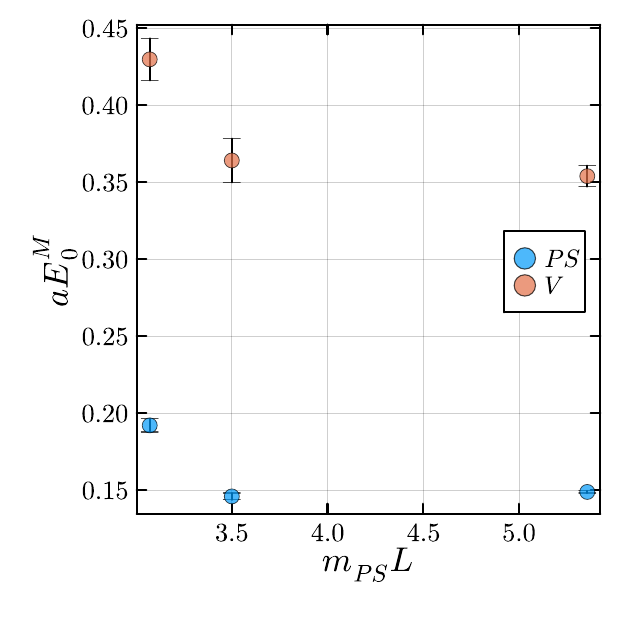}
    \caption{Zero-momentum, ground state energy levels, measured using single-meson operators, for the PNGBs, ${\rm PS}$ mesons transforming as $\textbf{5}$, and of the spin-1, ${\rm V}$ mesons transforming as $\textbf{10}$ of the global, unbroken $Sp(4)$ symmetry. The energies are expressed in lattice units, and are extracted from lattices with different choices of spatial volumes, as listed in Tab.~\ref{tab:ensembles}. Left to right: heavy, medium, and light ensembles. The lattice action parameters are those of  DB1M6 ($\beta=6.9$, $am_0=-0.92$), DB2M4 ($\beta=7.05$, $am_0=-0.863$), and DB2M5 ($\beta=7.05$, $am_0=-0.867$), respectively. }
    \label{fig:FV_effects}
\end{figure}

In order to assess the size of residual finite-volume effects, we use the three groups of ensembles in Table~\ref{tab:ensembles}, which allow comparing the measurements of the same spectroscopy data on different lattice volumes. As shown in Fig.~\ref{fig:FV_effects}, for the largest available volumes we see no significant evidence of such effects, given the size of our statistical errors. We report the  value of $m_{\rm PS}L$ in Tab.~\ref{tab_spec:extraction_1},  for all the ensembles used for this part of the spectroscopy study, showing that for some ensembles with light fermion masses we include in the analysis measurements with $m_{\rm PS}L\sim 3.5-7$, while for larger fermion masses all measurements have $m_{\rm PS}L\gtrsim 7$.  The measurements in  Fig.~\ref{fig:FV_effects} justify this choice. When the energy levels are displayed as a function of $m_{\rm PS}L$, the convergence to the physical (infinite) volume limit for the lighter ensembles appears to be fast. Chiral perturbation theory predicts that finite-volume correction, at leading order, are suppressed by a factor of ${\cal O}\left(\frac{m_{\rm PS}^2}{f_{\rm PS}^2}\right)$ at a given value of $m_{\rm PS}L$, besides the exponential suppression, of ${\cal O}\left(e^{-m_{\rm PS}L}/(m_{\rm PS}L)^{3/2}\right)$ \cite{Gasser:1986vb,Leutwyler:1987ak,Gasser:1987ah,Gasser:1987zq}. This observation might explain this trend, which 
has been observed also in lattice QCD---see, e.g.,  Fig.~13 in Ref.~\cite{Fodor:2012gf}. We conclude that we expect residual finite-volume effects to affect only negligibly our single-operator measurements, compared to other sources of uncertainty.

\subsection{Continuum extrapolation}
\label{app:continuum}

\begin{table}[t]
\caption{%
\label{tab_spec:fit_results_a2_am2}
Low energy constants obtained from continuum and massless extrapolations using the fitting functions in Eq.~(\ref{eq:con_fit_a2_am2}), inspired by W$\chi$PT, for the meson masses ($X=m$) and decay constants ($X=f$).
In respect to the results reported in the main body of the paper, in Tab.~\ref{tab_spec:fit_results-0}, the fitting ansatz differs by the additional inclusion of contributions ${\cal O}\left(\hat{a}\hat{m}_{\rm PS}^2\right)$.
We provide the reduced chi-square values, $\chi^2/{\rm N_{\rm d.o.f.}}$, as an indication of  the quality of the individual fits.
}
\begin{center}
\begin{tabular}{|c|c|c|c|c|c|c|}
\hline\hline
M & $\hat{m}_{M,\,\chi}^2$ & $L_{\rm M}^m$ & $W_{\rm M}^m$ & $R_{\rm M}^m$ & $C_{\rm M}^m$ & $\chi^2/{\rm N_{d.o.f}}$ \\
\hline\hline
$\rm V$ & $0.320(17)$ & $2.83(18)$ & $-0.175(40)$ & $0.027(27)$ & $-0.017(47)$ & $1.49$ \\
$\rm T$ & $0.331(20)$ & $2.78(17)$ & $-0.192(51)$ & $0.037(37)$ & $-0.040(68)$ & $1.62$ \\
$\rm S$ & $0.682(62)$ & $2.39(42)$ & $-0.09(15)$ & $-0.05(14)$ & $0.21(37)$ & $1.48$ \\
$\rm AV$ & $0.776(66)$ & $1.88(40)$ & $-0.03(17)$ & $-0.06(14)$ & $0.35(42)$ & $0.87$ \\
$\rm AT$ & $0.801(57)$ & $2.10(43)$ & $-0.07(15)$ & $-0.06(14)$ & $0.28(42)$ & $1.17$ \\
\hline\hline
\end{tabular}

\end{center}
\begin{center}
\begin{tabular}{|c|c|c|c|c|c|c|}
\hline\hline
M & $\hat{f}_{M,\,\chi}^2$ & $L_{\rm M}^f$ & $W_{\rm M}^f$ & $R_{\rm M}^f$ & $C_{\rm M}^f$ & $\chi^2/{\rm N_{d.o.f}}$ \\
\hline\hline
$\rm PS$ & $0.00561(34)$ & $3.76(22)$ & $-0.0035(12)$ & $0.00213(94)$ & $-0.0026(21)$ & $0.71$ \\
$\rm V$ & $0.0198(16)$ & $1.35(24)$ & $0.0027(42)$ & $-0.0003(30)$ & $-0.0036(64)$ & $1.01$ \\
$\rm AV$ & $0.0381(73)$ & $0.79(41)$ & $-0.055(23)$ & $0.033(17)$ & $-0.015(21)$ & $1.11$ \\
\hline\hline
\end{tabular}

\end{center}
\end{table}

\begin{figure}[t]
\begin{center}
\includegraphics[width=0.95\textwidth]{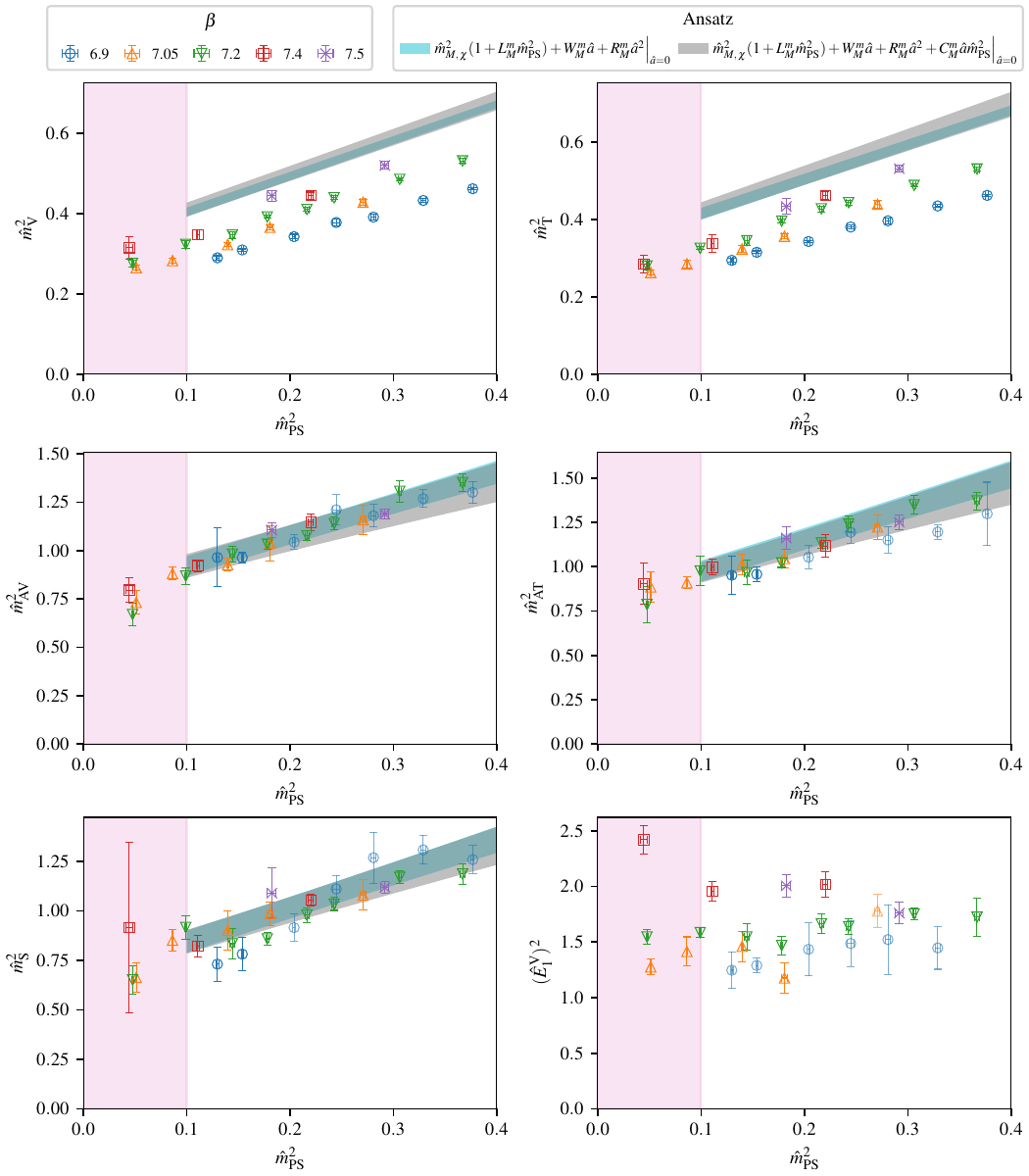}
\caption{%
\label{fig:spec_mass_con}%
Continuum extrapolation of the squared meson masses, $\hat{m}_M^2$, expressed in units of the gradient flow scale, $w_0$, plotted as a function of the squared pseudoscalar mass, $\hat{m}_{\rm PS}^2$.
Measurements have been performed using single-meson interpolating operators with Wuppertal and APE smearing.
From left to right, and top to bottom, we display the individual measurements of $\hat{m}^2_{\rm V}$, $\hat{m}^2_{\rm T}$, $\hat{m}^2_{\rm AV}$, 
$\hat{m}^2_{\rm AT}$,  $\hat{m}^2_{\rm S}$, and the first excited-state energy of vector meson, $(\hat{E}_1^{\rm V})^2$ (without continuum extrapolation). 
The symbols with different shapes (and colors) denote different $\beta$ values, as shown in the legend, while markers with reduced opacity indicate measurements satisfying $am_M > 1$, excluded from the extrapolations. The pink shaded region marks the small mass regime omitted in the extrapolation due to the limitation of single-meson operators in accessing the ground states. 
The cyan and gray bands represent the continuum-extrapolated results ($\hat{a}=0$) with using Eq.~(\ref{eq:fit_m}) and Eq.~(\ref{eq:con_fit_a2_am2}), respectively.
}
\end{center}
\end{figure}

\begin{figure}[t]
\begin{center}
\includegraphics[width=\textwidth]{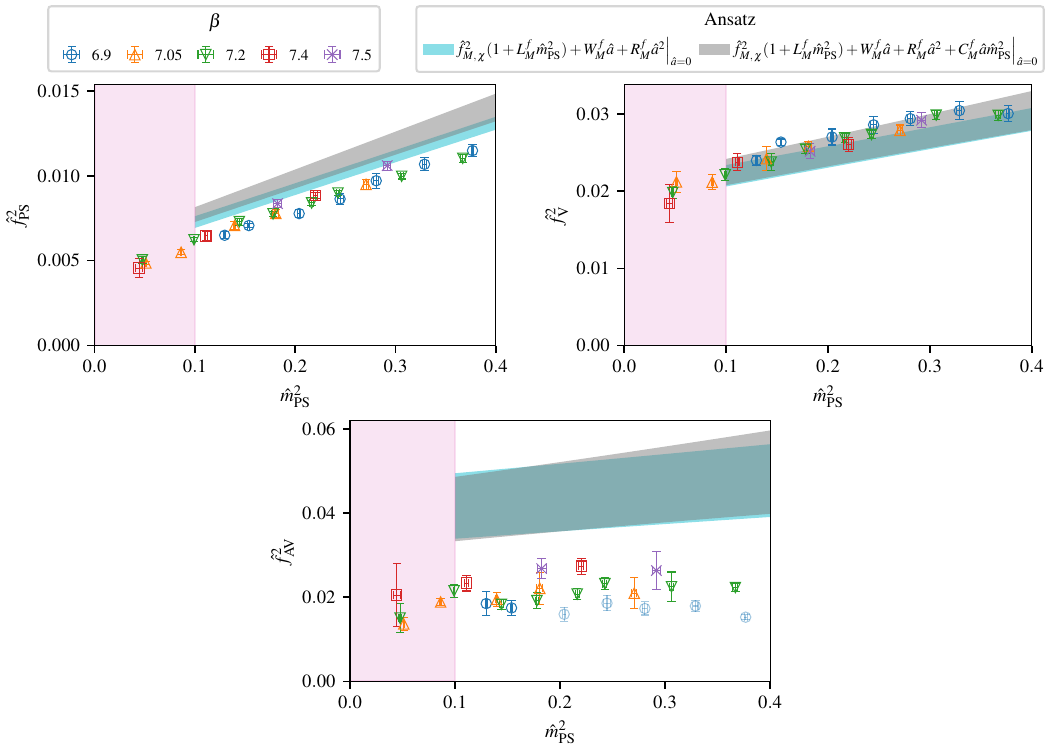}
\caption{%
\label{fig:spec_decay_con}%
Continuum extrapolation of the squared meson decay constants, $\hat{f}_M^2$, expressed in units of the gradient flow scale, $w_0$, as a function of the squared pseudoscalar mass, $\hat{m}_{\rm PS}^2$.
Measurements have been performed using single-meson interpolating operators with Wuppertal and APE smearing.
The top-left, top-right, and bottom panels display $\hat{f}^2_{\rm PS}$, $\hat{f}^2_{\rm V}$, and $\hat{f}^2_{\rm AV}$, respectively.
The symbols with different shapes (and colors) denote different $\beta$ values, as shown in the legend, while markers with reduced opacity indicate measurements satisfying $am_M > 1$ that were excluded from the extrapolations. The pink shaded region marks the small mass regime omitted in the extrapolation due to the limitation of single-meson operators in accessing the ground states. 
The cyan and gray bands represent the continuum-extrapolated results ($\hat{a}=0$) with using Eq.~(\ref{eq:fit_f}) and Eq.~(\ref{eq:con_fit_a2_am2}), respectively.
}
\end{center}
\end{figure}

To remove discretization artifacts from our measurement of meson masses and decay constants extracted from correlation functions involving only single-meson operators, as discussed in Sect.~\ref{Sec:spectra}, we perform continuum extrapolations with fitting ansätze inspired by W$\chi$PT~\cite{Sheikholeslami:1985ij,Rupak:2002sm,Sharpe:1998xm,Symanzik:1983dc, Luscher:1996sc}.
In this Appendix, we discuss the procedure in detail, in particular we explain our data selection criteria and ansatz choices.

Let us start by explaining our selection of ensembles entering the extrapolations.  Firstly, we exclude data points above the lattice cut-off, $am_{M} > 1.0$ This restriction affects the measurements of heavy mesons, as shown in Tab.~\ref{tab_spec:extraction_2}.  Second, we observe in Fig.~\ref{fig_spec:latt_space} that in the three lightest ensembles the mass of the ${\rm V}$ meson appears to be above the two-PNGB threshold. However, as discussed in the main text, the single-meson operator is not sufficient to access the true ground state for these light ensembles---see, e.g., the bottom panel of Fig.~\ref{fig:E_L_10_non_res_and_res} for ensemble DB2M5. Even for the ensembles in which the mass of the ${\rm V}$ meson is slightly below, yet close to the threshold, similar considerations apply, as shown in the top panel of Fig.~\ref{fig:E_L_10_non_res_and_res} for ensemble DB2M4. We therefore exclude such light ensembles in the continuum and massless extrapolations to avoid potential systematic errors arising from the limitation of the single-meson operators in accessing the correct ground state. More specifically, we restrict our analysis to $\hat{m}_{\rm PS}^2\geq 0.1$---see also Fig.~\ref{fig:mass_con_summary-0}. 

Even after imposing these conservative restrictions on our ensembles we have available an extended and improved set of measurements, in respect to the literature~\cite{Bennett:2019jzz}. We hence perform our extrapolation to the continuum limit by considering a more general ansatz for the functional form of the fitting functions, which  extends beyond the 
NLO, W$\chi$PT-inspired expressions  involving terms linear in the lattice spacing, $\hat{a}$, and the PNGB mass squared, $\hat{m}_{\rm PS}^2$. We have considered three possible corrections, appearing at ${\cal O}\left(\hat{m}_{\rm PS}^4\right)$, ${\cal O}\left(\hat{a}^2\right)$, and ${\cal O}\left(\hat{a}\hat{m}_{\rm PS}^2\right)$, respectively. 
 
Having restricted attention  to the range of mass with $\hat{m}_{\rm PS}^2\leq 0.4$, we find that the addition of terms such as ${\cal O}\left(\hat{m}_{\rm PS}^4\right)$ does not yield any meaningful changes in the fitting results, in particular the central value for these coefficients are compatible with zero, and the fitting function is unaffected over the whole range of available masses. 

Conversely, we find that we must retain an additional term ${\cal O}\left(\hat{a}^2\right)$. The 4-parameter fits of the eight observables of interest (meson masses and decay constants) are displayed in the main body of the paper,  in Sect.~\ref{Sec:spectra}. Along with our best-fit results, in Tab.~\ref{tab_spec:fit_results-0}  we show also the reduced chi-square values, $\chi^2/N_{\rm d.o.f.}$, which indicate that this ansatz provides a good description of the data, in the range of parameters explored with our ensembles.

We also show here the results of the analysis obtained  by adding corrections ${\cal O}\left(\hat{a}\hat{m}_{\rm PS}^2\right)$, which gives an ansatz depending on five parameters, for each of the eight observables, and we write as
\begin{align}
\label{eq:con_fit_a2_am2}
\hat{X}^{2, {\rm NLO}}_M = \hat{X}_{M,\,\chi}^2(1+L_{M}^X \hat{m}_{\rm PS}^2)+W_{M}^{X} \hat{a}+R_{M}^{X} \hat{a}^2 + C_M^X \hat{a}\hat{m}_{\rm PS}^2\,,
\end{align}
where $X = m,\,f$ denotes the mass and decay constant, respectively.
Our best fit results and uncertainties for these low energy constants are shown in Tab.~\ref{tab_spec:fit_results_a2_am2}, together with the corresponding reduced chi-square values.  Compared to the results in Table~\ref{tab_spec:fit_results-0}, determined by using the fit ansatz in Eqs.~(\ref{eq:fit_m}) and~(\ref{eq:fit_f}), we observe that the extrapolated masses and decay constants in the continuum-and-massless limit, $\hat{m}_{M,\chi}$ and $\hat{f}_{M,\chi}$, are consistent with each other.

The extrapolation procedures for masses and decay constants  are summarized in Figs.~\ref{fig:spec_mass_con} and~\ref{fig:spec_decay_con}, respectively.
The squared quantities, $\hat{m}_{\rm M}^2$  and $\hat{f}_{\rm M}^2$,  from each ensemble are plotted as a function of $\hat{m}_{\rm PS}^2$, with different symbols (and colors) corresponding to the $\beta$ values indicated in the legends. 
In the case of  heavy mesons with $am_M > 1$, which are excluded from the continuum extrapolations, we show the measurements with reduced opacity.  The pink shaded region in the small mass regime identifies ensembles excluded from the extrapolation analysis, due to the limitation of the single-meson operators, as discussed above. 
In Fig.~\ref{fig:spec_mass_con}, we also present the results for the first excited states of ${\rm V}$ meson, $(\hat{E}^{\rm V}_1)^2$, extracted from the GEVP analysis.  Because the measured values of $\hat{E}_1^{\rm V}$ not only suffers from  large systematic uncertainties intrinsic in the methodology, but are also found to be large, with values above the lattice cut-off, we do not include them in the  continuum extrapolation. 

In the figures, we also display the difference between the two fitting ansatz, over the whole range of parameters.
The cyan bands represent the continuum-extrapolated results ($\hat{a}=0$) obtained using Eqs.~(\ref{eq:fit_m}) and (\ref{eq:fit_f}), with the low-energy constants provided in Tab.~\ref{tab_spec:fit_results-0}. 
The gray bands are obtained by using  the low-energy constants in Tab.~\ref{tab_spec:fit_results_a2_am2}, extracted with the more general ansatz defined in Eq.~(\ref{eq:con_fit_a2_am2}), which includes the additional $\hat{a}\hat{m}_{\rm PS}^2$ term. It can be clearly seen that these two  ansatz yield compatible results, over the whole extent of the available mass range, which justifies our choices of identifying as our final results  the fit functions in Eqs.~(\ref{eq:fit_m}) and (\ref{eq:fit_f}).

\bibliographystyle{JHEP}
\bibliography{bibliography}

\end{document}